\documentclass[fleqn,usenatbib,useAMS]{mnras}
\extrafloats{100}

\usepackage{graphicx}	
\usepackage{amsmath}	
\usepackage{amssymb}	
\usepackage{multicol}        
\usepackage{bm}		
\usepackage{pdflscape}	

\usepackage{longtable}

\newcommand{\kms}{\,km\,s$^{-1}$} 
\newcommand{\cms}{\,(cm\,s$^{-2}$)} 
\newcommand{\cd}{d\,$^{-1}$} 

\usepackage[dvipsnames,usenames]{color}

\defcitealias{2021MNRAS.506.1073H}{Paper\,I}

\usepackage[T1]{fontenc}
\usepackage{ae,aecompl}

\usepackage{newtxtext,newtxmath}

\title[{\it TESS} Cycle\,2 roAp stars]{{\it TESS} Cycle\,2 observations of roAp stars with 2-min cadence data}

\author[D.\,L.\,Holdsworth et al.]{D.\,L.\,Holdsworth,$^{1,2}$\thanks{e-mail:dlholdsworth@uclan.ac.uk}
M.~S.~Cunha,$^{3}$
M.~Lares-Martiz,$^{4}$
D.~W.~Kurtz,$^{5,1}$
V.~Antoci, $^{6}$
S.~Barcel\'o Forteza,$^{7}$\and
P.~De Cat,$^{8}$
A.~Derekas,$^{9,10,11}$
C.~Kayhan,$^{12}$
D.~Ozuyar,$^{13}$
M.~Skarka,$^{14,15}$
D.~R.~Hey,$^{16}$
F.~Shi,$^{17,18}$\and
D.~M.~Bowman,$^{19,20}$
O.~Kobzar,$^{21}$
A.~Ayala G\'omez,$^{22}$
Zs.~Bogn\'ar,$^{23}$
D.~L.~Buzasi,$^{24}$
M.~Ebadi,$^{25}$\and
L.~Fox-Machado,$^{26}$
A.~Garc\'{\i}a Hern\'andez,$^{27}$
H.~Ghasemi,$^{28}$
J.~A.~Guzik,$^{29}$
R.~Handberg,$^{30}$
G.~Handler,$^{31}$\and
A.~Hasanzadeh,$^{32}$
R.~Jayaraman,$^{33}$
V.~Khalack,$^{21}$
O.~Kochukhov,$^{34}$
C.~C.~Lovekin,$^{35}$
P.~Miko\l{}ajczyk,$^{36,37}$\and
D.~Mkrtichian,$^{38}$
S.~J.~Murphy,$^{39}$
E.~Niemczura,$^{36}$
B.~G.~Olafsson,$^{1}$
J.~Pascual-Granado,$^{40}$
E.~Paunzen,$^{41}$\and
N.~Posi\l{}ek,$^{36}$
A.~Ram\'on-Ballesta,$^{42}$
H.~Safari,$^{43}$
A.~Samadi-Ghadim,$^{44}$
B.~Smalley,$^{45}$
\'A.~S\'odor,$^{23}$\and
I.~Stateva,$^{46}$
J.~C.~Su\'arez,$^{47}$
R.~Szab\'o,$^{23,48}$
T.~Wu,$^{49,50,51,52,53,54}$
E.~Ziaali,$^{55}$
W.~Zong$^{56,57}$
and S.~Seager$^{58,59,60}$
\\
Author affiliations shown in Appendix\,\ref{app:affiliations}.
}
\date{\today}

\pubyear{2024}

\begin{document}
\label{firstpage}
\pagerange{\pageref{firstpage}--\pageref{lastpage}}
\maketitle

\begin{abstract}
We present the results of a systematic search of the \textit{Transiting Exoplanet Survey Satellite} ({\it TESS}) 2-min cadence data for new rapidly oscillating Ap (roAp) stars observed during the Cycle\,2 phase of its mission. We find seven new roAp stars previously unreported as such and present the analysis of a further 25 roAp stars that are already known. Three of the new stars show multiperiodic pulsations, while all new members are rotationally variable stars, leading to almost 70\,per\,cent (22) of the roAp stars presented being $\alpha^2\,$CVn-type variable stars. We show that targeted observations of known chemically peculiar stars are likely to overlook many new roAp stars, and demonstrate that multi-epoch observations are necessary to see pulsational behaviour changes. We find a lack of roAp stars close to the blue edge of the theoretical roAp instability strip, and reaffirm that mode instability is observed more frequently with precise, space-based observations. In addition to the Cycle\,2 observations, we analyse {\it TESS} data for all known roAp stars. This amounts to 18 further roAp stars observed by {\it TESS}.  Finally, we list six known roAp stars that {\it TESS} is yet to observe. We deduce that the incidence of roAp stars amongst the Ap star population is just 5.5\,per\,cent, raising fundamental questions about the conditions required to excite pulsations in Ap stars. This work, coupled with our previous work on roAp stars in Cycle\,1 observations, presents the most comprehensive, homogeneous study of the roAp stars in the {\it TESS} nominal mission, with a collection of 112 confirmed roAp stars in total. 
\end{abstract}




\section{Introduction}

The \textit{Transiting Exoplanet Survey Satellite} \citep[{\it TESS};][]{2015JATIS...1a4003R} is providing a rich photometric data set for millions of stars in most regions of the sky. While the primary mission focus is on the detection of exoplanet transits, the acquired data are well suited for studies of stellar variability. {\it TESS} observes the sky in strips of 24$^\circ \times 96^\circ$ for about 27\,d each, referred to as \textit{sectors}. {\it TESS} is observing in 1-year {\it cycles}, each composed of 13 sectors. For the nominal mission comprising Cycles 1 \& 2, {\it TESS} recorded data in two cadences: 30\,min for the Full Frame Images (FFIs) and 2\,min for a selection of $\sim$20\,000 stars per sector. These 2-min cadence data provide a data set with which it is possible to perform a homogeneous search for high-frequency stellar variability. We have previously exploited this capability to search for high-frequency pulsations in the chemically peculiar Ap stars \citep[][hereafter Paper\,I]{2021MNRAS.506.1073H} in the Cycle\,1 data. Here we extend this study to encompass the 2-min Cycle\,2 observations.

The rapidly oscillating Ap (roAp) stars are a rare subset of the chemically peculiar, magnetic, Ap stars \citep{1982MNRAS.200..807K}. To date there are of the order 100 known in the literature \citep[see e.g., ][for catalogues and lists]{2015MNRAS.452.3334S,2016A&A...590A.116J,2019MNRAS.487.3523C,2021MNRAS.506.1073H,2021MNRAS.504.1370A}, discovered through targeted high-speed ground-based photometry and spectroscopy \citep[e.g., ][]{1991MNRAS.250..666M,2018RAA....18..135P,2008MNRAS.385.1402F,2013MNRAS.431.2808K} and space-based survey data \citep[e.g., ][]{2011A&A...530A.135G,2011MNRAS.413.2651B,2019MNRAS.488...18H,2019MNRAS.487.3523C}. 

The broader population of Ap stars is characterised by strong, global, magnetic fields \citep{2017A&A...601A..14M} that suppress near-surface convection in the stellar atmosphere \citep{1981A&A...103..244M,2009A&A...495..937L,2010A&A...516A..53A}, leading to gravitational settling of some light elements and radiative levitation of metals like P, Si, Ti, Fe, Cr and Mn \citep{2015IAUS..307..383K,2015MNRAS.453.3766L,2017MNRAS.471..926K,2018MNRAS.477.3390N,2022A&A...658A.105F} and rare earth elements like Pr, Nd, Eu and Y, which can reach atmospheric abundances of a million times that of the Sun \citep[e.g.,][]{2010A&A...509A..71L,2017A&A...601A.119C,2018MNRAS.477..882K}.  These levitated chemical elements create persistent atmospheric anomalies that may be stable for at least a century \citep[e.g.,][]{2016MNRAS.455.2567B}. These anomalies, i.e. chemical spots, modulate the observed flux, enabling the accurate determination of the stellar rotation period. Rotationally variable Ap stars are known as $\alpha^2$\,CVn stars.

The roAp stars are characterised by high overtone ($n\gtrsim15$), low degree ($\ell\lesssim3$) pressure (p) mode pulsations with frequencies in the range $0.7-3.6$\,mHz ($60-310$\,\cd; $P=4.7-23.6$\,min). In many non-magnetic stars, the pulsation axis is aligned with the rotation axis as this is the axis of maximum deformation from spherical symmetry. In the roAp stars, however, the pulsation axis is closely aligned with the magnetic axis \citep[e.g.,][]{2004ApJ...615L.149K} which in turn is inclined to the rotation axis. Such a geometry means the pulsations are viewed from varying aspects as the star rotates, leading to the oblique pulsator model \citep[e.g.,][]{1982MNRAS.200..807K,2004MNRAS.350..485S}, where the mode amplitude and phase appear modulated over the stellar rotation period. In a Fourier spectrum of a light curve, this results in a multiplet with $2\ell+1$ components for a non-distorted mode, with the components split from the central mode frequency by exactly the stellar rotation frequency. With a geometrically distorted mode, the multiplet shows more components \citep[e.g.,][]{2016MNRAS.462..876H} and the pulsation amplitude never reaches zero as a node crosses the line of sight. The distortion arises from the magnetic field's interaction with the pulsation mode's eigenfunction, which then requires higher order spherical harmonics to describe the mode \citep[see][and references therein for discussions]{1992MNRAS.259..701K,2018MNRAS.473...91H,2021MNRAS.506.5629S}. 

The driving mechanism for the pulsations in these stars is still uncertain. In many cases, theoretical models can recreate the observations by employing the opacity ($\kappa$) mechanism acting in the hydrogen ionisation layers in regions of the star where the magnetic field suppresses convection \citep{2001MNRAS.323..362B}. However, this method cannot explain all of the observed frequencies, especially those at very high frequency, where instead it is suggested that turbulent pressure plays a role in the mode excitation \citep{2013MNRAS.436.1639C}.

The most promising stars for asteroseismic insight are those that show multiperiodic variability. In a non-magnetic star, the high frequency pulsations occur in the asymptotic regime, forming a series of alternating odd and even degree modes \citep{1979PASJ...31...87S,1980ApJS...43..469T} that allow the use of asteroseismic techniques to constrain the stellar properties, e.g., \citet{2021A&A...650A.125D}. However, the situation is complicated in some roAp stars as the pulsations become magneto-acoustic in the outer layers of the star, and their frequencies suffer from magnetic perturbation \citep{2000MNRAS.319.1020C,2006MNRAS.365..153C,2004MNRAS.350..485S}. Detecting such phenomena in the roAp stars helps to drive forward the theoretical modelling of the class, and provide the opportunity to fully understand roAp stars.

Recently, the validity of the roAp class was questioned by \citet{2022MNRAS.510.5743B}, who suggested that all roAp stars are just $\delta$\,Sct stars. However, this is clearly not the case. While there is overlap in the frequency regime where both $\delta$\,Sct pulsations and roAp pulsations are found (in different stars), the roAp stars are distinct from the $\delta$\,Sct stars given their Ap chemical peculiarities, oblique pulsations, and strong, stable, magnetic fields. There are some cases where roAp stars do show low-frequency $\delta$\,Sct pulsations \citep[e.g.,][]{2020MNRAS.498.4272M}, and new results where a magnetic field has been confirmed in an Ap star potentially showing $\delta$\,Sct pulsations \citep[e.g.,][]{2015MNRAS.454L..86N,2016A&A...588A..71E,2023MNRAS.526L..83H}. While the former work rules out a companion to the roAp star, they place limits on the magnetic field strength that can suppress low-overtone pulsations. The latter work cautions that an undetected companion to the Ap star may provide the $\delta$\,Sct pulsations and that their derived field strength is close to the limit provided by \citet{2020MNRAS.498.4272M}, thus no certain conclusions can be drawn on the presence of $\delta$\,Sct modes in a magnetic Ap star. The suggestion, therefore, that the roAp classification should be abandoned is not useful and is rejected. It is only with careful study and interpretation of valid evidence that stars should be admitted to the roAp class of pulsating variable.

\section{Data sample and search strategy}
\label{sec:datasample}

To present a homogeneous sample of roAp stars observed by {\it TESS} during its nominal mission, we followed the same procedure as in \citetalias{2021MNRAS.506.1073H}. One team, Team\,1, took the 2-min target list for each sector of observations from {\it TESS}\footnote{https://tess.mit.edu/observations/target-lists/} which consisted of 20\,000 stars per sector. These lists were then crossmatched with version 8 of the {\it TESS} Input Catalogue \citep[TIC;][]{2019AJ....158..138S} with an 8\,arcsec radius, giving us access to the secondary information in the TIC to refine our list of targets. Using only the $T_{\rm eff}$  parameter, we selected stars with temperatures of 6000\,K or greater (including stars with a `null' value). This amounted to about 7500 stars per sector, with 93\,163 light curves in total, corresponding to 36\,209 unique stars (with 15\,675 stars observed in more than one sector). The light curves for each star in all available sectors were downloaded from the Mikulski Archive for Space Telescopes (MAST) server. These data have been processed with the Science Processing Operations Center (SPOC) pipeline \citep{2016SPIE.9913E..3EJ}, which are the most suitable data for the investigation of stellar variability \citep{2022A&A...666A.142S}. In the following we used the Pre-search Data Conditioning Simple Aperture Photometry (PDC\_SAP) data unless otherwise stated.

Initially, two independent analysis techniques were implemented for the search of pulsational variability. For one team, Team 1, each individual light curve was automatically prewhitened of all frequencies up to 0.23\,mHz (20\,\cd) to an amplitude limit of the approximate noise level between $2.3-3.3$\,mHz ($200-300$\,\cd) to remove instrumental artefacts and any low-frequency signal whose window function affects the noise at high frequency. If multiple sectors for a target were available, the prewhitened sectors were then combined. An amplitude spectrum of the light curve was calculated to the Nyquist frequency of 4.2\,mHz (360\,\cd), which also included a calculation of the false alarm probability (FAP). 

From these amplitude spectra, we selected stars that showed peaks with frequencies $>0.52$\,mHz ($>45$\,\cd) and a corresponding FAP of $<0.1$. The amplitude spectra of these stars were then plotted for visual inspection. This totalled 6713 stars, of which 2125 stars have multi-sector observations. 

A second team, Team 2,  used a complementary method which calculated the skewness of the amplitude spectrum \citep{2019MNRAS.485.2380M} of the MAST PDC\_SAP data above 0.46\,mHz (40\,\cd). Where multiple sectors of data were available, they were combined to a single light curve. If the skewness was greater than 5, then all peaks in the amplitude spectrum with FAP values below 0.05 were extracted and the star flagged as variable for later human inspection. This method produced a total of 189 variable star candidates.

These two lists were then combined to produce a master list.  An initial pass was made with the {\sc{echo}} code \citep[described in][]{2019MNRAS.490.4040A} to flag possible false positive detections with human inspection of all produced outputs (light curves, Fourier transform plots and detected peaks). A detection was determined to be false positive if the signal-to-noise ratio (S/N) of the high-frequency signal was below 4.0; if there was obvious contamination from a low-frequency harmonic series; or if the amplitude spectrum displayed obvious characteristics of either $\delta$\,Sct stars (with many modes in the $\approx0.2-1.0$\,mHz range, e.g., \citealt{2018MNRAS.476.3169B}), sdBV stars (stars with pulsations in low and high-frequency ranges) or pulsating white dwarfs (modes with amplitudes significantly greater than those known in roAp stars). The final sample consisted of 163 stars.

Further, we solicited input from the {\it TESS} Asteroseismic Science Consortium (TASC) Working Group 4 (WG4; AF-stars). This ensured that our study was not biased by the two methodologies outlined above. Contributions from the WG4 are as follows:

\begin{itemize}
\item {\it Team A} used the {\it TESS} Input Catalog \citep{2018AJ....156..102S} and {\it Gaia} DR2 Catalog \citep{2018A&A...616A...8A} to look for all stars with SIMBAD spectral types between B8 and F2 (stars with no classification were retained) in the Northern ecliptic hemisphere with effective temperatures between 6000\,K and 10\,000\,K, luminosity between 5 and 100\,L$_{\odot}$ and a contamination ratio lower than 0.01. Using the interpolation and frequency analysis routine of \citet{2015A&A...579A.133B}, the team found 121 candidates where the highest amplitude peak was in the frequency range $0.7-4.0$\,mHz      and without lower frequency modes in order to avoid the $\delta$\,Sct star pulsation regime \citep[e.g.,][and references therein]{2010aste.book.....A,2020A&A...638A..59B}.

\item {\it Team B} used the Cycle\,2 data to search for targets with a magnitude brighter than $T =10$\,mag and an effective temperature of >6000\,K. The resultant SAP light curves were used to derive rotation periods which were then removed from the data before pulsation frequencies were searched for using  the Lomb-Scargle method. This procedure resulted in 9 candidates.

\item {\it Team C} considered stars located close to the northern ecliptic pole (within $15^\circ$), which were brighter than $T=11$\,mag, with temperatures in the range $6000 <T_{\rm eff}< 10\,000$\,K \citep{2022A&A...666A.142S}. Frequency spectra, in the range $0-4.2$\,mHz ($0-360$\,\cd) were created from the PDC\_SAP data obtained from the MAST and searched for significant peaks. In total, 1600 stars showed peaks in the frequency range $0.9-4.2$\,mHz ($80-360$\,\cd) with a S/N~$>4$. However, only stars with peaks with S/N~$>5$ were selected as good candidates (8 stars).  

\item {\it Team D} provided 55 roAp candidates using the {\it TESS}-Ap procedure \citep{2021mobs.confE...7K}. A search for pulsations in the frequency range $1.2-4.2$\,mHz ($100-360$\,\cd) was conducted to limit the inclusion of $\delta$\,Sct stars. Only stars in the effective temperature range $6000-10\,000$\,K were considered. Peaks were considered real if the S/N~$> 4$. After the automatic selection, manual inspection was conducted to reject false positives. 

\item {\it Team E} used the TIC to select all stars in the temperature range $6000-10\,000$\,K for which 2-min cadence data were available. The light curves were downloaded from the MAST server, from which the PDC\_SAP flux data were extracted. Only the first $\sim12.5$\,d of data were selected to avoid the central downlink gap. Any other gaps are much smaller in size and a piecewise cubic Hermite polynomial was used to fill them with interpolated data at 2-min cadence. No detrending of low-frequency variability was performed. The sampling intervals were then regularised in order to obtain an evenly sampled time series that was analysed using an FFT to calculate the amplitude spectrum to the Nyquist frequency. A star is selected to be variable when a local maximum is found in the amplitude spectrum to have a S/N~$>5$ and a frequency greater than 0.5\,mHz ($\sim43$\,\cd). This procedure returned 1241 stars which were vetted by eye to produce a final list of 1134 variable candidates. 

\item Finally, 8 additional targets were submitted from members of WG4 where individuals had made private note of stars of interest.
\end{itemize}

All of these inputs from WG4 members were cross-checked with the results from Team 1 and Team 2 to avoid duplication. Team 1 then re-vetted all unique submissions to reject obvious non-roAp variable stars to produce a final sample list of 116 stars.This final sample was redistributed amongst the members of the TASC WG4 for detailed analysis to confirm the presence of a positive roAp detection, and to extract rotation periods from the light curves. 

This process identified seven new roAp stars previously unreported as such in the literature (Sec.\,\ref{sec:New_roAp}) and nine roAp stars discovered through {\it TESS} observations (Sec.\,\ref{sec:TESS_roAp}). There were also positive detections of pulsations in 12 roAp stars known prior to the launch of {\it TESS} (Sec.\,\ref{sec:Known_roAp}), with four roAp stars where {\it TESS} did not detect pulsational variability. These four stars were originally identified in {\it Kepler} data which has a lower noise threshold than the {\it TESS} data analysed here. Finally, we present seven roAp candidate stars where the observational and archival data do not allow us to conclude with certainty that the star is a true roAp star. 

\section{Spectroscopic observations}
\label{sec:spec}

\subsection{LAMOST data}

The Large Sky Area Multi-Object Fiber Spectroscopic Telescope (LAMOST) is a 4-m class telescope located at the Xinglong observatory (China) that is operated and managed by the National Astronomical Observatories, Chinese Academy of Sciences \citep{2012RAA....12.1197C}. Its scientific observations started in October 2011. With its quasi-meridian transit configuration, it is possible to observe northern stars with a declination above $\sim10^\circ$ from two hours before to two hours after meridian passage. This unique instrument combines a large aperture (primary mirror of $5.72\,\rm{m}\times4.40$\,m; secondary mirror of $6.67\,\rm{m}\times6.05$\,m) with a large field of view ($\sim$20\,deg$^2$). This circular field with a diameter of $5^\circ$ is homogeneously covered with 4000 fibres, connected to 16 spectrographs with 250 fibres each. Currently, observations can be done either in the full optical range ($3700-9000$\,\AA) in the low-resolution mode (LAMOST-LRS; $R\sim1800$; \citealt{2012RAA....12..723Z}) or in a blue ($4950-5350$\,\AA) and red ($6300-6800$\,\AA) wavelength range in the medium-resolution mode (LAMOST-MRS; $R\sim7500$; \citealt{2020arXiv200507210L}). The LAMOST-LRS observations are single shot observations while the LAMOST-MRS observations are done in two major groups of observations\citep[for further details see][]{2020arXiv200507210L,2020ApJS..251...15Z}. We checked the LAMOST database and found data for six of our targets. Four targets had one spectrum each (TIC\,21024812, TIC\,101624823, TIC\,118247716, and TIC\,467074220) in LAMOST-LRS and two targets had groups of observations in LAMOST-MRS (TIC\,26833276; 1 group of spectra, TIC\,259017938; 6 groups of spectra). The pipeline-reduced spectra \citep{2015RAA....15.1095L} from data release 9 (DR9, v1.0)\footnote{\url{http://www.lamost.org/dr9/}} were used to check if typical chemical peculiarities expected for Ap stars can be identified in the data.

\subsection{HERMES data}

The High Efficiency and Resolution Mercator Echelle Spectrograph (HERMES; \citealt{2011A&A...526A..69R}) is mounted to the 1.2-m Mercator telescope at the Observatorio del Roque de los Muchachos on La Palma in the Canary Islands, Spain, and is operated by KU Leuven\footnote{\url{http://www.mercator.iac.es/instruments/hermes/}}. HERMES is a fibre-fed prism-cross-dispersed echelle spectrograph with an effective wavelength coverage between 3770 and 9000\,\AA, and in its standard high-resolution mode has a spectral resolving power of $\sim$85\,000. Reduced spectra are automatically produced by an efficient {\sc Python} pipeline, {\sc HermesDRS} (v7.0), which includes bias subtraction and flat fielding, as well as wavelength calibration using a ThArNe spectrum, echelle order merging, and removal of cosmic rays. With its Northern hemisphere observing window, as of semester 2018b, we embarked upon a large programme (program 85; PI Bowman) to assemble high signal-to-noise (S/N > 100), multi-epoch (minimally two epochs), high-resolution HERMES spectra of all bright ($V \leq 10$~mag) Ap stars in the catalogue of \citet{2009A&A...498..961R}. The programme ran for two years and included more than 450 Ap stars in total with $\delta > -30^\circ$. Thus the majority of observed Ap stars are located in the Northern hemisphere ensuring complementarity with {\it TESS} Cycle\,2 observations. Three stars presented in this work had HERMES data available.

\subsection{Analysis}
Data from both LAMOST and HERMES were normalised using the {\sc{SUPPNet}} software \citep{2022A&A...659A.199R} with the default settings. Where we provide spectral classifications, comparison is made between the observed spectrum and a set of MK Standard stars. We aimed to find a match between the Balmer line profiles and the non-peculiar metals to determine the broad spectral type, then note those peculiar lines that are particularly abnormal in the observed spectrum.

The high resolution HERMES spectra were also analysed to provide estimates of the effective temperature ($T_{\rm eff}$), surface gravity ($\log g$) and the projected rotational velocity ($v\sin i$). Two separate teams carried out this task, with their procedures below. We combine their results in the relevant star section.

The first team used the software packages {\sc{iSpec}} \citep{2014A&A...569A.111B,2019MNRAS.486.2075B} and the extended version of {\sc{handy}}\footnote{\url{https://github.com/RozanskiT/HANDY}} (R\'o\.za\'nski, private communication). Firstly, with {\sc{iSpec}}, Balmer line profiles were fitted and an estimate of the effective temperature with spectral synthesis was made using the {\sc{synthe}} \citep{1993sssp.book.....K,2004MSAIS...5...93S} programme with the Castelli and Kurucz model atmospheres \citep{2003IAUS..210P.A20C} and the VALD line list \citep{2011BaltA..20..503K}. This initial result was then used in {\sc{handy}}-ext where synthetic Balmer lines were calculated using the same underlying codes/inputs to determine $\log g$, $T_{\rm eff}$ and $v\sin i$ via manually adjusting the parameters until a good (visual) match with the Balmer lines was obtained. Those parameters were then used to calculate a second, full, synthetic spectrum in {\sc{iSpec}} with which non-blended lines of Fe\,{\sc{i}} and {\sc{ii}} were found so that $\log g$, $v\sin i$ and $T_{\rm eff}$ were re-calculated, via spectral synthesis, with error estimates.

The second team used the `Balmer' code, written in {\sc{Python}}, based on the spectrum synthesis method (Posi\l{}ek, private communication). The atmospheric models and synthetic spectra were calculated with the {\sc{atlas9/synthe}} software \citep{2005MSAIS...8...14K}. The determination of $T_{\rm eff}$ and $\log g$ was performed using four Balmer lines (H$\alpha$, H$\beta$, H$\gamma$, H$\delta$). In the case of temperatures lower than about 8500\,K, the Balmer lines become insensitive to $\log g$, and only the temperature was determined, with $\log g$ assumed as 4.0\,\cms. Microturbulence was assumed to be $\xi = 1$\,\kms. An approximate value of $v\sin i$ was determined from the Mg\,4481\,\AA\ line and metal lines. Errors of estimated parameters were calculated as a weighted median standard deviation of the fit to each line considered.

The results of these analyses are presented in the relevant subsections for the three stars, TIC\,21024812, TIC\,435263600 and TIC\, 445493624.

\section{Results}
\label{sec:results}

Here we present the results of our search and analysis of new and known roAp stars in {\it TESS} Cycle\,2 data. We provide an overview of the targets to be discussed in Table\,\ref{tab:stars} and divide this section, and the table, into subsections addressing new discoveries reported in this work, previous discoveries reliant on the {\it TESS} data, and {\it TESS} observations of known roAp stars. Finally, we provide a list of candidate roAp stars where the currently available data do not allow us to confirm the nature of these stars.

\begin{table*}
    \centering
    \caption{Details of the stars analysed in this paper. The columns provide the TIC identifier and star name, the TIC\,v8.1 {\it TESS} magnitude, the spectral type as referenced in the stars discussion section, where `*' denotes a classification derived in this work, the effective temperature as provided in the TIC (with typical errors up to 250\,K), the sectors in which {\it TESS} observed the target in Cycle\,2. The final three columns provide the stellar rotation period derived in this work, with a $^\ddag$ indicating a more precise period is presented in the literature and a $^\dagger$ indicating this could be an orbital period, the pulsation frequency(ies) found in this work (note that sidelobes that arise from oblique pulsation are not listed), and the pulsation amplitude(s) seen in the {\it TESS} data. A`**' denotes rotation periods derived from rotationally split sidelobes.}
    \label{tab:stars}
    \begin{tabular}{lcccclrcc}
        \hline
         \multicolumn{1}{c}{TIC} & {HD/TYC}    	& {\it TESS} & {Spectral} 	 & $T_{\rm eff}^{\rm TIC}$   & {Sectors} &  \multicolumn{1}{c}{{$P_{\rm rot}$}}  	& \multicolumn{1}{c}{Pulsation frequency}	&  \multicolumn{1}{c}{Pulsation amplitude}     \\
                       		             &   {name}          	&  mag & {type}            &   {(K)}                               & {} &  \multicolumn{1}{c}{(d)}         		&   \multicolumn{1}{c}{(mHz)}			& \multicolumn{1}{c}{(mmag)}  \\
        \hline

         \multicolumn{6}{l}{\textit{New {\it TESS} roAp stars found in this work}}\\	
	
101624823 & 100598	& 8.18	& Ap\,Cr(SrEu)*			& 6780	& 22	& $4.050\pm0.001$		&	$1.44176\pm0.00001$	& $0.124\pm0.006$	\\

165052884 & 51561		& 8.81	& A5			& 6900	& 20	& $\sim22$	       	&	$2.176721\pm0.000007$ & $0.243\pm0.008$ 	\\

233200244 & 161846	& 8.81	& A3		& 8220	& 14-19, 21-26	& $7.25002\pm0.00006$	& $1.2614617\pm0.0000016$ & $0.030\pm0.003$ 	\\
		&			&		&			&		&	&					&	 $1.4390374\pm0.0000005$ & $0.091\pm0.003$ \\
		
259017938 & 210684	& 7.15	& F0		& 7570	& 15,16	& $5.0226\pm0.0005$	&	$1.352116	\pm0.000006$ & $0.064\pm0.003$ 	\\

335457083 & 48409		& 8.20	& A3		& 7510	& 20		& $2.9139\pm0.0001$	& 	$2.08819\pm0.00002$ &	$<0.015$ \\	
		&			&		&		&		&		&					&	$2.20575\pm0.00003$ &	$0.049\pm0.006$ \\
		&			&		&		&		&		&					&	$2.24835\pm0.00003$ &	$0.043\pm0.006$ \\
		&			&		&		&		&		&					&	$2.29691\pm0.00005$ &	$0.029\pm0.006$ \\

445493624 & 11948	& 7.7		& A5\,SrEu(Cr)*		& 8060	& 18		&$7.288\pm0.002$		&	$1.22522\pm0.00003$ & $0.06\pm0.01$ 	\\

467074220 & 40142 & 7.55	& A8\,SrCr(Eu)*		& 7410	& 19		& $5.4195\pm0.0004$	&	$0.70462\pm0.00006$ & $0.030\pm0.007$\\
			&			&		&		&		&		&					&	$0.72476\pm0.00003$ & $0.059\pm0.007$\\

 \multicolumn{6}{l}{\textit{roAp stars previously discovered by {\it TESS}}}\\
        
21024812 & 14522		& 8.59	& A8\,SrEu(Cr)*		& 8070	& 18	& $\sim 24$	&	$1.10259\pm0.00002$ &	$0.096\pm0.007$	\\

25676603 & 2194-2347-1	& 8.50	& F0\,SrEu	& 6680	& 15	& >26		&	$1.787321\pm0.000027$ &$0.072\pm0.008$ \\
		&			&		&			&		&	&			&	$1.819500\pm0.000002$ & $0.864\pm0.008$ \\
		&			&		&			&		&	&			&	$1.846724\pm0.000008$ & $0.249\pm0.008$ 	\\

26833276 & 10682		& 7.51	& A3			& 6970	& 17	& >25 		&	$1.58317\pm0.00002$ & $0.081\pm0.005$\\
		&			&		&			&		&	&			&	$1.60714\pm0.00003$ & $0.046\pm0.005$\\
		&			&		&			&		&	&			&	$2.74058\pm0.00004$ & $0.033\pm0.005$\\
		&			&		&			&		&	&			&	$2.80741\pm0.00004$ & $0.032\pm0.005$\\
		
72392575 & 225578		& 8.83	& A\,Sr:		& 7570	& 14	& $3.9016\pm0.0008$	&	$1.57343\pm0.00003$ & $0.09\pm0.01$\\

118247716 & 12519		& 8.13	& A7\,SrEuCr*	& 7060	& 17	& No signature			& 	$1.94247\pm0.00004$  & $0.040\pm0.006$\\
		&			&		&			&		&	&					&	$1.97093\pm0.00002$  & $0.098\pm0.006$\\
		&			&		&			&		&	&					&	$2.03280\pm0.00002$  & $0.070\pm0.006$\\

120532285 & 213258	& 7.40	& Fp			& 7210	& 16	& No signature			&	$2.171017\pm0.000011$  & $0.095\pm0.004$\\
		&			&		&			&		&	&					&	$2.198746\pm0.000008$  & $0.128\pm0.004$\\
		&			&		&			&		&	&					&	$2.226270\pm0.000006$  & $0.159\pm0.004$\\

129820552 & 2322-1440-1 & 9.90 	& Ap			& 7400	& 18 & $6.230\pm0.002$		&	$2.01305\pm0.00002$ & $0.23\pm0.01$ 	\\

394860395 & 2666-85-1	& 9.23	&		& 7170	& 14		& $22\pm2$**			&	$1.18050\pm0.00005$ & $0.09\pm0.01$	\\

435263600 & 218439 & 7.59 	& A5\,SrSi	&	& 17, 24		&$3.0739\pm0.0002$	&	$1.097186\pm0.000003$ & $0.063\pm0.010$\\
		&			&		&		&		&		&					&	$1.097049\pm0.000003$ & $0.059\pm0.010$\\
		&			&		&		&		&		&					&	$1.117976\pm0.000003$ & $0.039\pm0.005$\\
		&			&		&		&		&		&					&	$1.124019\pm0.000003$ & $0.037\pm0.005$\\

\hline
        
            \end{tabular}
\end{table*}

\begin{table*}
    \centering
    \contcaption{}
    \label{tab:stars_cont_2}
    \begin{tabular}{lcccccrcc}
        \hline
         \multicolumn{1}{c}{TIC} & {HD/KIC/TYC}    	& {\it TESS} & {Spectral} 	 & $T_{\rm eff}^{\rm TIC}$   & {Sectors} &  \multicolumn{1}{c}{{$P_{\rm rot}$}}  	& \multicolumn{1}{c}{Pulsation frequency}	&  \multicolumn{1}{c}{Pulsation amplitude}     \\
                       		             &   {name}          	&  mag & {type}            &   {(K)}                               & {} &  \multicolumn{1}{c}{(d)}         		&   \multicolumn{1}{c}{(mHz)}			& \multicolumn{1}{c}{(mmag)}  \\
        \hline
        \multicolumn{5}{l}{\textit{Known roAp stars prior to {\it TESS} launch}}\\

26418690 	& K11296437		& 11.38	&	A9\,EuCr		& 7010		& 14,15 	& $7.11\pm0.02$ & $1.409717\pm0.000007$ & $0.32\pm0.02$ \\

26749633		& K11031749		& 12.39	&	F1\,SrCrEu	& 7000 		& 14,15 	& No signature	& No signal in {\it TESS} data & <0.18\\

27395746		& K11409673		& 12.58	&	A9\,SrEu		& 7500		& 14,15 	& No signature	& No signal in {\it TESS} data & <0.15\\

77128654		& 97127			& 9.06	&	F3\,SrEu(Cr)	& 6710		& 22		& No signature	& $1.233855\pm0.000004$ & $0.489\pm0.009$ \\
			&				&		&				&			&		&			& $1.248661\pm0.000031$ & $0.061\pm0.009$ \\
			
123231021	& K7582608		& 11.43	&	A7\,EuCr		& 	8400		& 14,26 	& $19.805\pm0.005^\ddag$ & $2.1034011\pm0.0000002$ & $1.32\pm0.02$\\ 

158216369	& K7018170		& 12.89	&	F2\,(SrCr)Eu	&	6950		& 14,26 	& No signature	& No signal in {\it TESS} data & <0.28\\

158271090	& K10195926		& 10.31	& 	F0\,Sr		&	7390		&14,15,26	& $5.6845\pm0.0002$ & $0.974579\pm0.000002$ & $0.019\pm0.009$\\

158275114	& K8677585		& 10.00	&	A5\,p		&	7460		& 14,26 	&	No signature	&$1.5853813\pm0.0000009$ & $0.107\pm0.009$\\
			&				&		&				&			&		&				& $1.6217146\pm0.0000019$ & $0.053\pm0.009$\\ 
			
159392323	& 3547-2692-1		& 11.92	&	F3\,SrEuCr 	&	6390		& 15 		& No signature & $1.47658\pm0.00005$ & $0.21\pm0.05$\\
			&				&		&				&			&		&			& $1.49021\pm0.00002$ & $0.66\pm0.05$ \\
			&				&		&				&			&		&			& $1.49610\pm0.00003$ & $0.34\pm0.05$ \\

169078762	& 225914			& 9.00	& 	A5\,SrCrEu	&	8100		& 14 		& $5.181\pm0.003^\ddag$ & $0.71130\pm0.00005$ & $0.05\pm0.01$\\

171988782	& 258048			& 10.02	&	F4\,EuCr(Sr)		&	6370		& 20 		& No signature	& $1.962238\pm0.000005$ & $0.55\pm0.01$\\
			&				&		&				&			&		&			& $1.975095\pm0.000023$ & $0.13\pm0.01$ \\
			
264509538	& K10685175		& 11.79	&	A4\,Eu		&	7970		& 14,15 	& $3.1029\pm0.0005^\ddag$ & $2.216610\pm0.000006$ & $0.57\pm0.03$\\

272598185	& K10483436		& 11.16	&	A7\,SrEuCr	&	7980		& 14,15  	& $4.2857\pm0.0006$ & No signal in {\it TESS} data & <0.05 \\

273777265	& K6631188		& 13.38	&	F0\,Sr		&	7500		& 14,15  	& $2.519\pm0.002^\ddag$ & $1.49349\pm0.00001$ & $1.45\pm0.16$ \\

286992225	& 2553-480-1		& 11.24	& 	A9\,SrEu		&	7490		& 23 		& No signature			& $2.69128\pm0.00002$ & $0.32\pm0.03$\\
			&				&		&				&			&		&				        & $2.72616\pm0.00001$ & $0.54\pm0.03$ \\
			
302602874	& 2488-1241-1		& 10.86	&	A6\,SrEu		& 	8160		& 21 		& $3.0900\pm0.0003$ 	&$2.283125\pm0.000002$ & $2.12\pm0.02$\\
      \\
    \multicolumn{5}{l}{\textit{Candidate roAp}}\\
    
9171107	& 2479-429-1	& 9.36	& Unknown		& 6770	& 20	& No signature			&	$1.50971\pm0.00004$ &	$0.053\pm0.009$\\
		&			&		&			&		&	&					&	$1.52902\pm0.00002$ &	$0.108\pm0.009$\\
		&			&		&			&		&	&					&	$1.53502\pm0.00002$ &	$0.094\pm0.009$ \\

91224991 & 191380		& 8.49	& F8			& 6710	& 14	& No signature			&	$1.159876\pm0.000002$ &$0.961\pm0.008$ 	\\

275642037 & 4330-716-1	& 9.52	& F0		& 7710	& 18		& $2.8745\pm0.0003$	&	$1.13943\pm0.00003$ & $0.08\pm0.01$ 	\\
		&			&		&		&		&		&					&	$1.16939\pm0.00002$ & $0.17\pm0.01$ 	\\

298052991 & 4554-625-1	& 8.8		& A2		& 7600	& 14, 21, 21, 26& $10.5852\pm0.0001$ & $2.217569\pm0.000003$ & $0.023\pm0.004$ 	\\
		&			&		&		&		&		&					&	$2.246466\pm0.000002$ & $0.038\pm0.004$ 	\\
		&			&		&		&		&		&					&	$2.291038\pm0.000001$ & $0.051\pm0.004$ 	\\

405892692 & 3567-1092-1 & 8.99	& F0		& 7230	& 14, 15		& $0.3786\pm0.0001^\dagger$		&	$1.17082\pm0.00002$ & $0.038\pm0.006$	\\

429251527& 2185-478-1	& 10.20	&			& 7920	& 15		& No signature    	&	$1.77798\pm0.00001$ & $0.45\pm0.02$	\\

445796153 & 34740	& 7.22	& A0p		& 8510	&	19, 26 	& $5.5721\pm0.0004$&	$0.601908\pm0.000002$ & $0.022\pm0.003$\\
		&			&		&		&		&		&					&	$0.675849\pm0.000002$ & $0.035\pm0.003$ 	\\
 
        \hline 
    \end{tabular}
\end{table*}
As with \citetalias{2021MNRAS.506.1073H}, we present the numerical results of the reference team to provide a homogeneous data set both within this paper and across the two papers. The results have been verified with the WG4 members, and discrepancies were resolved with secondary checks. 

Furthermore, we use the same approach as in \citetalias{2021MNRAS.506.1073H} which briefly comprises of: for each star, we combined available sectors to produce a single light curve for analysis. We calculated a Discrete Fourier Transform, following \citet{1985MNRAS.213..773K}, using the PDC\_SAP data obtained from the MAST servers. Initially we searched in the low-frequency range (up to 0.11\,mHz; 10\,\cd) to identify signs of rotational variability. Where present, we fitted a harmonic series by non-linear least squares to determine the stellar rotation frequency, and used the highest amplitude peak to determine the rotation frequency, in some cases this was the second harmonic of the rotation frequency. Where no rotational variability was found, we also checked the SAP light curve for signs of rotation as the SPOC pipeline can remove low-frequency astrophysical signal. 

Subsequently, we removed low-frequency peaks (rotational or instrumental) via iterative pre-whitening of the light curve to a noise level determined in the range $2.3-3.3$\,mHz ($200-300$\,\cd). A new amplitude spectrum was calculated to the Nyquist frequency, and any signals found were extracted through non-linear least squares fitting. Both the derived rotation period and the pulsation frequency(ies) for each star are listed in Table\,\ref{tab:stars}. We provide only the {\it{pulsation mode}} frequencies and not rotationally split sidelobes; any sidelobes detected are mentioned in the text, and indicated on the appropriate figure. Where multiplets due to oblique pulsation exist, we apply the relations of \citet{1990MNRAS.247..558K} to investigate the geometry of the pulsation mode as derived by the multiplet amplitudes. For a dipole mode, the multiplet amplitudes give a relation between the inclination angle, $i$, and the angle of obliquity, $\beta$, that is the angle between the rotation and pulsation axes. When conducting these tests we set a zero point in time for the linear least squares fit such that the phases of the $\pm\nu_{\rm rot}$ sidelobes are equal. It is expected that all peaks in a multiplet have equal phase for a non-distorted mode, thus applying the oblique pulsator model provides insight to the mode structure.

For stars that have previously been discussed at length in the literature, with use of the {\it TESS} data, we do not present an in-depth analysis unless significant differences in the results were found. We do, however, include these stars in Table\,\ref{tab:stars} to provide a complete inventory of {\it TESS} observations of roAp stars in Cycle\,2. For all stars newly classified as roAp stars in this work, and roAp stars that have been discovered in {\it TESS} data, we provide plots of their pulsation frequencies. For the known roAp stars, we provide plots in Appendix\,\ref{app:known} unless there is new information in the  {\it TESS} data.

Where available, we provide the Str\"omgren-Crawford indices for a given object. We used these values, in conjunction with the calibrations of \citet{1975AJ.....80..955C,1979AJ.....84.1858C} to determine if the values are indicative of a chemically peculiar star.


\subsection{New {\it TESS} roAp stars}
\label{sec:New_roAp}
We classified the following stars as new roAp stars based on the presence of high frequency pulsation in the target and either an Ap spectral classification, the presence of multiplets in the amplitude spectrum split by the stellar rotation frequency, or Str\"omgren-Crawford indices that indicate a chemically peculiar star. While some stars already appear in the literature as showing rapid oscillations, they were not admitted to the roAp class. Here we provide a careful consideration of the available evidence to rectify previous oversights. 

\subsubsection{TIC\,101624823}

TIC\,101624823 (HD\,100598) has a spectral classification in the literature of A2\, \citep{1919AnHar..94....1C}. \citet{2006A&A...450..735M} estimate the effective temperature to be 6800\,K, based on $V$ and 2MASS IR photometry, and \citet{2019AJ....158..138S} estimate $\log g = 3.8$\,\cms. The surface gravity was not obtained via spectroscopy, but calculated from the reported mass and radius. In addition, the value for the effective temperature of 7000\,K and for the surface gravity $\log g = 3.85$\,\cms\ are extracted from the \citet{2022yCat.1355....0G} {\it Gaia} DR3 catalogue. These values are extracted from GSP-Phot Aeneas best library using Bp/Rp spectra, apparent G magnitude and parallax. The final result is the median of the MCMC samples. The literature values of radial velocity for this star are $-15.45\pm0.33$\,\kms\, measured by \citet{2018yCat.1345....0G}, with a more recent value from the \citet{2022yCat.1355....0G} {\it Gaia} DR3 catalogue of $-15.64\pm0.19$\,\kms, thus both in agreement. This star appears in the Str\"omgren-Crawford measurements of \citet{2015A&A...580A..23P} with values of $b-y=0.223$, $m_1=0.157$, $c_1=0.672$, H$\beta=2.714$, which give $\delta m_1 = 0.018$ and $\delta c_1=0.105$. Both of these values indicate a chemically normal star, given the H$\beta$\, value. There are no magnetic field measurements for this star in the literature. 

\begin{figure}
\centering
\includegraphics[width=\columnwidth]{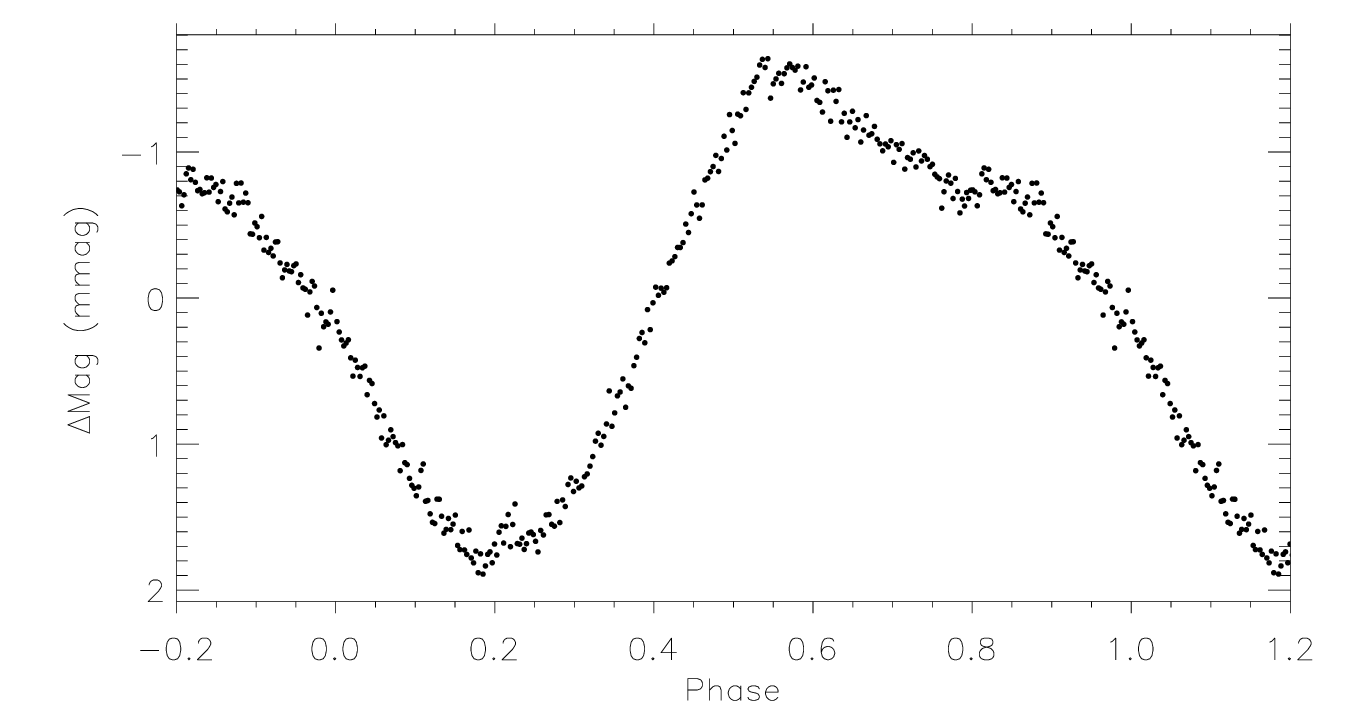}
\caption{Phase folded light curve of TIC\,101624823, phased on a period of $4.050\pm0.001$\,d. The data are binned 50:1.}
\label{fig:101624823_rot}
\end{figure}

\begin{figure}
\centering
\includegraphics[width=\columnwidth]{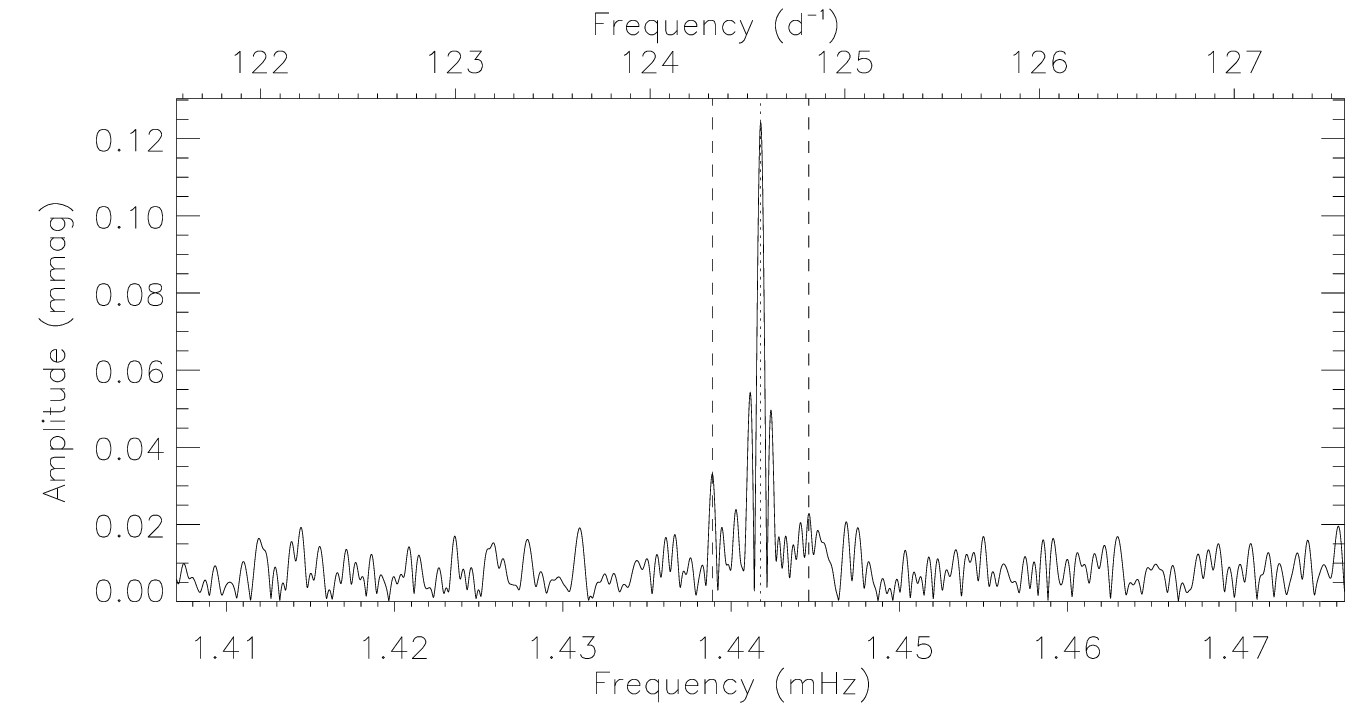}
\caption{Amplitude spectra of TIC\,101624823. The pulsation mode is identified by the vertical dotted line, with the dashed lines showing the locations of the rotationally split sidelobes.}
\label{fig:101624823_ft}
\end{figure}

{\it TESS} data for TIC\,101624823 used in this work were collected during sector 22. The data show a clear rotational signature of $4.050\pm0.001$\,d (Fig.\,\ref{fig:101624823_rot}). This is in agreement with the value reported by \citet{2022MNRAS.510.5743B} using the same data, although the previous author did not include an error on the measurement. After removing the rotation signal from the data, we analysed the high frequency peak ($1.44176\pm0.00001$\,mHz; $124.5678\pm0.0009$\,\cd), which was also noted by \citet{2022MNRAS.510.5743B} but the star was not admitted to the roAp star class in that work. Here we rectify that oversight with the identification of a rotationally split sidelobe at $\nu-\nu_{\rm rot}$ (at S/N $\sim 4.5$) as is expected from oblique pulsations in roAp stars, which is shown in Fig.\,\ref{fig:101624823_ft} by the vertical dotted line. A second sidelobe is identifiable (at $\nu+\nu_{\rm rot}$), albeit at low S/N ($\sim3.1$), indicating the presence of a triplet, and a dipole mode. 

With these two sidelobes, we are able to apply the oblique pulsator model (as described above) to derive $\tan i\tan\beta=0.369\pm0.071$. We also found that the mode is slightly distorted when investigating the phase relations between the mode and the sidelobes. We found that pulsation maximum occurs at rotation phase 0.46 in Fig.\,\ref{fig:101624823_rot}, hence the pulsation pole is located close to the principal abundance spot that gives rise to the rotational light variations. Hence this star is typical of the roAp star oblique pulsators in that the pulsation pole lies close to the magnetic pole.

To further cement its membership of the roAp class, we have obtained a spectrum from LAMOST.  While the data are of low S/N, there is clear presence of Cr\,{\sc{ii}} in the spectrum (Fig.\,\ref{fig:101624823_spec}), with the possible presence of Sr\,{\sc{ii}}  and Eu\,{\sc{ii}}, but at a lower confidence. Therefore, we classified this star as an Ap\,Cr(SrEu) star. This shows that Ap stars with Str\"omgren-Crawford indices that are only slightly deviant from normal stars, can be roAp stars.


\subsubsection{TIC\,165052884}

TIC\,165052884 (HD\,51561) has a spectral classification of A5 \citep{1918AnHar..92....1C}. \citet{2019AJ....158..138S} estimate $T_{\rm eff} = 6900$\,K and $\log g = 4.18$\,\cms. The effective temperature is derived from dereddened {\it Gaia} Bp-Rp colour. As with the previous star, this value is calculated using the mass and radius estimates. The {\it Gaia} DR3 \citep{2022yCat.1355....0G} gives $T_{\rm eff} = 6850$\,K and $\log g = 4.22$\,\cms.

{\it TESS} observed TIC\,165052884 in sector 20 of Cycle\,2, with the data analysed by \citet{2022MNRAS.510.5743B} who measured a rotation period (using sector 20 and 44 data) of $P_{\rm rot} = 17.857$\,d and a pulsation frequency of $2.177$\,mHz ($\nu$ = 188.069\,\cd). However, our analysis cannot reproduce the rotation results of \citet{2022MNRAS.510.5743B}, even if we include the sector 44 data. We further investigated this discrepancy by analysing the SAP data, which does indeed show variability that can be attributed to rotation. However, the signal indicated a rotation period about the same length of the data ($P_{\rm rot}\sim22$\,d compared to a data set length of $\sim26$\,d), so cannot be precisely determined.

\begin{figure}
\centering
\includegraphics[width=\columnwidth]{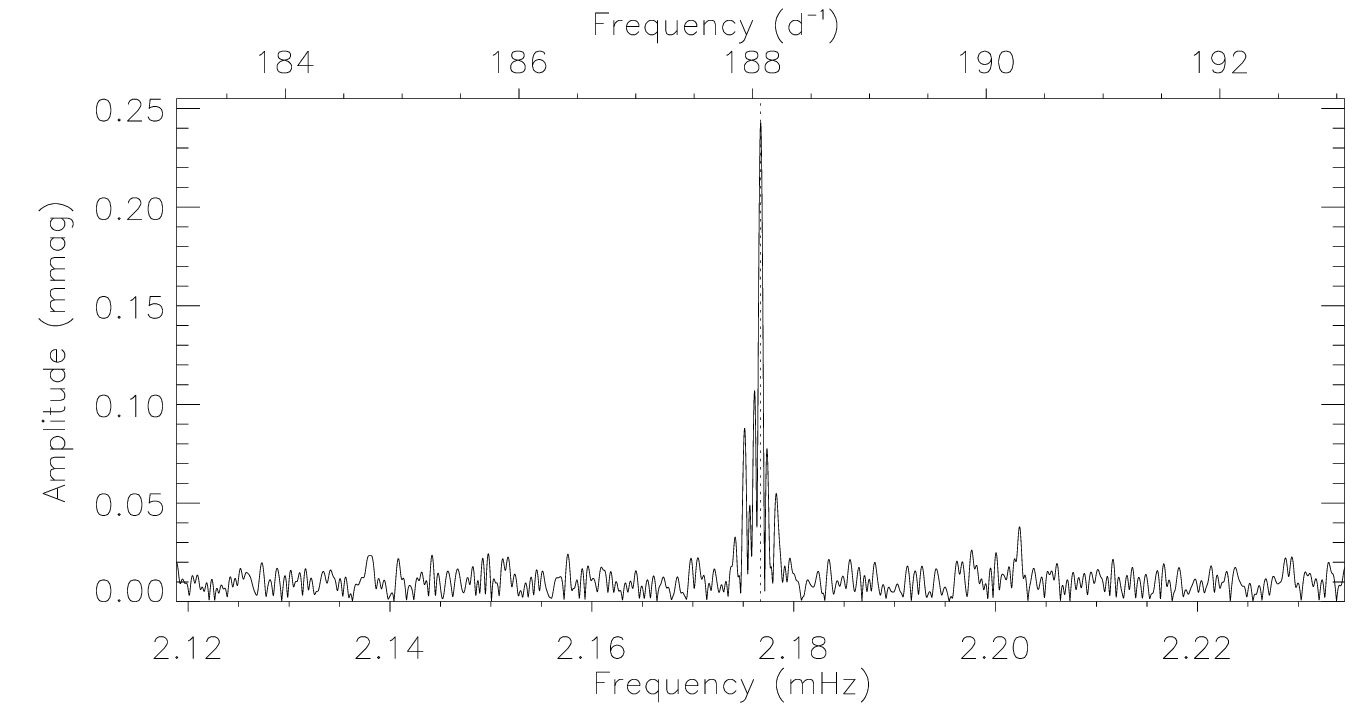}
\caption{Amplitude spectrum of TIC\,165052884 with the pulsation mode identified with the vertical dotted line. }
\label{fig:165052884_ft}
\end{figure}

The high frequency pulsation mode in TIC\,165052884 is found at $2.176721\pm0.000007$\,mHz ($188.0687\pm0.0006$\,\cd; Fig.\,\ref{fig:165052884_ft}), which is in agreement with \citet{2022MNRAS.510.5743B}. After pre-whitening this signal, there is an indication of a much lower amplitude mode at $2.17566\pm0.00004$\,mHz ($187.977\pm0.003$\,\cd) which is separated by the principal peak by a frequency consistent with half of the presumed rotation frequency ($\sim 1.06$\,$\muup$Hz; $0.09$\,\cd; 10.9\,d). Again, more data are required to confirm this detection.

We admitted TIC\,165052884 to the roAp class based on the pulsation and the tentative detection of a rotational sidelobe to the pulsation mode.


\subsubsection{TIC\,233200244}

TIC\,233200244 (HD\,161846) was classified as A3 in the Henry Draper Catalogue \citep{1922AnHar..97....1C}. From photometry, \citet{2010PASP..122.1437P} give a similar spectral type of A5\,V. The TIC gives $T_{\rm eff} = 8200\pm150$\,K, with similar values of 8050\,K and 8100\,K obtained by \citet{2017MNRAS.471..770M} and \citet{2018ApJ...867..105T}, respectively, while the {\it Gaia} DR3 lists $T_{\rm eff}$ as 8000\,K. The star was analysed by \citet{2022MNRAS.510.5743B} who classified it as a new variable A star, with a rotation period of $P_{\rm rot} = 7.299$\,d (without an error) and a pulsation frequency of $1.44$\,mHz ($124.471$\,\cd), which also showed rotational sidelobes.The star also appears in the catalogue by \citet{2022A&A...666A.142S} who classified it as a rotationally modulated star, with a period of 7.246\,d, and a candidate roAp star after studying all A and F stars in {\it TESS}'s northern continuous viewing zone. 

\begin{figure}
\centering
\includegraphics[width=\columnwidth]{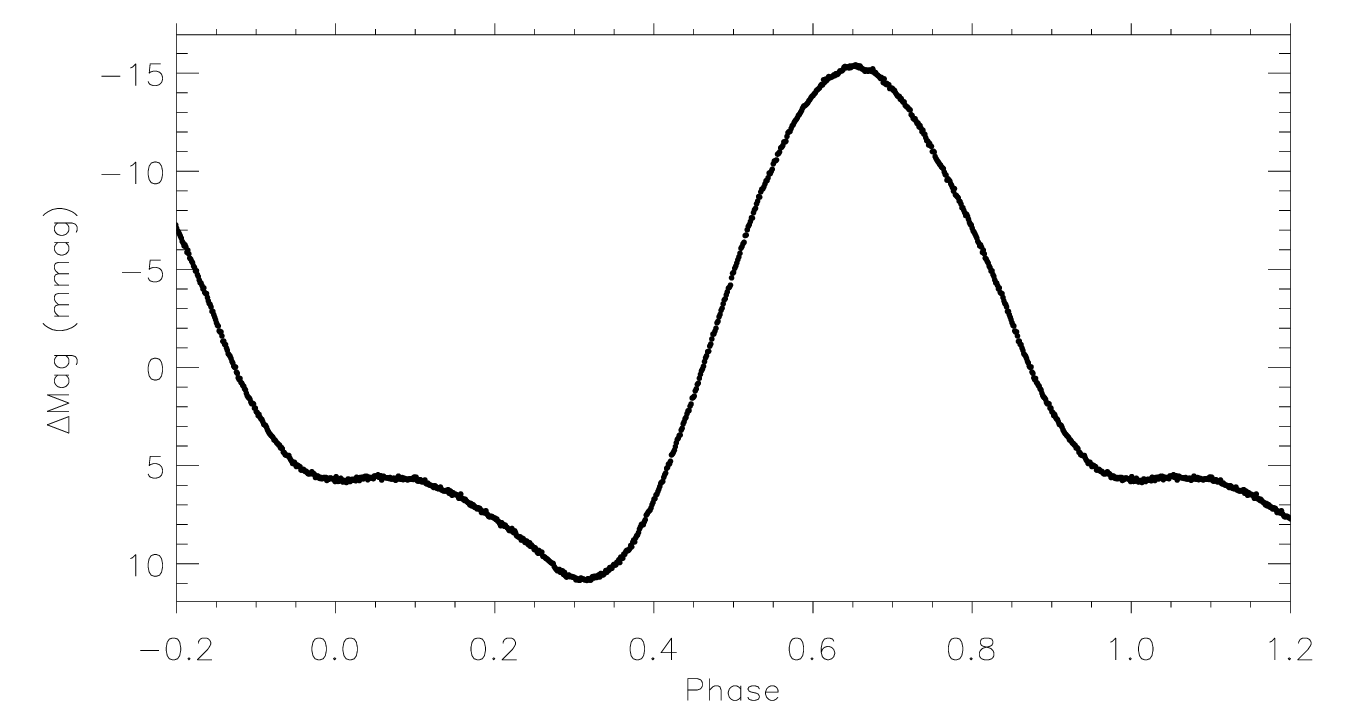}
\caption{Phase folded light curve of TIC\,233200244, phased on a period of $7.25002\pm0.00006$\,d. The data are binned 200:1.}
\label{fig:233200244_rot}
\end{figure}

\begin{figure}
\centering
\includegraphics[width=\columnwidth]{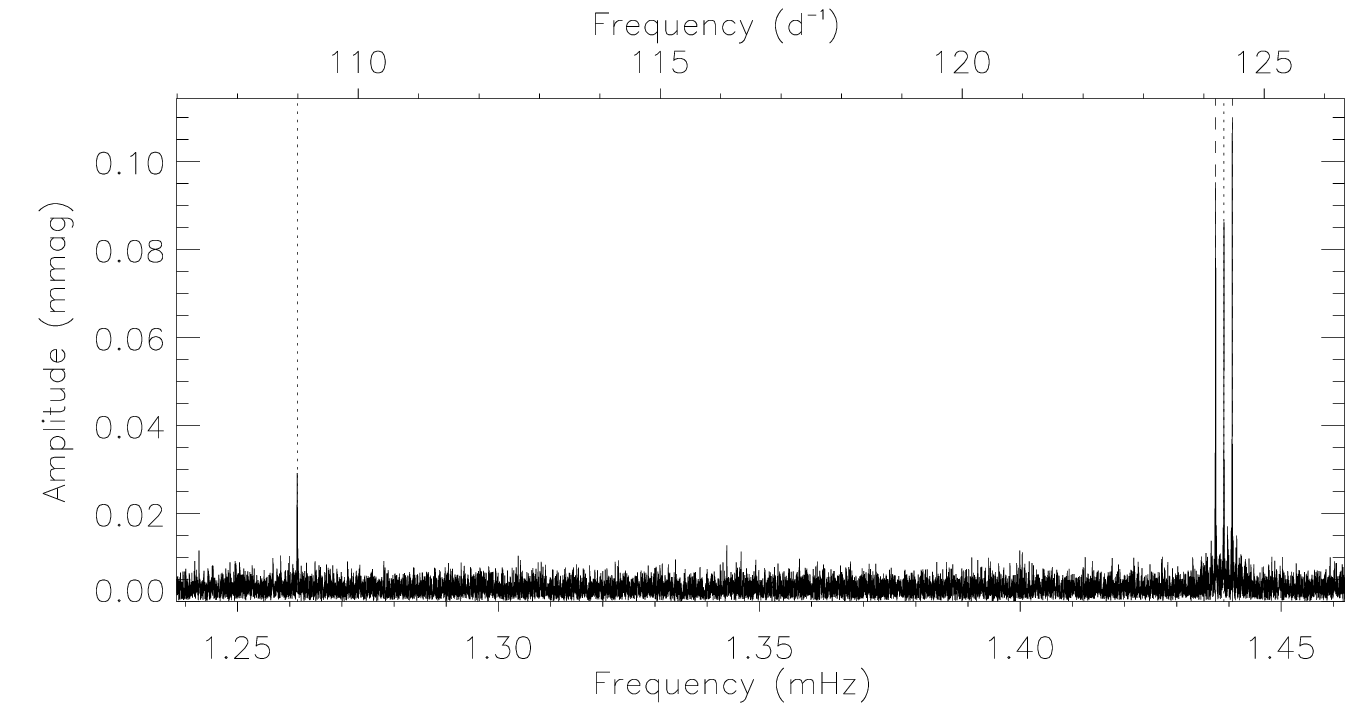}
\caption{Amplitude spectrum of TIC\,233200244 showing the two pulsation modes (marked by the vertical dotted lines) and the rotationally split sidelobes of the higher frequency mode. }
\label{fig:233200244_ft}
\end{figure}

{\it TESS} observed the star in all sectors of Cycle\,2 except sector 20. This allows us to derive a precise rotation period of $7.25002\pm0.00006$\,d. A folded light curve is shown in Fig.\,\ref{fig:233200244_rot}. We proceeded to prewhiten this signal from the individual sectors before analysing the pulsation modes in this star, with the resultant amplitude spectrum shown in Fig.\,\ref{fig:233200244_ft}. We found two pulsation modes at $1.261462\pm0.000002$\,mHz ($108.9903\pm0.0001$\,\cd) and $1.4390374\pm0.0000005$\,mHz ($124.33283\pm0.00005$\,\cd) separated by $177.576\pm0.002\,\muup$Hz, which could be three times the large frequency separation\footnote{defined as the frequency difference between two consecutive modes of the same degree} ($\sim60\,\muup$Hz), when scaling from the solar value. 

The higher frequency mode is flanked by rotational sidelobes split by exactly the rotation frequency, indicating the mode is an $\ell=1$ dipole mode. Given their presence, we applied the oblique pulsator model to test the phase relations and determine whether the mode is distorted. We found there is a little distortion to the mode, with the sidelobes having a phase of $-0.553\pm0.030\,$rad and the pulsation mode having $-0.910\pm0.031\,$rad. The time zero point for this calculation corresponds to rotation phase 0.68 in Fig.\,\ref{fig:233200244_rot}, i.e., at {\it TESS} light maximum. 

Given the detection of the pulsations and the sidelobes, coupled with a clearly stable rotational light curve, we admitted this star to the class of roAp stars.


\subsubsection{TIC\,259017938}

TIC\,259017938 (HD\,210684) has been classified F0 \citep{1924AnHar..99....1C}. \citet{2019A&A...623A..72K} revealed the existence of a physical companion orbiting TIC\,259017938 from the influence on the star's proper motion. Assuming $M_1=2.03\pm0.10$\,M$_{\odot}$ \citep{2000A&AS..141..371G} and $R_1=2.66\pm0.13$\,R$_{\odot}$ \citep{2004A&A...426..297K}, they found the mass of the companion and the radius of its orbit to be $M_2=80_{-10}^{+23} $\,M$_{\rm{Jup}}$ and 1.892 au, respectively. Str\"omgren-Crawford indices are given by  \citet{2015A&A...580A..23P} as: $b-y=0.160$, $m_1=0.196$, $c_1= 0.824$ and H${\beta} = 2.570$. We derive $\delta m_1 = 0.066$ and $\delta c_1=0.619$. These do not indicate an Ap star, however we note that the $b-y$ and H$\beta$ indices do not agree. \citet{2015A&A...580A..23P} combined two consistent values of $b-y$, while only reporting one value of H$\beta$. When using $b-y$ as a proxy for H$\beta$ \citep[corresponding to 2.786; ][]{1975AJ.....80..955C} we found $\delta m_1 = 0.002$ and $\delta c_1=0.070$ which do not change our conclusion.

\begin{figure}
\centering
\includegraphics[width=\columnwidth]{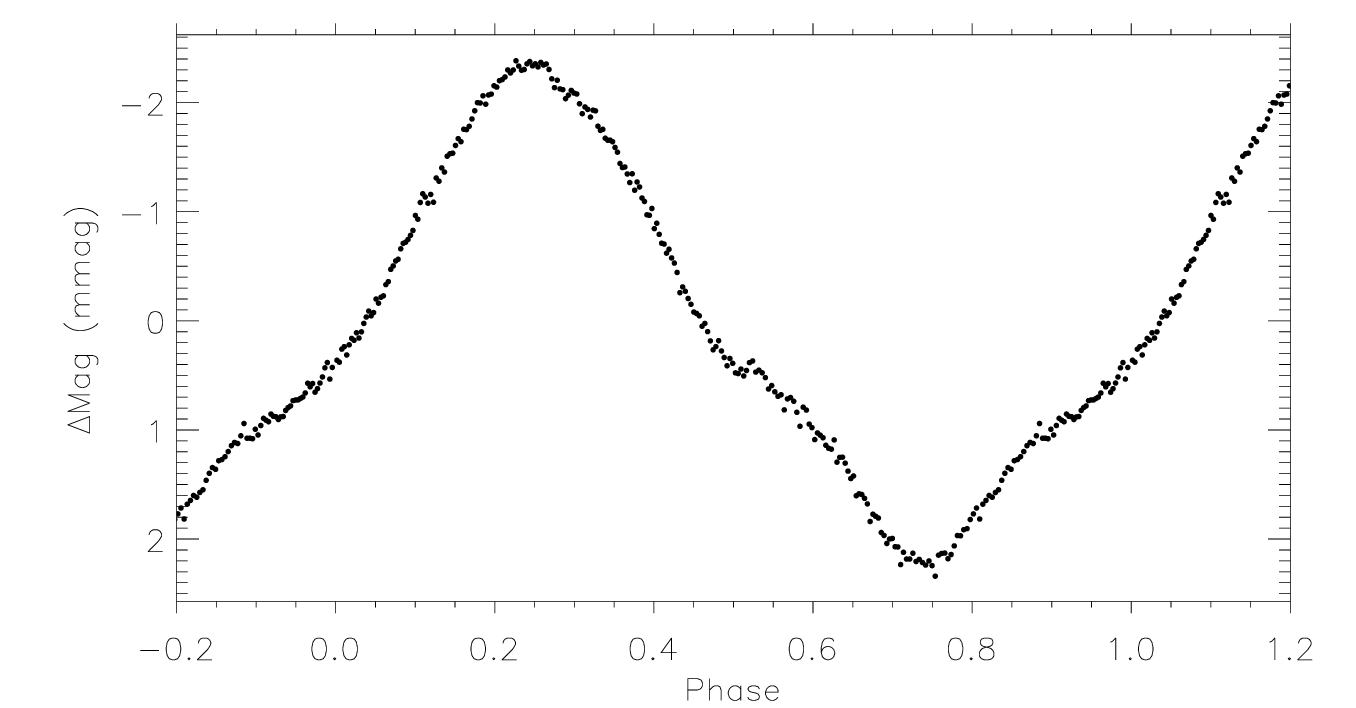}
\caption{Phase folded light curve of TIC\,259017938, phased on a period of $5.0226\pm0.0005$\,d. The data are binned 100:1.}
\label{fig:259017938_rot}
\end{figure}

\begin{figure}
\centering
\includegraphics[width=\columnwidth]{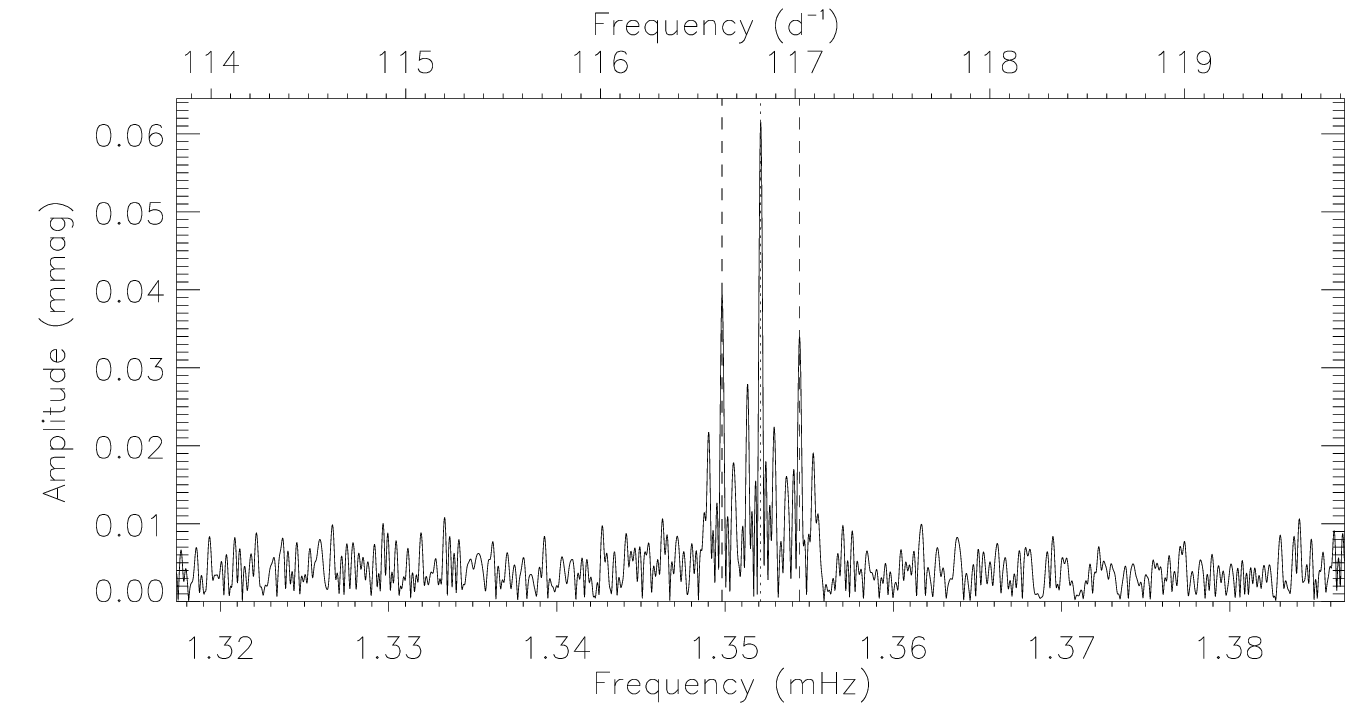}
\caption{Amplitude spectrum of TIC\,259017938 showing the pulsation mode (marked by the vertical dotted line) and the rotationally split sidelobes. }
\label{fig:259017938_ft}
\end{figure}

{\it TESS} observed TIC\,259017938 during sectors 15 and 16 and revealed the star to be a rotationally and pulsationally variable star \citep{2022MNRAS.510.5743B}. We determined a rotation period of $5.0226\pm0.0005$\,d (Fig.\,\ref{fig:259017938_rot}), which differs from the previously published value (of $P=5.201$), although that value has no associated error. After removing the signature of rotation, we identified a single pulsation mode in the star at $1.352116\pm0.000006$\,mHz ($116.8228\pm0.0005$\,\cd) that showed a rotationally split multiplet (Fig.\,\ref{fig:259017938_ft}). We again applied the oblique pulsator model to this dipole triplet to investigate the mode distortion and found the phase of the pulsation mode ($-2.48\pm0.05$\,rad) differs from the sidelobes ($-2.17\pm0.09$\,rad) by $3\,\sigma$, indicating possible slight mode distortion. The pulsation maximum corresponds to rotation phase 0.21 in Fig.\,\ref{fig:259017938_rot}, i.e., just before {\it TESS} light maximum.

We have also obtained three medium resolution spectra from the LAMOST database to search for chemical peculiarities (Fig.\,\ref{fig:259017938_LAMOST}). We identified lines of Pr\,{\sc{iii}}, Nd\,{\sc{ii}} \& {\sc{iii}} and La\,{\sc{ii}} amongst others, which indicate the star to be an Ap star, thus providing further evidence to admit the star to the roAp class.


\subsubsection{TIC\,335457083}

TIC\,335457083 (HD\,48409), was classified as A3 in the Henry Draper Catalogue \citep{1918AnHar..92....1C}. The TIC gives $T_{\rm eff} = 7500\pm140$\,K and {\it Gaia} DR3 7500\,K with a small uncertainty. The literature values, however, range from 9100\,K \citep{2018AJ....156..102S}  down to 6900\,K \citep{2019ApJS..241...12S}, with intermediate values of 8700\,K, 8750\,K and 7650\,K \citep{2003AJ....125..359W, 2006ApJ...638.1004A, 2017MNRAS.471..770M}, respectively. Based on its Str\"omgren-Crawford indices, of $b-y=0.213$, $m_1=0.308$, $c_1=0.517$ and H$\beta=2.838$  \citep{1998A&AS..129..431H}, which give $\delta m_1=-0.100$ and $\delta c_1=-0.332$, and a large $\Delta p$ ``peculiarity'' parameter of 6.779 (defined by \citet{1998A&AS..128..265M}), the star was selected to be observed by the Nainital-Cape Survey \citep{2006A&A...455..303J}, but no variability was found in this targeted ground-based photometric study. Previously, \citet{2022MNRAS.510.5743B} reported a rotation period of $2.907$\,d and a pulsation frequency of $2.084$\,mHz (180.075\,\cd) in this star using the {\it TESS} data.

\begin{figure}
\centering
\includegraphics[width=\columnwidth]{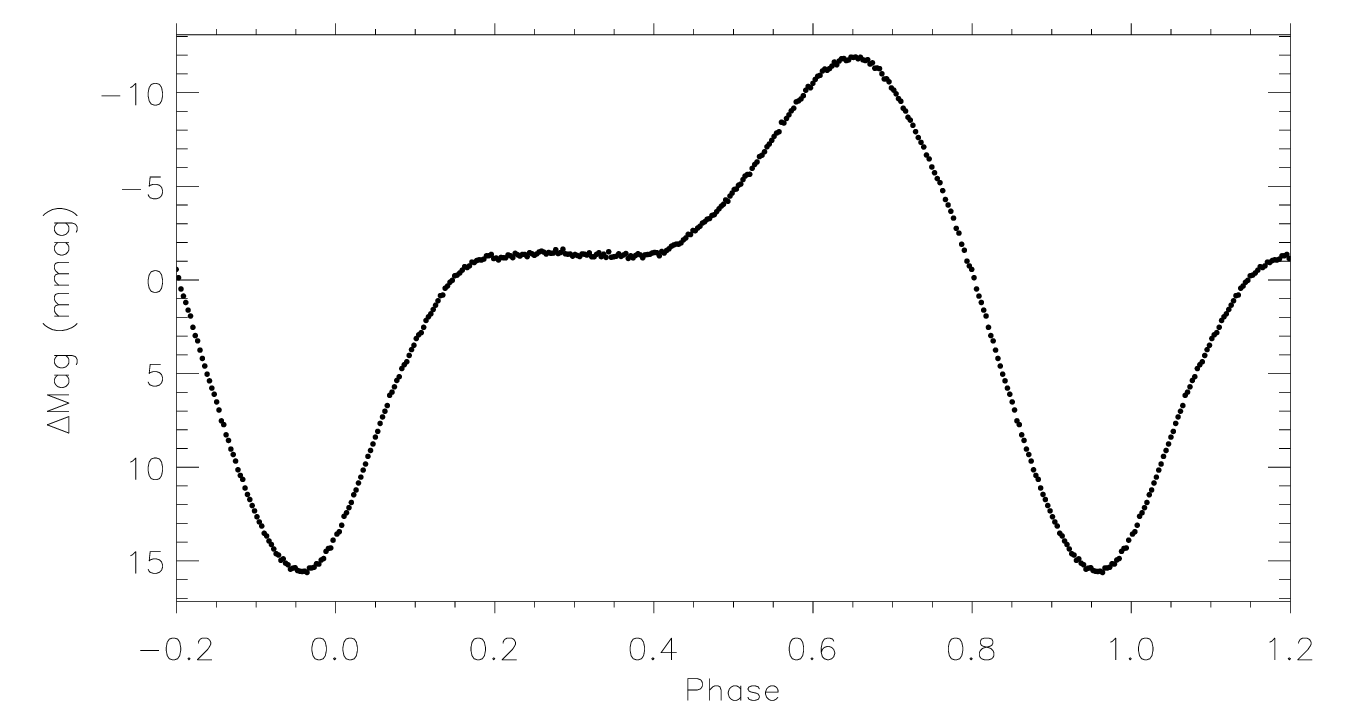}
\caption{Phase folded SAP light curve of TIC\,335457083, phased on a period of $2.9139\pm0.0001$\,d. The data are binned 50:1.}
\label{fig:335457083_rot}
\end{figure}

\begin{figure}
\centering
\includegraphics[width=\columnwidth]{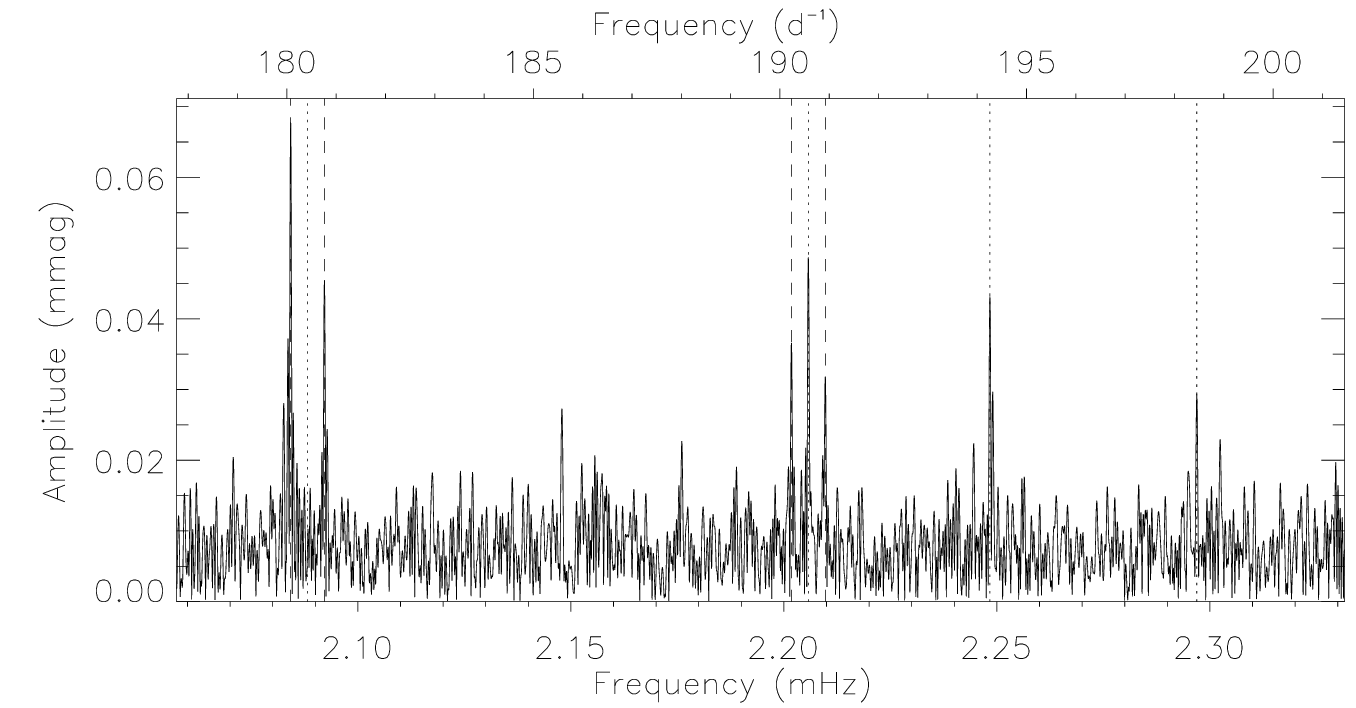}
\caption{Amplitude spectrum of TIC\,335457083 showing the pulsation modes (marked by the vertical dotted lines) and the rotationally split sidelobes (marked with the vertical dashed lines). }
\label{fig:335457083_ft}
\end{figure}

{\it TESS} observed TIC\,335457083 in sector 20 from which we derived a rotation period of $2.9139\pm0.0001$\,d (Fig.\,\ref{fig:335457083_rot}). At high frequencies, we inferred the presence of four pulsation modes (Fig.\,\ref{fig:335457083_ft}). In increasing frequency order, we found a doublet and inferred the presence a mode from the observation of two peaks separated by twice the rotation frequency (i.e., a mode at $2.08819\pm0.00002$\,mHz; $180.419\pm0.002$\,\cd), a triplet (centred on a mode at $2.20575\pm0.00003$\,mHz; $190.577\pm0.002$\,\cd), followed by two single peaks at $2.24835\pm0.00003$\,mHz ($194.258\pm0.003$\,\cd) and $2.29691\pm0.00005$\,mHz ($198.453\pm0.004$\,\cd). The doublet and triplet show very different ratios of the pulsation mode amplitude and the amplitude of the rotationally split sidelobes. For both modes to be dipole modes, we would require two pulsation axes in this star. Therefore, we assumed the doublet to be a quadrupole mode, and the triplet to be a dipole mode. Under this assumption, we calculated the values of $i$ and $\beta$ following \citet{1990MNRAS.247..558K} such that $i=56^\circ$ and $\beta=45^\circ$. This leads to $i+\beta=100^\circ$ which is consistent with the partial double wave of the rotation curve, supporting our choice of mode identifications. Of the observed, or inferred modes, there is no regular spacing that can be interpreted as the large frequency separation. However, there are further peaks above the noise in the amplitude spectrum of this star too, but further data are required to reduce the noise level to allow confident extraction of those possible mode frequencies. These additional modes may lead to the identification of the large separation frequency in this star.

Given both the strong indication that this star is chemically peculiar from the Str\"omgren-Crawford photometry, and the presence of modes with rotational splitting, we admitted this star to the class of roAp stars. 


\subsubsection{TIC\,445493624}

TIC\,445493624 (HD\,11948) has several spectral classifications in the literature: F0p \citep{1965ApJS...12..215H}, A5 Am/p \citep{1980A&AS...39..205O}, A5\,SrEu \citep{1988A&AS...76..127R}, A5\,SrCrEu \citep{1998A&A...336..953G}, all of which indicate this to be a chemically peculiar star. In addition, Str\"omgren-Crawford indices have also been measured several times \citep[e.g.,][]{1982ApJS...50..451P} with values being consistent with: $b-y=0.115$, $m_1=0.243$, $c_1=0.880$ and H$\beta =2.873$ \citep{2015A&A...580A..23P}, which are also consistent with a chemically peculiar A star (with $\delta m_1 =-0.041$ and $\delta c_1=-0.036$). The effective temperature of the star is reported consistently across the literature, e.g., $T_{\rm eff}=8050$\,K \citep{2012MNRAS.427..343M}, $T_{\rm eff} = 8250$\,K  \citep{2016A&A...590A.116J} and the TIC providing $8050$\,K. \citet{2019A&A...623A..72K} report a mass and radius of $M=2.20$\,M$_\odot$ and $R=2.15$\, R$_\odot$. The {\it Gaia} DR3 parallax provides a distance of $180\pm1$\,pc, which is at odds to the report that TIC\,445493624 is a member of the NGC\,744 open cluster \citep{1988AcA....38..225N}.

The star has a constant radial velocity in the literature of 32\,\kms\ (\citealt{2012A&A...546A..61D}, \citealt{2019A&A...623A..72K}, \citealt{2021ApJS..254...42B}), and has a root-mean-squared longitudinal magnetic field of $\langle B_\ell\rangle= 500\pm50$\,G with extreme values of the longitudinal component being $B_\ell$ $-550$ and $+300$\,G \citep{2008AstBu..63..139R}.

The star was searched for pulsations during the Nainital-Cape survey \citep{2006A&A...455..303J}, but none was found. Since then the {\it TESS} data have been used by \citet{2022MNRAS.510.5743B} to identify a rotation period of $7.333$\,d, and a pulsation frequency of $1.225$\,mHz (105.857\,\cd). In that study, the star was classified as a possible roAp star.

\begin{figure}
\centering
\includegraphics[width=\columnwidth]{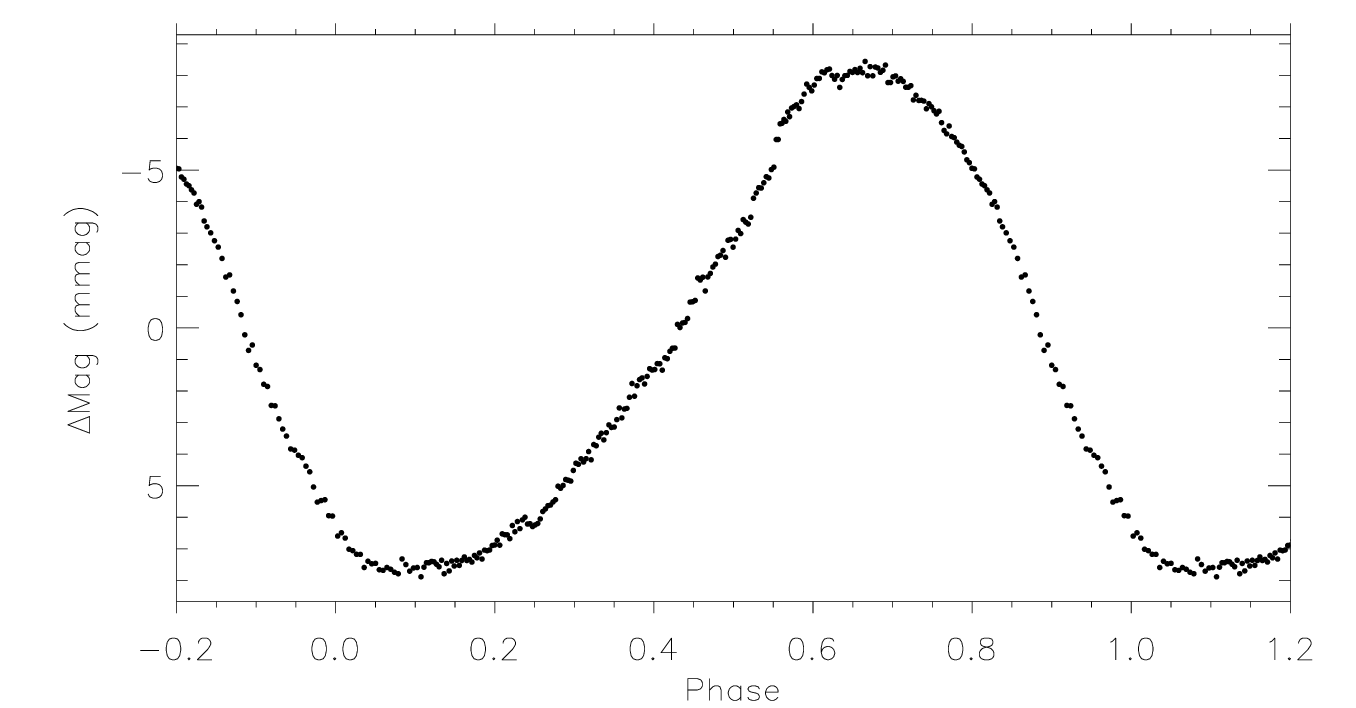}
\caption{Phase folded SAP light curve of TIC\,445493624, phased on a period of $7.288\pm0.002$\,d. The data are binned 50:1.}
\label{fig:445493624_rot}
\end{figure}

\begin{figure}
\centering
\includegraphics[width=\columnwidth]{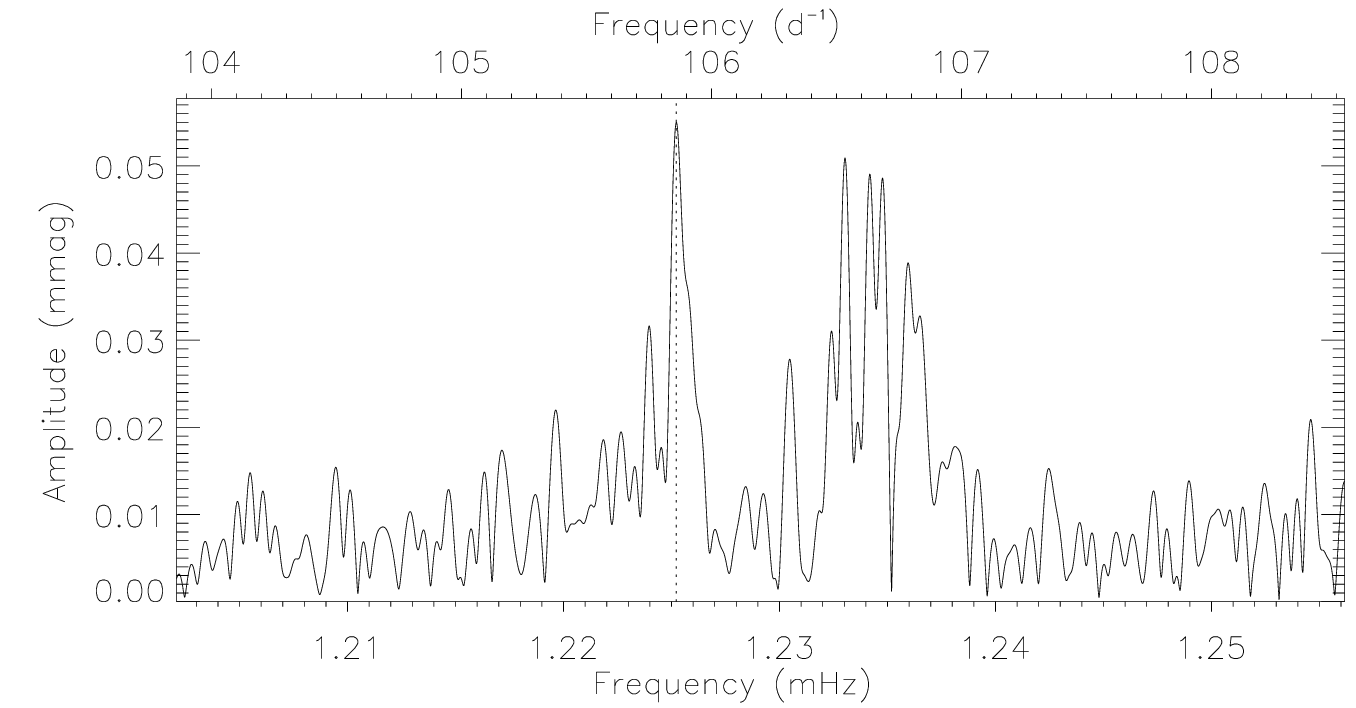}
\caption{Amplitude spectrum of TIC\,445493624. We have extracted the pulsation mode marked with the vertical dotted line. The other variability clustered around 1.235\,mHz is unresolved in this short set of data. }
\label{fig:445493624_ft}
\end{figure}

{\it TESS} observed TIC\,445493624 during sector 18. To derive the rotation period, we cleaned the SAP data of outlying points (which constituted about 5\,per\,cent of the data) since the signal in the PDC\_SAP data had been altered by the pipeline corrections. We derived a rotation period of $P_{\rm rot}=7.288\pm0.002$\,d, as shown in Fig.\,\ref{fig:445493624_rot}. To inspect the high frequency variability, we returned to the PDC\_SAP data and removed the low-frequency peaks to the noise level in the frequency range of interest. Here we found significant evidence of variability (Fig.\,\ref{fig:445493624_ft}). There are many peaks with amplitudes significantly above the noise level, however many are unresolved. We extracted a single mode, which itself does not resemble the window function, at a frequency of $1.22522\pm0.00003$\,mHz ($105.859\pm0.002$\,\cd). Since the frequency resolution for the data set is 0.4 $\umu$Hz, which is significantly smaller than the expected small separation in roAp stars, the evidence suggests amplitude and/or frequency variability of the mode. This is not unprecedented in the Ap stars, even on short time scales, (e.g., HD\,60435; \citealt{2019MNRAS.487.2117B}).

To confirm the chemical peculiarity in this star, we have obtained a spectrum using the HERMES spectrograph (Figs.\,\ref{fig:HERMES_BLUE} \& \ref{fig:HERMES_RED}). We found clear presence of Sr\,{\sc{ii}}  and Eu\,{\sc{ii}}, as well as enhancements of other rare earth elements in the region around 6150\,\AA. Using the procedures outlined above, we derived the following parameters from the spectrum: $T_{\rm eff}=8370\pm90$\,K, $\log g=4.0\pm0.3$\,\cms\ and $v\sin i =14\pm2$\,\kms.

With the evidence we have gathered, and that in the literature, we admitted  TIC\,445493624 to the class of roAp stars.


\subsubsection{TIC\,467074220}

TIC\,467074220 (HD\,40142) has several classifications in the literature: Ap\,Sr-Eu \citep{1965PASP...77..184C}; F0p \citep{1993AJ....105.1903N}; F0 \citep{1918AnHar..92....1C}. Its estimated effective temperature is $T_{\rm eff}=7200$\,K from {\it Gaia} DR2 \citep{2018A&A...616A...1G, 2018yCat.1345....0G} and 6950\,K from \citep{2012MNRAS.427..343M}, while \citet{2022ApJS..259...63S} determined 7580\,K from LAMOST spectra. {\it Gaia} DR2 also provided $\log L/{\rm L_\odot}=1.37$. \citet{1998A&AS..129..431H} provide Str\"omgren-Crawford indices  of: $b-y=0.216$, $m_1=0.175$, $c_1=0.869$ and H$\beta=2.814$, giving $\delta m_1= 0.031$ and $\delta c_1=0.064$, indicating a normal star.

This star has a magnetic field first reported by \citet{2008AstBu..63..139R} with a root mean value of $\langle B_\ell\rangle = 700\pm50$\,G with three magnetic field observations from the 6-m telescope of the Special Astrophysical Observatory of the Russian Academy of Sciences. The extremes measured were $-780$\,G and $+780$\,G.

Pulsations have been searched for in this star before by \citet{1993AJ....105.1903N} using ground-based $V$ and $B$ photometric data, but with none found. No other searches have been published. 

\begin{figure}
\centering
\includegraphics[width=\columnwidth]{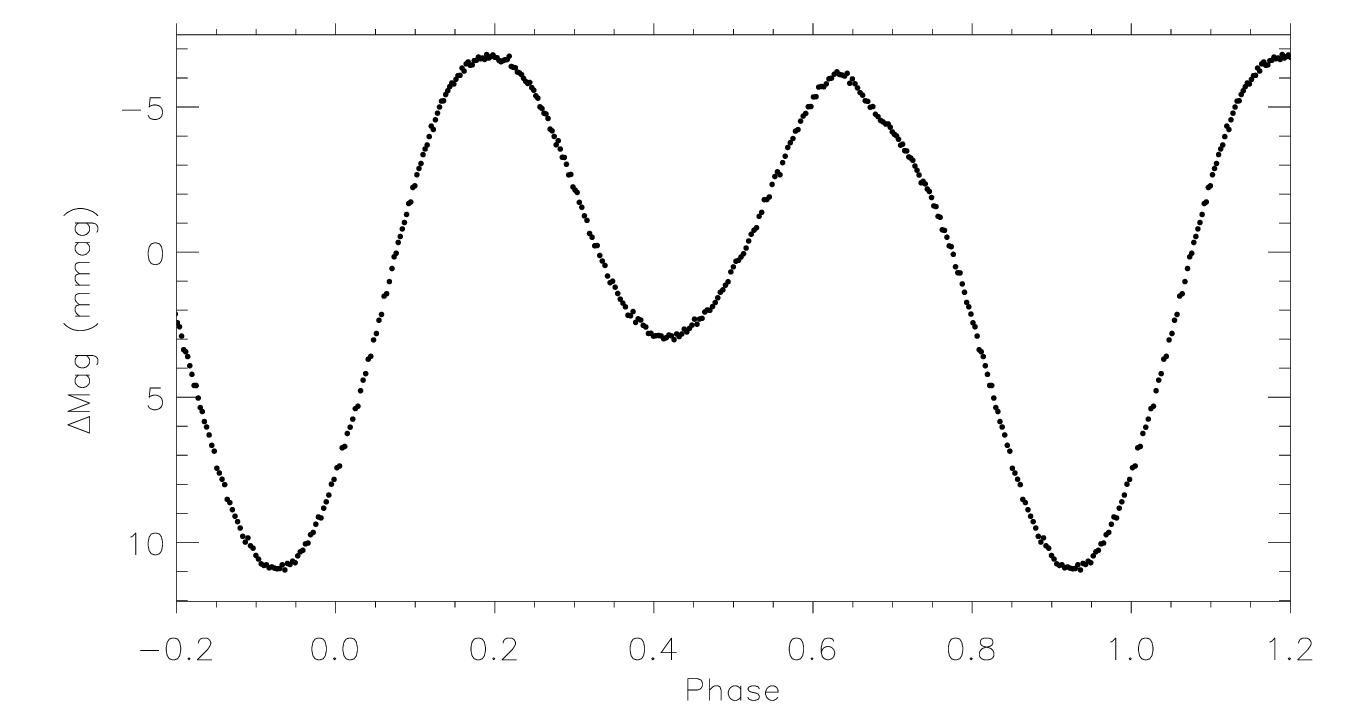}
\caption{Phase folded light curve of TIC\,467074220, phased on a period of $5.4195\pm0.0004$\,d. The data are binned 50:1.}
\label{fig:467074220_rot}
\end{figure}

\begin{figure}
\centering
\includegraphics[width=\columnwidth]{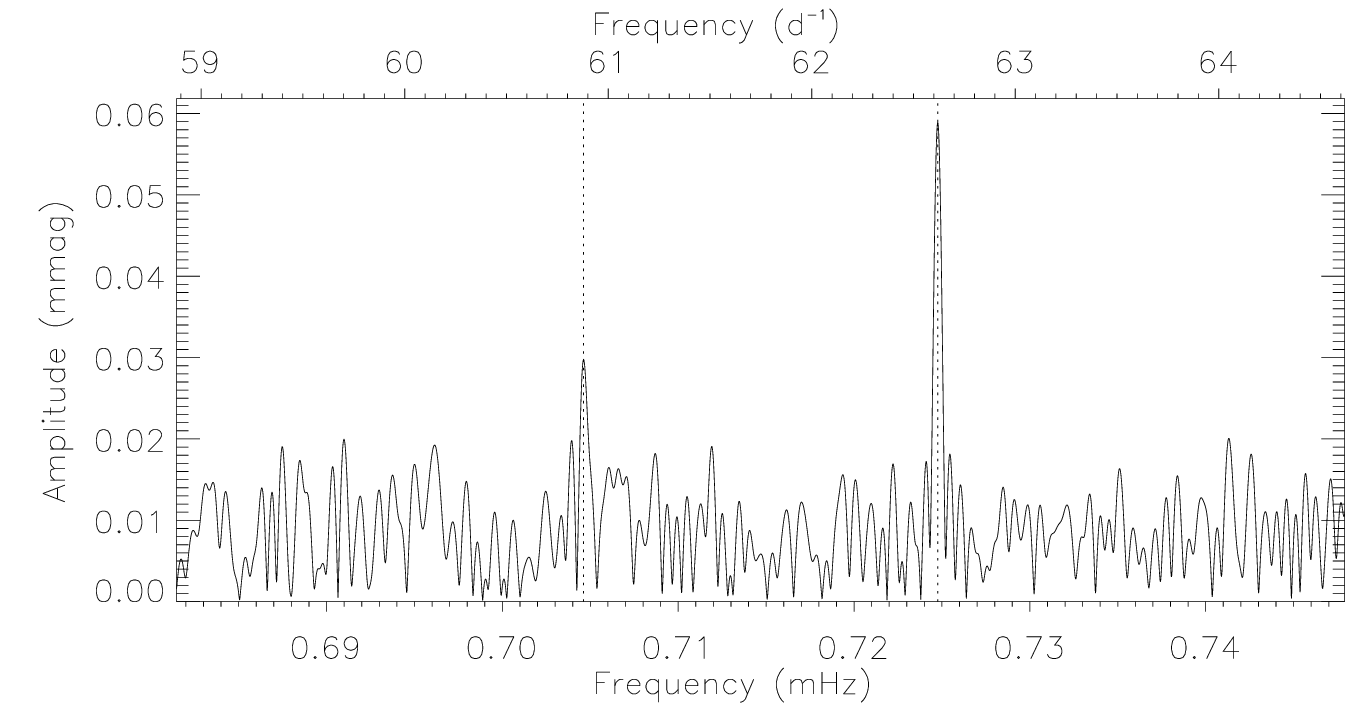}
\caption{Amplitude spectrum of TIC\,467074220. The two pulsation modes marked with vertical dotted lines.}
\label{fig:467074220_ft}
\end{figure}

{\it TESS} observed this star during sector 19. The data show a clear rotation signature with a period of $P_{\rm rot}=5.4195\pm0.0004$\,d (Fig.\,\ref{fig:467074220_rot}). This is the first report of this signature in the literature. Further, we detected two pulsation modes (Fig.\,\ref{fig:467074220_ft}) at $0.70462\pm0.00006$\,mHz ($60.879\pm0.005$\,\cd) and $0.72476\pm0.00003$\,mHz ($62.619\pm0.002$\,\cd). These modes are similar in frequency to $\delta$\,Sct modes, but there are only two modes detected which is unusual for $\delta$\,Sct stars. At this slightly higher frequency, there are more examples of roAp stars with similar pulsations in the literature (e.g., HD\,177765; \citet{2012MNRAS.421L..82A,2016IBVS.6185....1H}).

We have extracted a spectrum of TIC\,467074220 from the LAMOST archive to confirm the spectral classification (Fig.\,\ref{fig:467074220_spec}). There is obvious over enhancements of Sr\,{\sc{ii}} and Cr\,{\sc{ii}}, with an indication of Eu\,{\sc{ii}}.

With a combination of the rotational modulation in the star, the detected magnetic field and observed chemical peculiarities, we admitted this star to the class of roAp.

\subsection{roAp stars previously discovered by {\it TESS}} 
\label{sec:TESS_roAp}
In this section, we present the roAp stars detected in {\it TESS} data by our method but which have already been published as roAp stars in the literature. 

\subsubsection{TIC\,21024812}

TIC\,21024812 (HD\,14522) was originally identified as a chemically peculiar magnetic star by \citet{1986A&A...161..203S}, and \citet{2009A&A...498..961R} classified it as A2\,SrEu. \citet{2013A&A...553A..95M} estimated the star's effective temperature using principal component analysis techniques to compare it to a homogeneous library of stellar spectra, and derived $T_{\rm eff} = 7750\,K$, in general agreement with the TIC value of $8050\pm130$\,K \citep{2019AJ....158..138S} and the {\it Gaia} DR3 value of 7950\,K \citep{2020yCat.1350....0G}. {\it Gaia} DR3 also derived a distance of 282\,pc, $\log g =  3.93$\,\cms\, and [Fe/H] = -0.11, while the TIC estimated the stellar radius to be $2.49\pm0.08$\,R$_{\sun}$ and mass to be $1.96\pm0.30$\,M$_{\sun}$.

\citet{2020MNRAS.493.3293B} made use of archival photometric data from a variety of ground-based surveys to estimate the rotational period to be $26.36\pm0.01$\,d, with an amplitude in $V$ of about 0.01 mag, while \citet{2006A&A...455..303J} conducted a 1.4-hr search for pulsations using a three-channel photometer on the ARIES 1.04-m Sampurnanand telescope, but found no variability.

TIC\,21024812 was reported as an roAp star by \citet{2022MNRAS.510.5743B} who derived a rotation period of $P_{\rm rot}=13.158$\,d and a pulsation frequency of $1.103$\,mHz ($95.265$\,\cd) using {\it TESS} sector 18 data. However, our more detailed analysis of the same sector 18 observations suggests that the rotation period is about twice that reported. We used the SAP light curve, which showed significant variability consistent with stellar rotation, to find that the signal is non-repeating over the light curve duration of $24.356$\,d. This is consistent with the rotation period derived by \citet{2020MNRAS.493.3293B} of $26.36\pm0.01$\,d. Further observations of this star are required to constrain the rotation period with precision.

\begin{figure}
\centering
\includegraphics[width=\columnwidth]{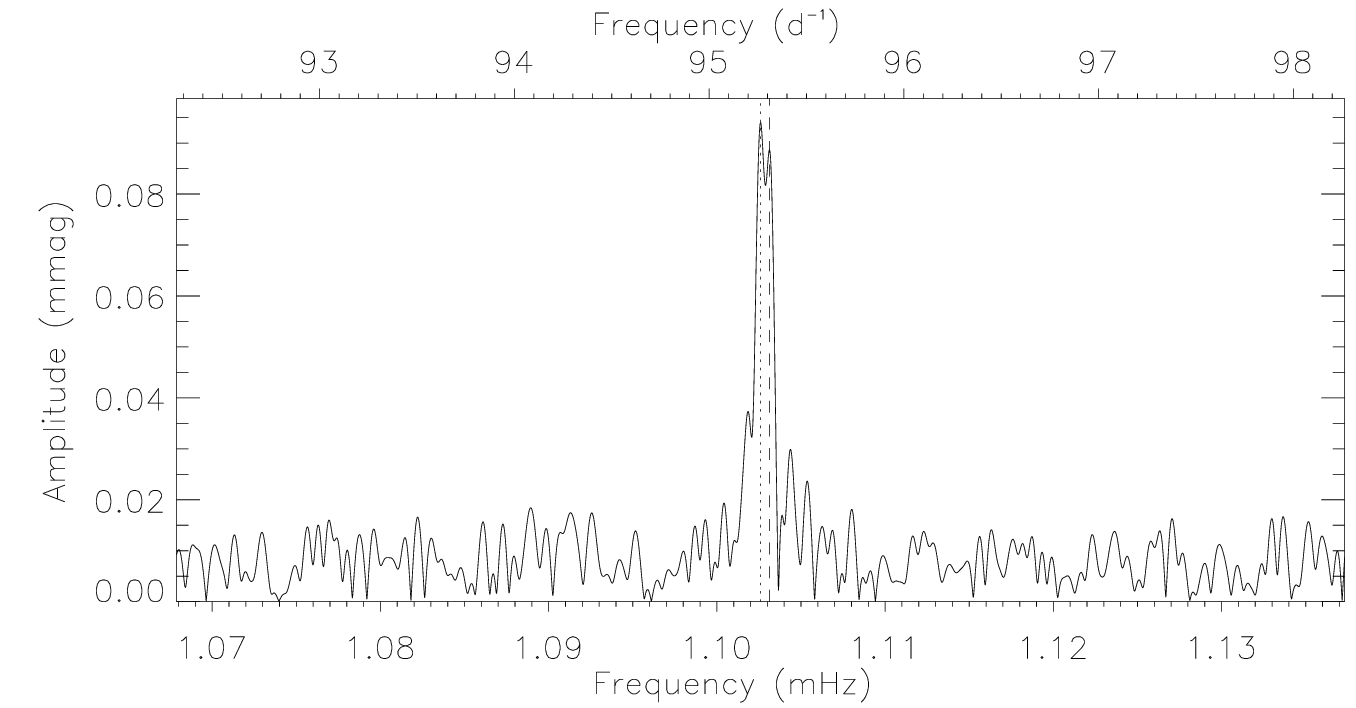}
\caption{Amplitude spectrum of TIC\,21024812, with the pulsation mode indicated with the vertical dotted line, and the suspected positive rotational sidelobe with the dashed line. The peak is seen to be broad, indicating amplitude/phase modulation, or unresolved sidelobes.}
\label{fig:21024812_ft}
\end{figure}

The short period variability seen in the amplitude spectrum of this star (Fig.\,\ref{fig:21024812_ft}) is seen at a frequency of $1.10259\pm0.00002$\,mHz ($95.263\pm0.002$\,\cd) with the indication of a sidelobe on the higher frequency side of the pulsation. The peak in the amplitude spectrum appears broad which may be caused by amplitude/phase modulation, or the presence of unresolved sidelobes.

Further to the classification of TIC\,21024812 being a chemically peculiar star in the literature, we have two sources of spectra of this star: LAMOST and HERMES. The LAMOST spectrum was obtained in its low resolution mode, and can be seen in Fig.\,\ref{fig:21024812_LAMOST}, while the HERMES spectrum is shown in Figs.\,\ref{fig:HERMES_BLUE} \& \ref{fig:HERMES_RED}. From the LAMOST spectrum, we classified this star as A8\,SrEu(Cr). Using the procedures outlined above, we derived the following parameters from the HERMES spectrum: $T_{\rm eff}=8250\pm70$\,K, $\log g=4.0\pm0.3$\,\cms\ and $v\sin i =8\pm3$\,\kms.


\subsubsection{TIC\,25676603}

TIC\,25676603 (TYC\,2194-2347-1) is a slowly rotating Fp star, with particularly strong absorption in the Sr\,{\sc{ii}} 4077\,\AA\ line and $v\sin i <1.1$\,\kms\ \citep{1991A&A...244..335N}. There are not many mentions of this star in the literature where detailed studies have been conducted. The {\it Gaia} DR3 gives an effective temperature of $T_{\rm eff}=6650$\,K and a $\log g=4.1$\,\cms. These are consistent with the spectral types of Ap\,SrEu \citep{1966VA......8...53B} and the Fp class.

TIC\,25676603 was first reported as an roAp star by \citet{2022MNRAS.510.5743B} who claimed to identify the rotation period to be $P_{\rm rot}=5.495$\,d and a pulsation mode at a frequency of $1.820$\,mHz ($157.205$\,\cd). However, our analysis of the {\it TESS} sector 15 data do not show a clear signal of rotation. We analysed the SAP data to investigate if a long rotation period has been removed by the PDC pipeline. Indeed, we found some variability that is consistent with a long rotation period, as would be expected from the previous measurement of the upper limit on $v\sin i$. We therefore concluded that the rotation period is longer than the current {\it TESS} data set, and the value published by \citet{2022MNRAS.510.5743B}  should be disregarded as noise or of instrumental origin.

\begin{figure}
\centering
\includegraphics[width=\columnwidth]{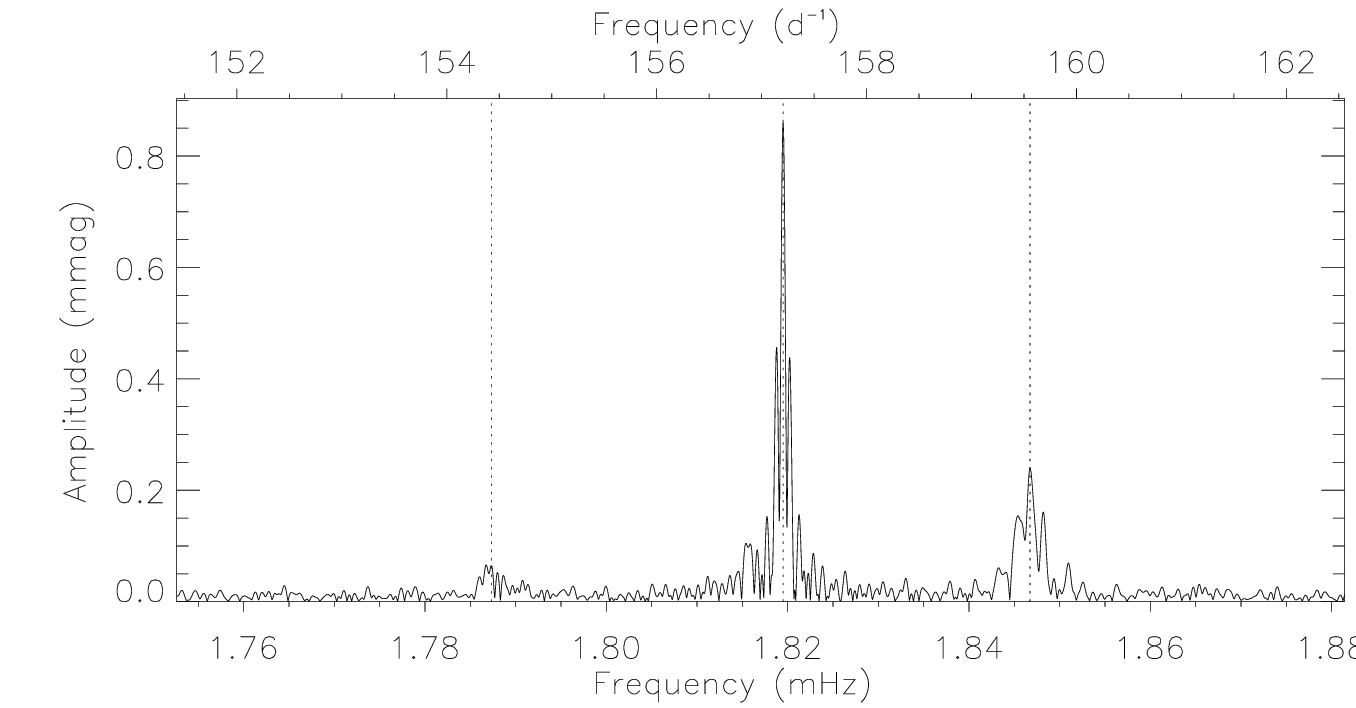}
\caption{Amplitude spectrum of TIC\,25676603, with the pulsation modes indicated with the vertical dotted lines.}
\label{fig:25676603_ft}
\end{figure}

We identified three pulsation modes in the PDC\_SAP sector 15 data for this star, at frequencies of $1.787321\pm0.000027$\,mHz ($154.4245\pm0.0024$\,\cd), $1.819500\pm0.000002$\,mHz ($157.2048\pm0.0002$\,\cd) and $1.846724\pm0.000008$\,mHz ($159.5569\pm0.0007$\,\cd). All three peaks show structure (Fig.\,\ref{fig:25676603_ft}), suggesting that the rotation period is not fully resolved in 27\,d of {\it TESS} data, confirming our previous conclusion. The three modes are not separated by the same frequency spacing ($32.18\pm0.03\,\muup$Hz and $27.22\pm0.01\,\muup$Hz), but since the modes are not fully resolved this view may change with the addition of future observations. 

\subsubsection{TIC\,26833276}

TIC\,26833276 (HD\,10682) is not well studied in the literature. The only classification giving an indication that this is a chemically peculiar star is from \citet{1979RA......9..479C} with F0\,Vp\,Sr. However, this is at odds with the classification of A3 \citep{1918AnHar..91....1C}. Catalogue values of the effective temperature range from 6750\,K \citep{2012MNRAS.427..343M} to 8700\,K \citep{2003AJ....125..359W}, while {\it Gaia} DR3 gives 7300\,K \citep{2022yCat.1355....0G}. The hottest temperature is based on the spectral type alone, while the others are derived from measurements. Therefore we suggest the effective temperature is closer to 7000\,K which supports the classification of an F0 star.

\citet{2022MNRAS.510.5743B} classified the star as an roAp star given the presence of a pulsation mode at a frequency of $1.583$\,mHz ($136.786$\,\cd), as well as a rotation period of $10.417$\,d, as seen in the {\it TESS} sector 17 data. However, a detailed view of the same {\it TESS} data here does not support the conclusion of the rotation period. This is another case where the PDC pipeline has altered the rotation signal since it appears to be longer than the data set. When inspecting the SAP data, it is clear there is a partial rotation curve that does not repeat. We therefore conclude that the rotation period is greater than the sector length, i.e., >25\,d, and the previous value presented in the literature should be ignored. 

\begin{figure}
\centering
\includegraphics[width=\columnwidth]{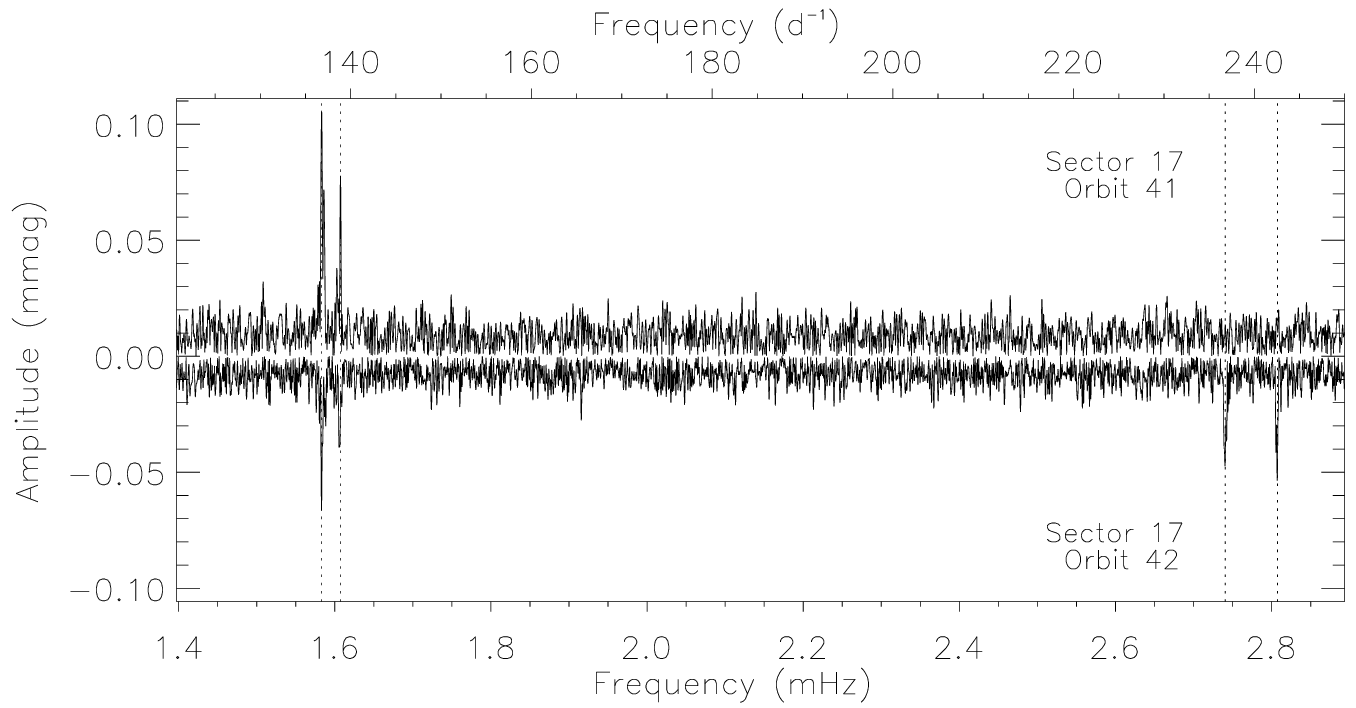}
\caption{Amplitude spectra of TIC\,26833276 calculated from the two {\it TESS} orbits when it was observed. The amplitude spectrum with positive amplitude is from orbit 41, with two pulsation modes around 1.6\,mHz. The amplitude spectrum with negative amplitude is from orbit 42, showing the emergence of two further modes at high frequency (around 2.8\,mHz).}
\label{fig:26833276_ft}
\end{figure}

The pulsations in this star are irrefutable. There are two groups of two modes; the lower frequency pair appear at $1.58317\pm0.00002$\,mHz ($136.786\pm0.001$\,\cd) and $1.60714\pm0.00003$\,mHz ($138.857\pm0.002$\,\cd) and are present in all of the sector 17 observations. The two modes at higher frequency are transient in nature and only appear in the second orbit of sector 17 (Fig.\,\ref{fig:26833276_ft}). These modes are at $2.74058\pm0.00004$\,mHz ($236.786\pm0.003$\,\cd) and $2.80741\pm0.00004$\,mHz ($242.560\pm0.003$\,\cd). We are unsure of the nature of the high-frequency pair.

This star has LAMOST MRS data (Fig.\,\ref{fig:26833276_LAMOST}) available in the archive which we analysed to identify chemical peculiarities and to resolve the discrepancy between the spectral types. We found an overabundance of Nd\,{\sc{iii}} and La\,{\sc{ii}}, amongst others, confirming the star to be chemically peculiar. By fitting the H$_\alpha$ line, we determine an effective temperature of $7200\pm200$\,K. Therefore, we can confirm the star is an F0p chemically peculiar star.


\subsubsection{TIC\,72392575}

TIC\,72392575 (HD\,225578) has an uncertain spectral classification; it is given as ApSr: by \citet{1964PASP...76..119C}, while it is classified as an A5 star in the Henry Draper Catalog and Extension \citep{1923AnHar..98....1C}. More recently, \citet{2009A&A...498..961R} gave the spectral classification for this star as A5\,Sr. 

\citet{2019MNRAS.483.2300S} and \citet{2019AJ....158..138S} estimate the effective temperature and $\log g$ values to be 7550\,K and 4.14\,\cms, respectively.  The effective temperature is determined from the dereddened Bp-Rp colour, while the $\log g$ value is derived from the given mass and radius. Other estimations of $T_{\rm eff}$ are 7650\,K \citep{2006ApJ...638.1004A} and 7850\,K \citep{2019A&A...628A..94A}

\citet{2022MNRAS.510.5743B} listed this star as a new roAp star discovered using {\it TESS} data, and also classified it as an $\alpha^{2}$\,CVn variable with a rotation period of $P_{\rm rot}=3.922$\,d.  He also gave an estimate of the luminosity $\log L/\rm{L}_{\odot}$= 1.07 using {\it Gaia} EDR3 parallaxes \citep{2016A&A...595A...1G, 2021A&A...649A...1G} in conjunction with interstellar reddening corrections from \citet{2017AstL...43..472G} and a bolometric correction calibration by \citet{2013ApJS..208....9P}.  

A previous search for pulsational variability did not yield a positive detection down to 1.6\,mmag in the amplitude spectrum \citep{2012A&A...542A..89P}.  However, \citet{2022MNRAS.510.5743B} found roAp variability in the {\it TESS} light curve, with the highest amplitude frequency at $1.573$\,mHz ($135.944$\,\cd). There are no magnetic field measurements for this star in the literature.  

\begin{figure}
\centering
\includegraphics[width=\columnwidth]{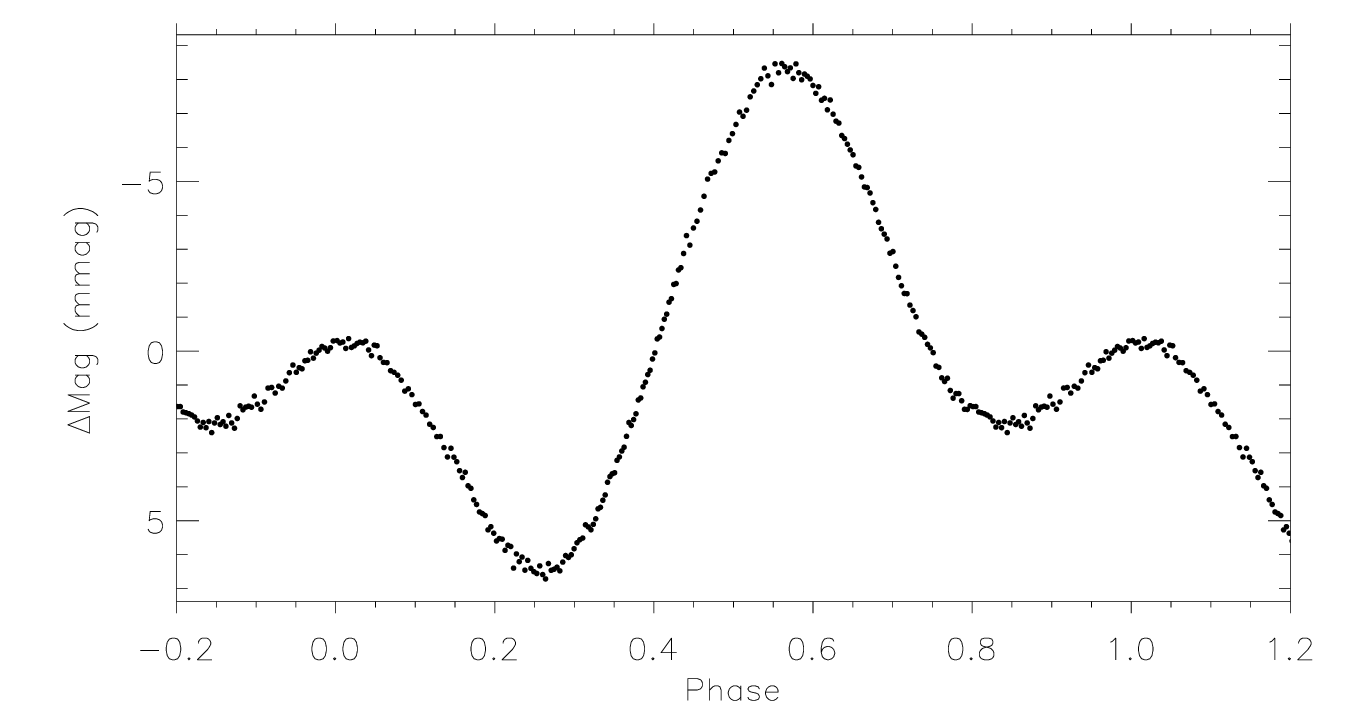}
\caption{Phase folded light curve of TIC\,72392575, phased on a period of $3.9016\pm0.0008$\,d. The data are binned 50:1.}
\label{fig:72392575_rot}
\end{figure}

\begin{figure}
\centering
\includegraphics[width=\columnwidth]{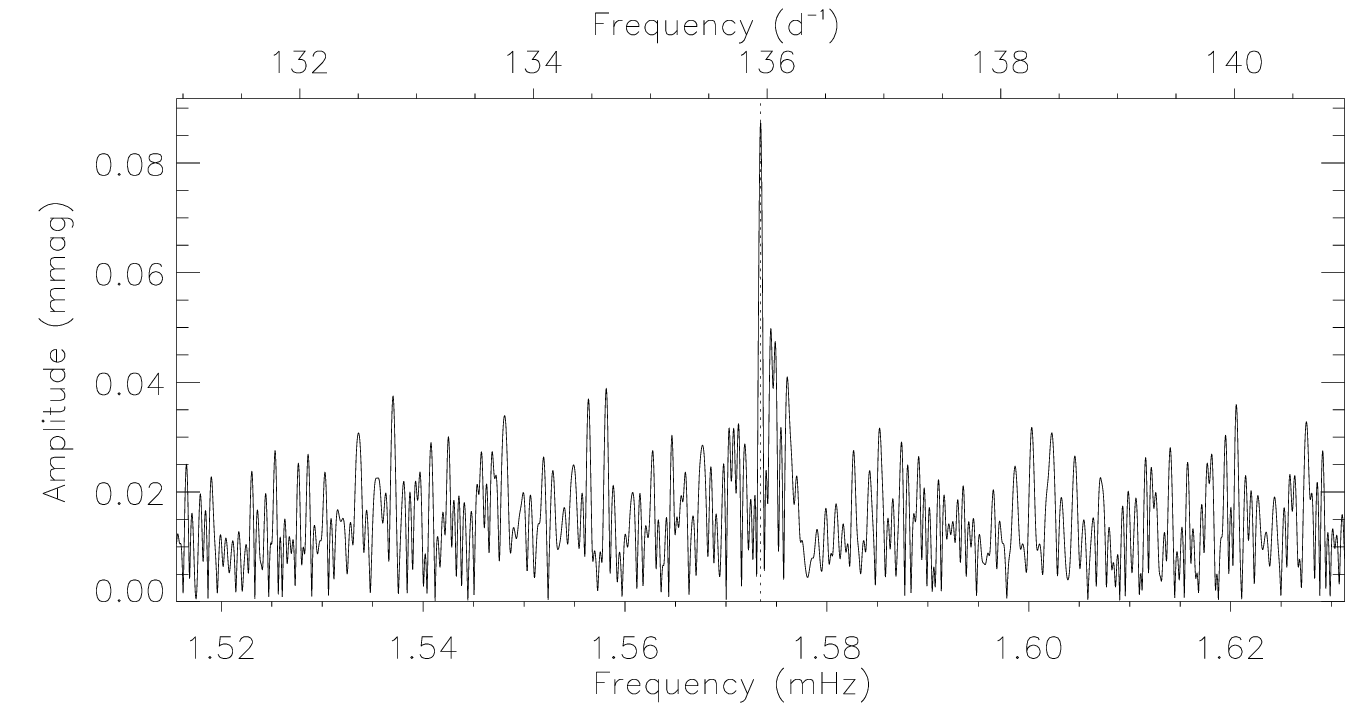}
\caption{Amplitude spectrum of TIC\,72392575, showing the pulsation mode. At slightly higher frequency there is a power excess which might be a rotational sidelobe (the single sharp peak).}
\label{fig:72392575_ft}
\end{figure}

Our analysis of the {\it TESS} light curve, from sector 14, confirms the rotation period as $3.9016\pm0.0008$\,d (Fig.\,\ref{fig:72392575_rot}), and the single high frequency pulsation at $1.57343\pm0.00003$\,mHz ($135.944\pm0.003$\,\cd; Fig.\,\ref{fig:72392575_ft}). Despite the short rotational period of this star, we do not detect rotational sidelobes to the pulsation, suggesting a low value of $i$ or $\beta$ for this star. However, there is a slight power excess at slightly higher frequency which may be a signature of a rotational sidelobe, but this is unclear.

\subsubsection{TIC\,118247716}

TIC\,118247716 (HD\,12519) has several spectral classifications in the literature: A5 \citep{1918AnHar..91....1C}, A4pSrEuCr \citep[with a comment of `composite?';][]{1979RA......9..479C} and A2pSrSi \citep{1999A&AS..137..451G}.  \citet{2019AJ....158..138S} estimate $T_{\rm eff} = 7050$\,K, and $\log g = 4.04$\,\cms. The effective temperature value is derived from a dereddened Bp-Rp colour. The surface gravity is calculated from the reported mass and radius of the star. {\it Gaia} DR3 estimates an effective temperature of 7100\,K and $\log g =  4.11$\,\cms\,\citep{2022yCat.1355....0G}. There are two consistent radial velocity estimates for this star: $-5.9\pm3.3$\,\kms\, \citep{2006AstL...32..759G} and more recently $-8.9\pm0.6$\,\kms\, from the {\it Gaia} DR3 data \citet{2022yCat.1355....0G} 

This star was targeted as part of the Nainital-Cape survey \citep{2006A&A...455..303J} which searched for photometric variability. The star was included in the target list given its Str\"omgren-Crawford indices of $b-y=0.238$, $m_1=0.289$, 	$c_1=0.546$ and H$\beta=2.803$ \citep{2015A&A...580A..23P}, giving $\delta m_1= -0.086$ and $\delta c_1=-0.239$, being indicative of a chemically peculiar star, and its $\Delta p$ peculiarity parameter of 5.989 \citep{1998A&AS..128..265M}. 

\begin{figure}
\centering
\includegraphics[width=\columnwidth]{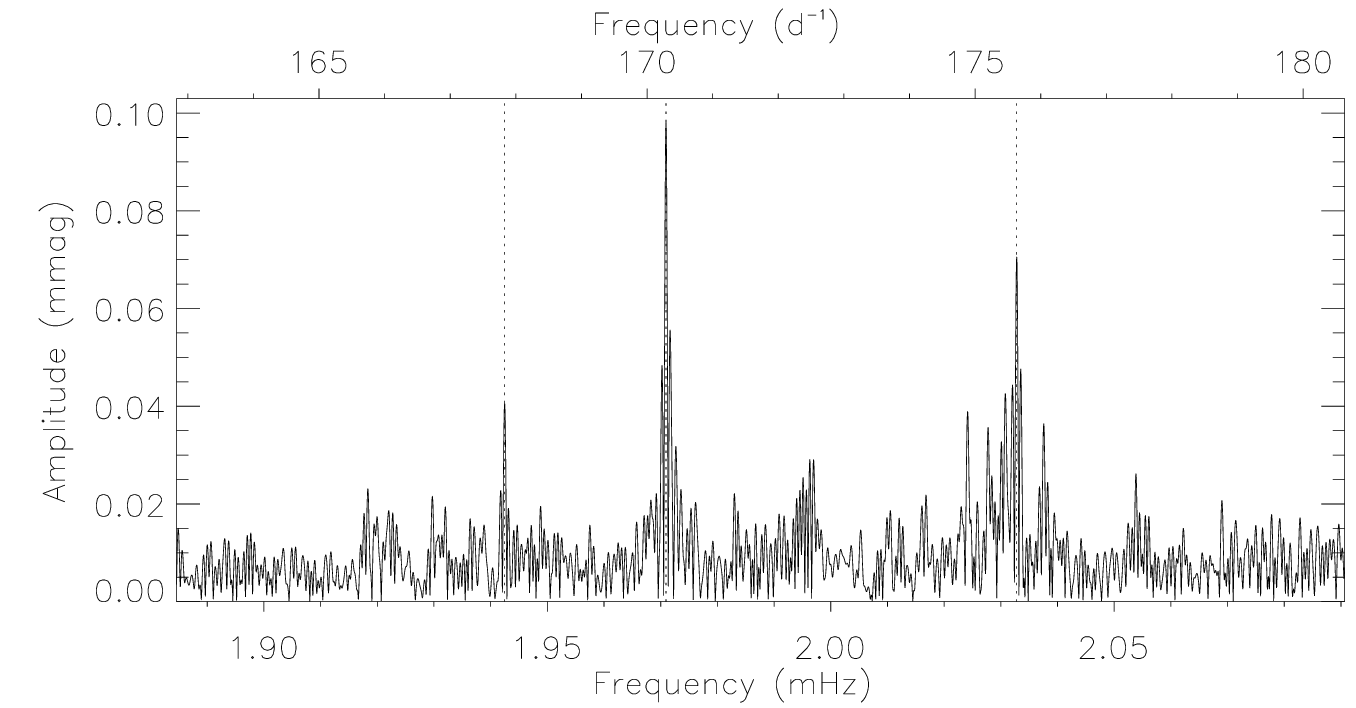}
\caption{Amplitude spectrum of TIC\,118247716, with three modes clearly identifiable. Further modes may be present close to the noise level.}
\label{fig:118247716_ft}
\end{figure}

TIC\,118247716 was first classified as an roAp star using {\it TESS} data \citep{2022MNRAS.510.5743B}. We corroborate this with the detection of three pulsation modes in the typical roAp frequency range (Fig.\,\ref{fig:118247716_ft}): $1.94247\pm0.00004$\,mHz ($167.829\pm0.003$\,\cd), $1.97093\pm0.00002$\,mHz ($170.2884\pm0.001$\,\cd), and $2.03280\pm0.00002$\,mHz ($175.634\pm0.002$\,\cd). There appear to be further modes in this star, but close to the noise level. Thus, we did not extract them, but suggest this star is revisited when more data are available. We see no rotationally split sidelobes to the pulsations which is consistent with the lack of rotational modulation in the light curve. 

This star has a LAMOST low resolution spectrum available in the archive, which we used to confirm its spectral type of A7p\,SrEuCr (Fig.\,\ref{fig:118247716_spec}). This cooler classification is justified by the lack of convincing Si lines in the spectrum, which are associated with hotter CP stars, and that the Balmer lines are well matched with that of an A7 normal star. This cooler classification also agrees better with the literature effective temperatures.


\subsubsection{TIC\,120532285}

TIC\,120532285 (HD\,213258) is a high proper motion star \citep{2020yCat.1350....0G}, with a measured radial velocity of $-86.8$\,\kms\ \citep{1991A&A...244..335N}. It is classified as an A3 star by \citet{1924AnHar..99....1C}, and an F\,str\,$\lambda$\,4077 star\footnote{These stars are spectroscopically similar to Am stars, but with an abnormally strong Sr\,{\sc{ii}} 4077\,\AA\ line \citep{1981AJ.....86..553B,1987A&A...186..191N}} by \citet{1985AJ.....90..341B}. {\it Gaia} DR2  estimates an effective temperature in the range 6750\,K to 7350\,K, a radius range of $1.75-2.09$\,R$_{\odot}$, and luminosity range of 8.07-8.12\,L$_{\odot}$ \citep{2018A&A...616A...1G}. \citet{2015A&A...580A..23P} provides Str\"omgren-Crawford indices of $b-y=0.222$, $m_1=0.187$ and $c_1=0.659$. Most recently, \citet{2023A&A...670A..72M} measured a mean magnetic field modulus of this star to be $\langle B\rangle\sim3.8$\,kG and a longitudinal field of about $-0.9$\,kG, whilst also concluding that the star must have a rotation period of the order 50\,yr. They also suggested that the star could be host to a substellar companion with an orbital period suspected to be shorter than 100\,d. They were also the first authors to report this star as an roAp star, with three pulsation modes, after analysing {\it TESS} data.

\begin{figure}
\centering
\includegraphics[width=\columnwidth]{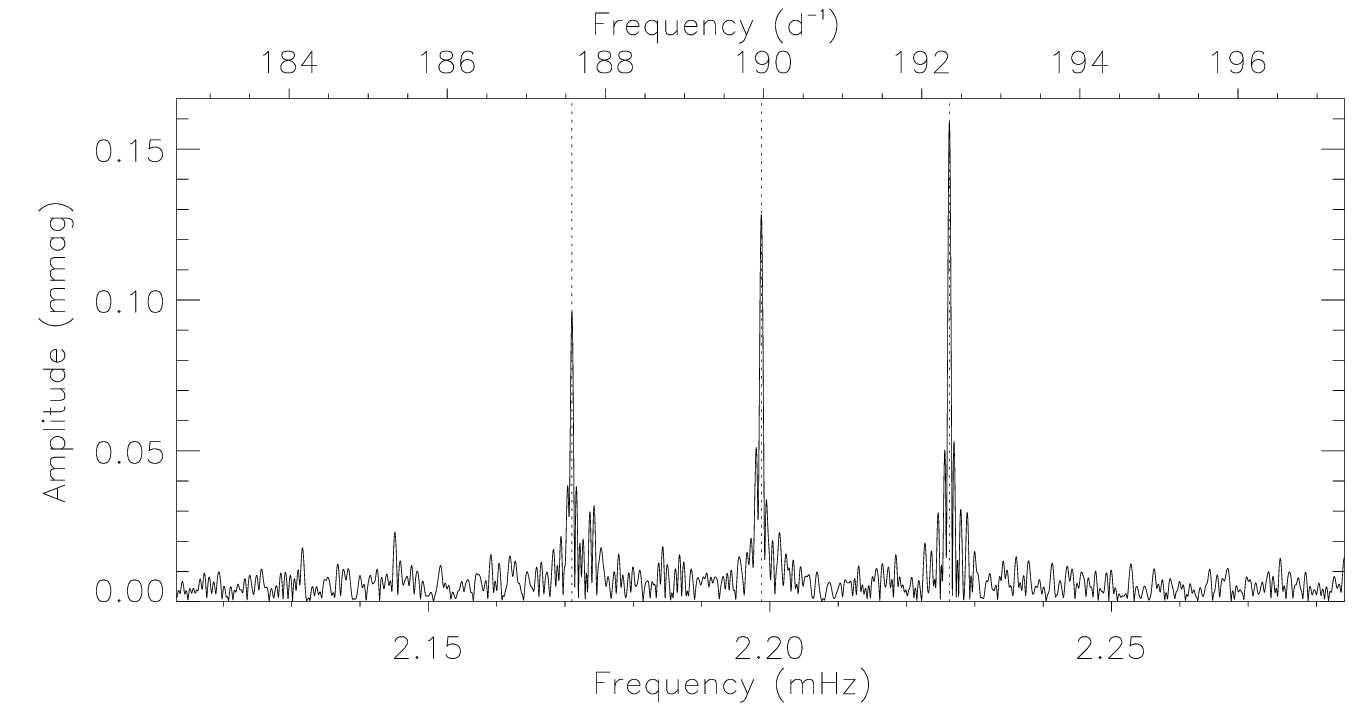}
\caption{Amplitude spectrum of TIC\,120532285. The pulsation modes are highlighted by vertical dotted lines, while the peak at about 2.145\,mHz is not significant but it does follow separation pattern of the other modes.}
\label{fig:120532285_ft}
\end{figure}

{\it TESS} observed TIC\,120532285 during sector 16. The light curve showed no signature of rotation, as expected given the conclusions of \citet{2023A&A...670A..72M}, but does show high frequency variability as previously reported. We detect three modes in the amplitude spectrum (Fig.\,\ref{fig:120532285_ft}) at frequencies of $2.171017\pm0.000011$\,mHz ($187.5759\pm0.0009$\,\cd), $2.198746\pm0.000008$\,mHz ($189.9716\pm0.0007$\,\cd), and $2.226270\pm0.000006$\,mHz ($192.3497\pm0.0006$\,\cd). The separation between the modes (in increasing frequency) is $27.73\pm0.01\,\umu$Hz and $27.52\pm0.01\,\umu$Hz which are plausibly half of the large frequency separation, given the stellar parameters in the literature. We note a fourth, but not significant, peak in the amplitude spectrum that also follows the separation pattern, at the lower frequency side of the extracted three peaks. 

\subsubsection{TIC\,129820552}

TIC\,129820552 (TYC\,2322-1440-1) is classified in the literature as an Ap\,SrEu star \citep{1998PASP..110..270B}. The {\it Gaia} collaboration, via DR2,  provide the following parameters: $T_{\rm eff}$ in the range $7200-7400$\,K; a radius in the range of $1.63-1.72$\,R$_{\odot}$, and a luminosity range of $7.06-7.41$\, L$_{\odot}$  \citep {2018A&A...616A...1G}. Previously, \citet{2017MNRAS.468.2745N} determined the rotation period of the star to be 6.221\,d. This star also appeared in the work of \citet{2022MNRAS.510.5743B} who provided a rotation period of 6.211\,d and identified a high frequency pulsation signal at a frequency of $2.013$\,mHz ($173.928$\,\cd) based on the {\it TESS} observations.

\begin{figure}
\centering
\includegraphics[width=\columnwidth]{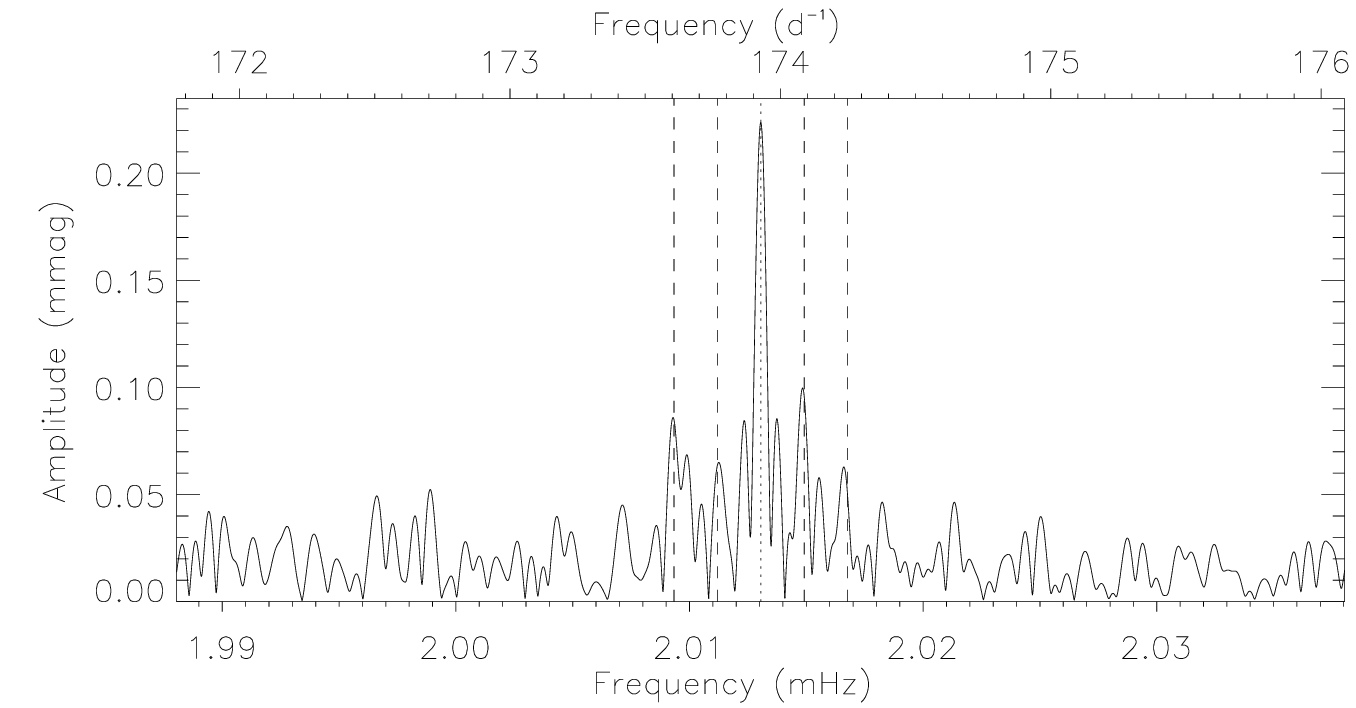}
\caption{Amplitude spectrum of TIC\,129820552 with the pulsation mode identified with the vertical dotted line, with the dashed lines indicating the rotationally split sidelobes. }
\label{fig:129820552_ft}
\end{figure}

Using those same data, obtained in sector 18, we derived a rotation period of $6.230\pm0.002$\,d which differs from the previous determinations (although no errors are given on previous work). We were also able to identify the pulsation mode in this star at a frequency of $2.01305 \pm 0.00002$\,mHz ($173.928\pm0.001$\,\cd; Fig.\,\ref{fig:129820552_ft}) and note that there are four rotational sidelobes present in the amplitude spectrum. By selecting the appropriate zero-point in time, we were able to fit the quintuplet such that the phases of the peaks were all in agreement, indicating this is a non-distorted quadrupole mode, and found $\tan i\tan\beta=2.95\pm0.61$, which implies $i\approx78^\circ$ and $\beta\approx32^\circ$, or vice versa.

\subsubsection{TIC\,394860395}

TIC\,394860395 (TYC\,2666-85-1) has been classified as Ap\,SrEu type peculiar star \citep{1985AJ.....90..341B}. Several studies have been carried out to reveal its magnetic structure. \citet{2006MNRAS.372.1804K} calculated the longitudinal component of the magnetic field from spectral observations of the depression at 5200\,\AA. Their measurements indicated that the star has displayed a magnetic field with the observed $B_\ell$ values ranging from $-520$ to $+540$ G. \citet{2009MNRAS.394.1338B} calculated the averaged quadratic longitudinal magnetic field as $\langle B_\ell \rangle =  380\pm111$\,G based on the measurements by \citet{2006MNRAS.372.1804K} using the metallic spectral lines. Recently, \citet{2020AstBu..75..294R} measured the longitudinal magnetic field and radial velocity values  by using the data taken in 2012. Accordingly, they found  $B_\ell= -340\pm40$\,G and  $V_{\rm{R}}=-32.5\pm3.5$\,\kms\ at JD\,2456140.484 and $B_\ell= 100\pm50$\,G and  $V_{\rm{R}}=-31.4\pm3.0$\,\kms\ at JD\,2456233.261. 

\citet{2013IBVS.6058....1P} reported that the star has shown a prominent peak at about 5\,mHz in the periodogram of Johnson $B$ photometric data and noted that such a high frequency had never been detected in any roAp star. However, \cite{2015A&A...575A..24P} could not find any variability at the previously reported frequency. They gave the effective temperature and interstellar reddening to be $T_{\rm{eff}}=6800\pm400$\,K and  E(B-V) of 0.08 mag, respectively. \citet{2013A&A...553A..95M} performed a principal component analysis to characterise the effective temperature from low-resolution spectroscopic data and calculated a temperature value of 7300\,K. 

\begin{figure}
\centering
\includegraphics[width=\columnwidth]{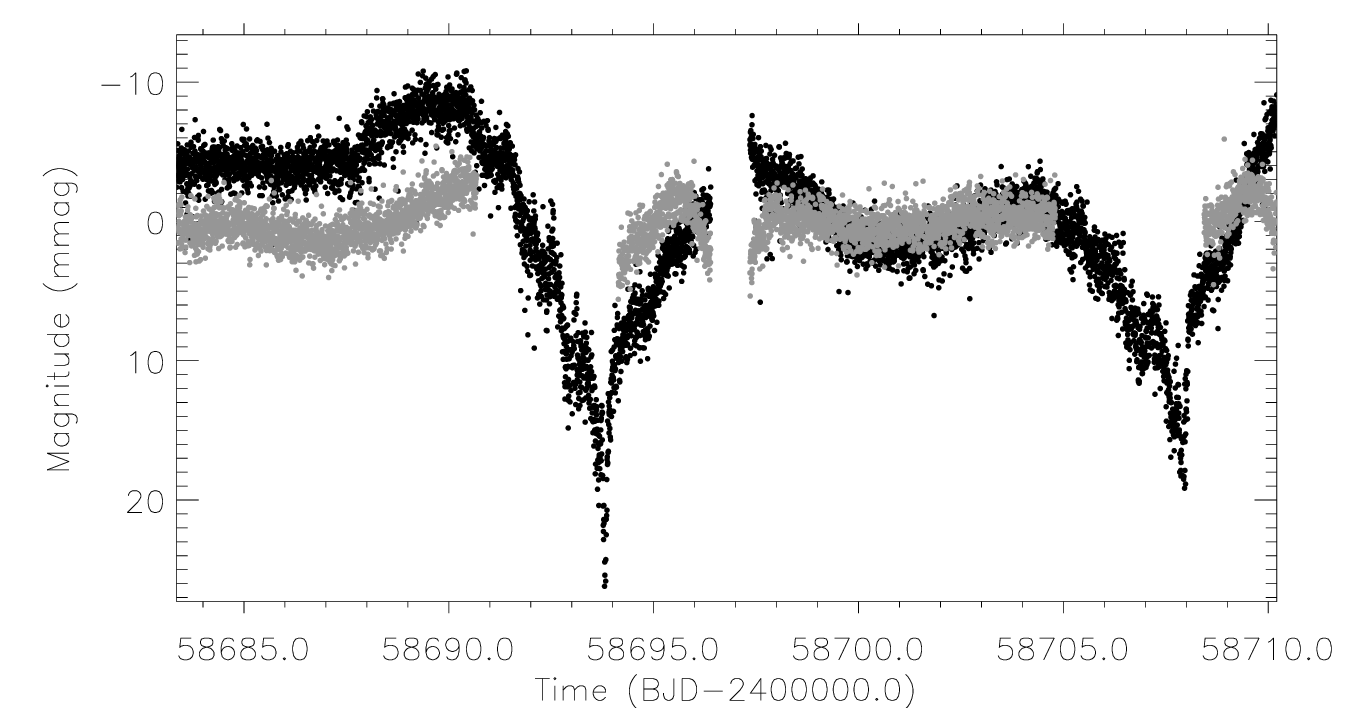}
\caption{Comparison of the PDC\_SAP light curve (grey) and the SAP light curve (black) for TIC\,394860395. Both light curves show variability, but we argue neither is indicative of a rotation period. The dimming in the SAP data are flagged as bad points in the PDC\_SAP data, and are excluded from the final light curve.}
\label{fig:394860395_lc}
\end{figure}

This star was reported to be a rotationally variable roAp star by \citet{2022MNRAS.510.5743B}. He measured a rotation period of $P_{\rm rot}=6.667$\,d and a pulsation frequency of $1.180$\,mHz ($101.949$\,\cd), using {\it TESS} data. Here we use the {\it TESS} sector 14 data to refute the previously determined rotation period. There is no clear indication in the PDC\_SAP data that the period above is of an astrophysical nature. We inspected the SAP data and found significant variability, although that does not wholly appear astrophysical either. We show both light curves in Fig.\,\ref{fig:394860395_lc}.

\begin{figure}
\centering
\includegraphics[width=\columnwidth]{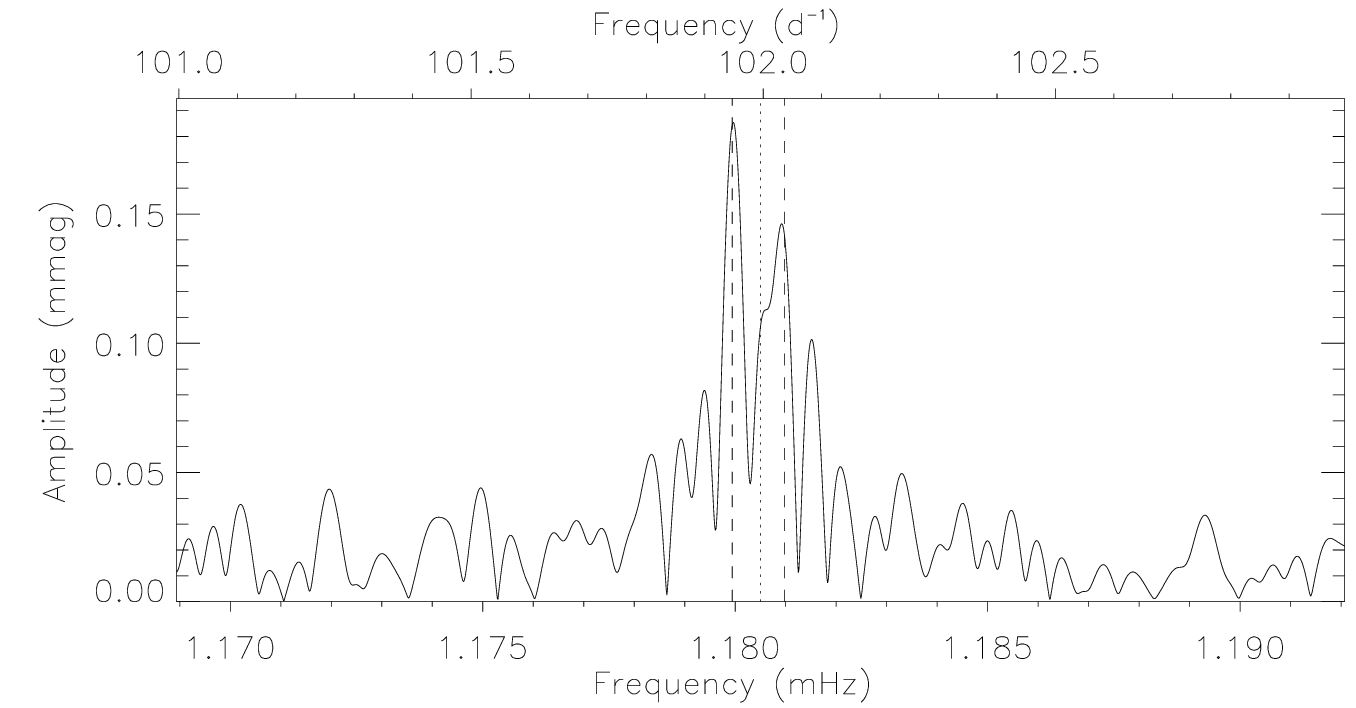}
\caption{Amplitude spectrum of TIC\,394860395 showing the pulsation modes (marked by the vertical dotted lines) and the rotationally split sidelobes (marked with the vertical dashed lines). }
\label{fig:394860395_ft}
\end{figure}

The pulsation signature in this star is also somewhat complicated (Fig.\,\ref{fig:394860395_ft}). At first inspection, it appeared there were two modes, but after removing those there remained a third peak exactly between the two previous ones. Therefore, we have taken the pulsation frequency to be the central peak at $1.18050\pm0.00005$\,mHz ($101.995\pm0.004$\,\cd) and deduced that the two outer peaks are in fact sidelobes due to rotation. Those sidelobes indicate that the rotation period is on the order of $23\pm2$\,d. This is an example of where caution should be exercised when deducing rotation periods with weak signals in the low frequency range in Ap stars.

\subsubsection{TIC\,435263600}

TIC\,435263600 (HD\,218439) was first classified as an A2p: star with Sr, Cr, and Si peculiar features \citep{1959ApJ...129...88S}, and more recently classified as an A5\,IVpSrSi star \citep{1985ApJS...59...95A}. The star is a known visual binary \citep[e.g.,][]{1987AJ.....93..688M, 1990AJ.....99..965M, 1994A&AS..105..503B, 1999AJ....117.1905G, 2006MNRAS.367.1170S}, with a close companion ranging in distance from the primary between 0.47\,arcsec to 0.62\,arcsec (depending on the epoch), with a magnitude difference of 0.5\,mag \citep{1976PASP...88..325H}. The mass of the primary is $1.6$\,M$_\odot$ \citep{2019A&A...623A..72K}, with the orbital elements of: $P_{\rm orb}= 415\pm 21$\,yrs, $e=0.51\pm0.05$, and the inclination of $i=165^\circ\pm3^\circ$ \citep{2010AJ....139.1521L}. 

Str\"omgren-Crawford indices were measured by \citet{1998A&AS..129..431H} to be: $b-y=0.187$, $m_1=0.138$, $c_1=1.008$ and H$\beta=2.841$, giving $\delta m_1= 0.070$ and $\delta c_1=0.154$. These indices suggest the star is perhaps metal deficient, but we note that a binary companion will affect the values and they should not be used to draw conclusions. These indices also suggest a temperature of about 8200\,K using relations of \citet{1985MNRAS.217..305M}, while an effective temperature of 8300\,K was measured through the intensity of the Ca\,{\sc{ii}}\,K line \citep{1981A&A...101..176F}.

\begin{figure}
\centering
\includegraphics[width=\columnwidth]{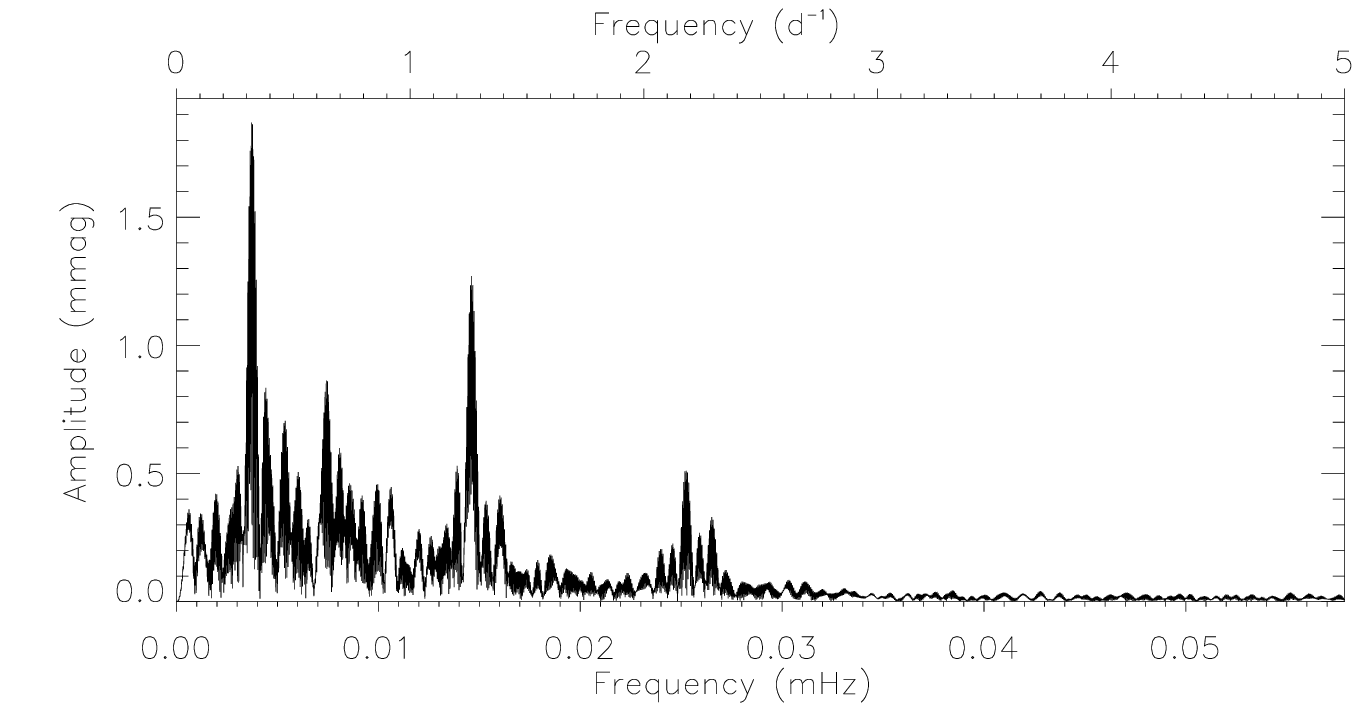}
\includegraphics[width=\columnwidth]{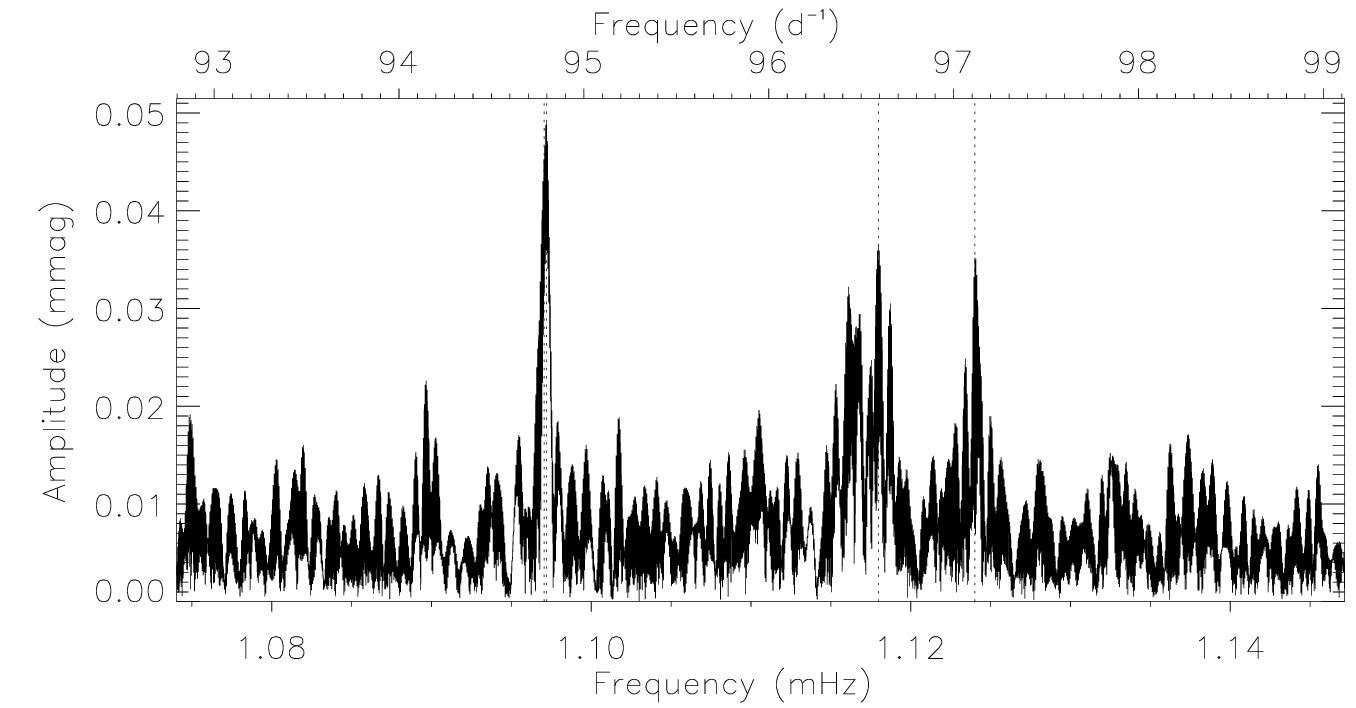}
\includegraphics[width=\columnwidth]{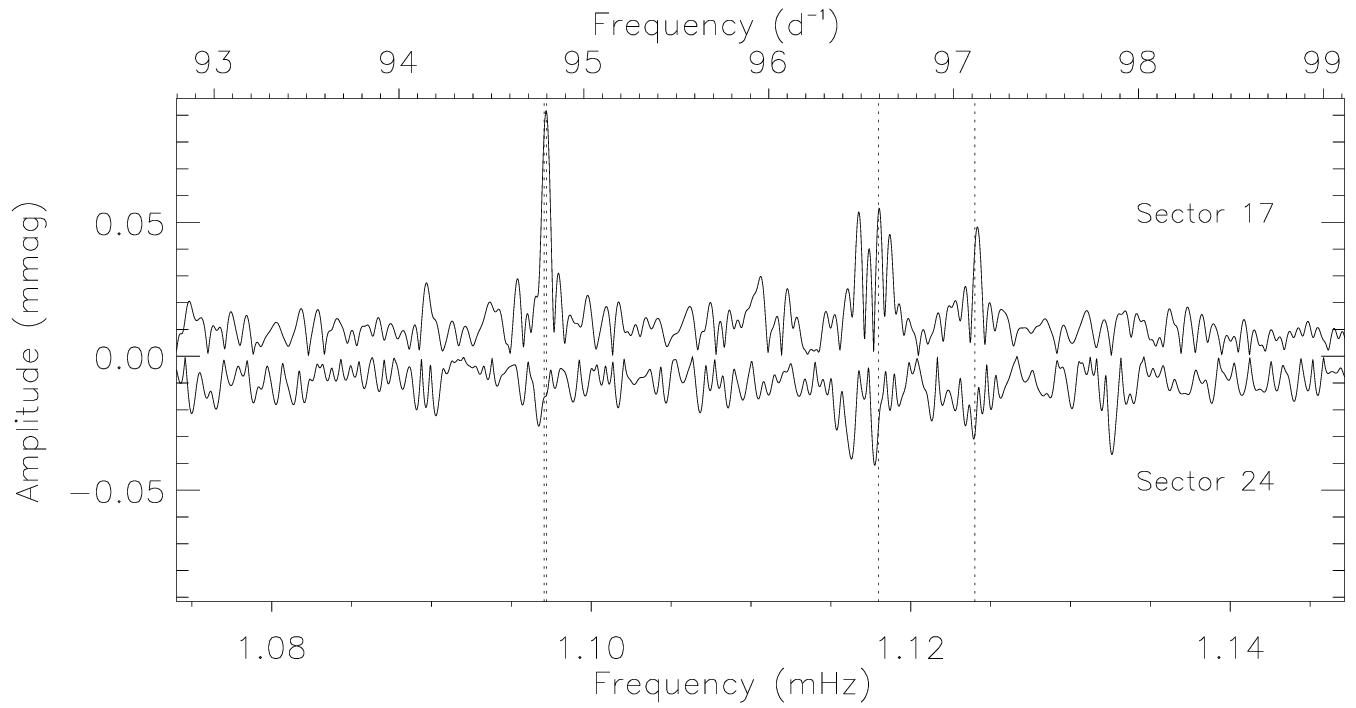}
\caption{Amplitude spectra of TIC\,435263600 showing (top) the low frequency range that indicated either rotational variation, g-mode pulsations, an orbital companion or a combination these phenomena. The middle panel shows the pulsation modes (marked by the vertical dotted lines). All modes show signs of frequency variability making precise determination of the frequencies difficult. The bottom panel shows the amplitude spectra for the same frequency range, but plotting the amplitude spectra for sector 17 (positive) and 24 (negative) data separately. }
\label{fig:435263600_ft}
\end{figure}

Variability in this star was first reported by \citet{2022MNRAS.510.5743B} who analysed the available {\it TESS} data. He classified it as an $\alpha^2\,$CVn star with both $\gamma$\,Dor and roAp pulsations. We corroborate his conclusion here, although note the complexities in the amplitude spectrum of the combined sector 17 and 24 {\it TESS} data. The top panel of Fig.\,\ref{fig:435263600_ft} shows the low frequency variability in this star. We assumed the highest amplitude peak, which also has a harmonic, corresponds to the rotation period of the star ($P_{\rm rot}=3.0739\pm0.0002$\,d). We then assumed the remaining peaks, above 0.01\,mHz are g-mode pulsations. However, given the star is a known binary it is unclear as to the source of the variability, or indeed if one or more of the peaks in the amplitude spectrum correspond to an orbital period. After removal of the dominant peaks, significant power remained in this low-frequency range. A longer timebase of observations is required to resolve these peaks.

At high frequency (Fig.\,\ref{fig:435263600_ft} middle and lower panels) we identified clear signs of pulsations commensurate with those seen in roAp stars. In this frequency range there is significant frequency and amplitude modulation of the pulsations. We show in the bottom panel of Fig.\,\ref{fig:435263600_ft} the amplitude spectra of the two sectors of data, which demonstrate this. It is possible to see the variability on shorter timescales too, on the order of 13-d {\it TESS} orbits or shorter. Such a phenomenon is becoming commonplace in the roAp stars given the long timebase, high precision data sets that are now available \citep[e.g.,][]{2021FrASS...8...31H}.

Finally, for this star we have retrieved a spectrum from the HERMES archive. There were two epochs of data, separated by 5\,d, which showed very different signatures. Fig.\,\ref{fig:435263600_spec} shows the spectra in the blue spectral region and H$_\alpha$. The change in the line profiles suggests there is a close companion with an orbital period much shorter than the period of 415\,yr provided by \citet{2010AJ....139.1521L}, perhaps making this a triple system. Thus we recommend this star be spectroscopically monitored to study the system in detail. Using the procedures outlined above, we derived the following parameters from the spectrum where primarily only one star is seen : $T_{\rm eff}=7950\pm100$\,K, $\log g=4.0\pm0.5$\,\cms\ and $v\sin i =65\pm9$\,\kms. We note that these values should be treated cautiously given the binary nature of the star.


\subsection{Known roAp stars prior to {\it TESS} launch}
\label{sec:Known_roAp}
In this section, we present an analysis of the roAp stars which were known prior to the launch of the {\it TESS} mission. As with the new roAp star discoveries, we present only new information on stars which have already been discussed in the literature. There are 23 stars which are known to be roAp stars in the northern ecliptic hemisphere. Of these, 16 were observed by {\it TESS} in 2-min cadence in Cycle\,2. Of the 7 that were not observed, 6 did not fall on a chip, while one star $\beta$\,CrB (TIC\,383521659; HD\,137909) was only observed at 30-min cadence in FFIs of sector 24.


\subsubsection{TIC\,26418690}

TIC\,26418690 (KIC\,11296437) was discovered to be an roAp star by \citet{2019MNRAS.488...18H} after searching for high frequency pulsations in {\it Kepler} long cadence data using super-Nyquist asteroseismology \citep{2013MNRAS.430.2986M}. The star was later studied by \citet{2020MNRAS.498.4272M} who showed it to be the first $\delta$\,Sct-roAp hybrid star. We refer the reader to the latter study for a detailed analysis. The rotation period reported here, $P=7.11\pm0.02$\,d from the analysis of data from sectors 14 and 15, is in agreement with the more precise value determined from {\it Kepler} observations, and the pulsation mode is recovered at the expected frequency (Table\,\ref{tab:stars}). However, the rotational sidelobes are of too low amplitude to be recovered in the {\it TESS} data. Fig.\,\ref{fig:26418690} shows the amplitude spectrum of this star, in the two regions of known pulsational variability. The spectral type of this star, A9\,EuCr (Table\,\ref{tab:stars}), is from \citet{2019MNRAS.488...18H}.

The non-detection of the low amplitude signals in the {\it TESS} data is not surprising. Using a pulsating eclipsing binary system, \citet{2019MNRAS.490.4040A} estimated that the redder passband of {\it TESS} accounts for an approximate 25\,per\,cent reduction in the relative amplitude when compared to data from the {\it Kepler} mission. While temperature dependent, this, coupled with the higher noise level in {\it TESS} data compared to {\it Kepler} data, explains the non-detection of previously observed peaks.


\subsubsection{TIC\,26749633}

TIC\,26749633 (KIC\,11031749) is another roAp star discovered by \citet{2019MNRAS.488...18H} in long cadence {\it Kepler} data. This star showed no rotation signal in the two {\it TESS} sectors in which it was observed, sectors 14 and 15, as was the case with the {\it Kepler} observations. There is a tentative detection of the pulsation mode at $1.37275\pm0.00003$\,mHz ($118.605\pm0.003$\,\cd) with an amplitude of $0.17\pm0.04$\,mmag, as shown in Fig.\,\ref{fig:26749633}. This tentative detection is likely a result of the reduced sensitivity of the {\it TESS} data over {\it Kepler} data, as discussed above. The spectral type of this star, F1\,SrCrEu (Table\,\ref{tab:stars}), is from \citet{2019MNRAS.488...18H}.


\subsubsection{TIC\,27395746}

TIC\,27395746 (KIC\,11409673) is a further roAp star discovered by \citet{2019MNRAS.488...18H}. This is another target that does not show the rotation or pulsation signal as reported from the {\it Kepler} observations. However, with knowledge of the rotation period, $P_{\rm rot} = 1.06365 \pm 0.00002$\,d,  from \citet{2019MNRAS.488...18H}, we are able to conclude we detect a harmonic of the rotation frequency of the star, but without that prior knowledge, would have dismissed the signal as instrumental given both its low frequency and marginal significance, hence we have not listed it in Table\,\ref{tab:stars}. The combination of the angles $i$ and $\beta$ in this star result in a low amplitude mode with sidelobes of the pulsation having higher amplitude. Therefore, we are unable to detect the pulsation mode, but identified the $\nu-\nu_{\rm rot}$ sidelobe at a frequency of $2.49995\pm0.00003$\,mHz ($215.996\pm0.002$\,\cd) with an amplitude of $0.18\pm0.04$\,mmag, as shown in Fig.\,\ref{fig:27395746}. The spectral type of this star, A9\,SrEu (Table\,\ref{tab:stars}), is from \citet{2019MNRAS.488...18H}. 

\begin{figure}
\centering
\includegraphics[width=\columnwidth]{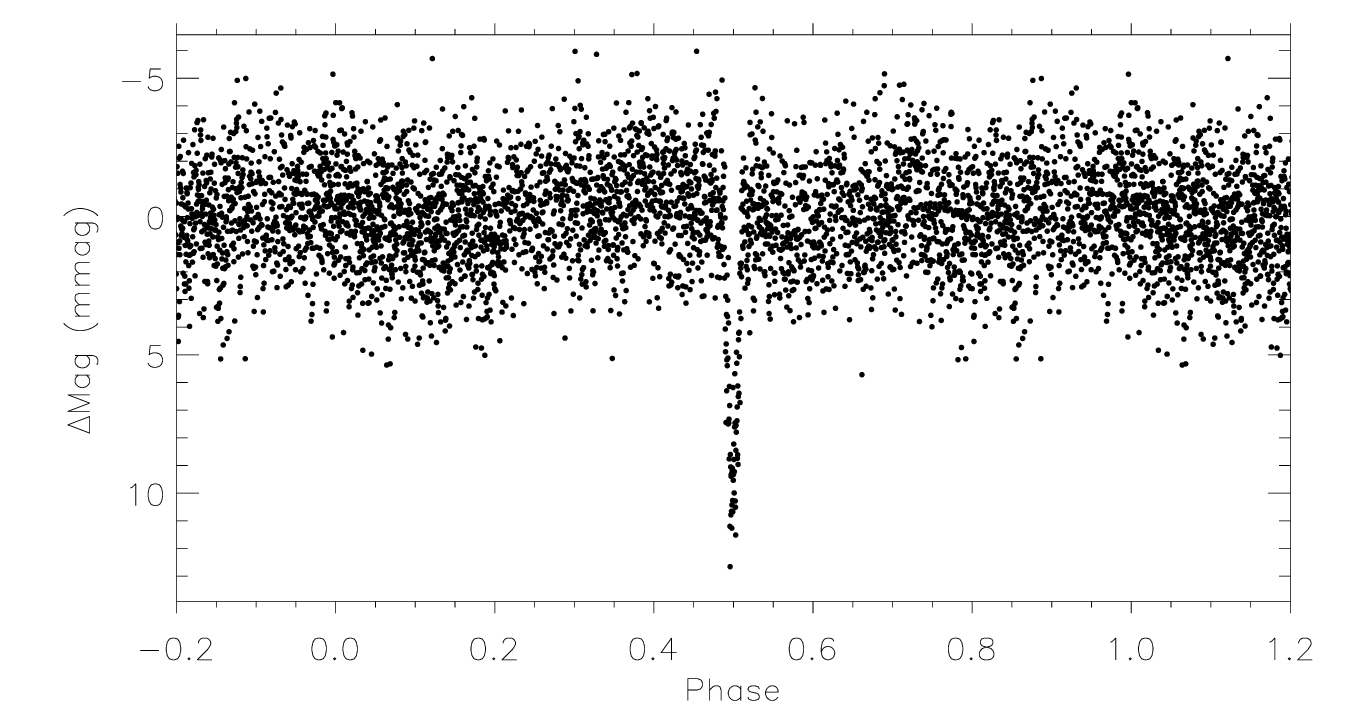}
\caption{Phase folded light curve of TIC\,27395746 showing an eclipse-like signature with a period of $10.27\pm0.01$\,d. The origin of the eclipse is unclear.}
\label{fig:27395746_elc}
\end{figure}

What is noted in the sector 14 and 15 light curve of TIC\,27395746 is the presence of eclipse signatures with a period of $10.27\pm0.01$\,d (Fig.\,\ref{fig:27395746_elc}) which are not present in the {\it Kepler} data. Inspection of the target pixel file (TPF) revealed a background eclipsing binary star included in the standard photometric aperture. While this star does appear in many catalogues concerned with eclipsing binary stars in the {\it Kepler} data \citep[e.g.,][]{2014AJ....147...45C}, although those sources quote the rotation period as the orbital period and have thus misclassified the star based on its surface spots. 

\subsubsection{TIC\,77128654}

TIC\,77128654 (HD\,97127) was discovered in SuperWASP data to be an roAp star \citep{2014MNRAS.439.2078H}, and classified as an F3\,SrEu(Cr). The amplitude spectrum of the sector 22 data for this star does not show any signature of rotation, but does reveal two significant pulsation frequencies (Fig.\,\ref{fig:77128654}). The dominant mode frequency agrees with that in the literature ($1.233855\pm0.000004$\,mHz; $106.6050\pm0.0003$\,\cd) while the second lower amplitude mode at $1.248661\pm0.000031$\,mHz ($107.8843\pm0.0027$\,\cd) is newly reported here. The separation of these two modes is $14.806\pm0.003\,\umu$Hz which is about one quarter of the expected large frequency separation for this star. 


\subsubsection{TIC\,123231021}

TIC\,123231021 (KIC\,7582608) was discovered to be an roAp star through a search of the SuperWASP photometric archive \citep{2014MNRAS.439.2078H}, and later studied in detail by \citet{2014MNRAS.443.2049H}. This star is known to show significant frequency/phase variability in its sole pulsation mode \citep{2014MNRAS.443.2049H,2021FrASS...8...31H} which makes precise determination of the pulsation mode frequency difficult.

The {\it TESS} observations of this star, obtained in sectors 14 and 26, add no additional information over that already presented in the literature (Fig.\,\ref{fig:123231021}). Each individual sector of data spans just over one rotation cycle of the star with the sectors separated by 300\,d. This leads to a less precise, and inaccurate, determination of the rotation period in the {\it TESS} data of $19.805\pm0.005$\,d \citep[compared to $20.4401\pm0.0005$\,d from ][]{2014MNRAS.443.2049H}. The {\it TESS} data also do not resolve the rotationally split sidelobes observed in this star with {\it Kepler} data. Therefore, we refer the reader to \citet{2014MNRAS.443.2049H} for a detailed discussion on this star. The spectral type of this star, A7\,EuCr (Table\,\ref{tab:stars}), is from \citet{2014MNRAS.439.2078H}.

\subsubsection{TIC\,158216369}

TIC\,158216369 (KIC\,7018170) is another roAp star identified by \citet{2019MNRAS.488...18H} in LC {\it Kepler} data. The sectors 14 and 26 data do not show signs of variability due to spots (as was the case with the {\it Kepler} data) nor pulsation (Fig.\,\ref{fig:158216369}). \citet{2019MNRAS.488...18H} suggested an intrinsic amplitude of 0.5\,mmag for the dominant mode in this star which is above the 0.28\,mmag upper limit we place on variability here. However, with the redder {\it TESS} bandpass, it is not surprising that the mode is not detected above the noise here. The spectral type of this star, F2\,(SrCr)Eu (Table\,\ref{tab:stars}), is from \citet{2019MNRAS.488...18H}.


\subsubsection{TIC\,158271090}

TIC\,158271090 (KIC\,10195926) was discovered to be an roAp star in the {\it Kepler} field by \citet{2011MNRAS.414.2550K}. {\it TESS} observed the star during sectors 14, 15 and 16. Analysis of the {\it TESS} light curve at low frequency allowed for the determination of the rotation period to be $5.6845\pm0.0002$\,d which is in agreement with that in the literature. 

{\it Kepler} observations revealed two pulsation modes in this star, however the {\it TESS} observations only showed the highest frequency mode (which also has the highest amplitude) to be above the noise level  (Fig.\,\ref{fig:158271090}). The pulsation mode detected in this star is similar to that in TIC\,27395746 where the geometry leads to higher amplitude sidelobes than the pulsation signal. Therefore, the pulsation frequency listed in Table\,\ref{tab:stars} is derived from the average of the rotationally split sidelobes. The spectral type of this star, F0\,Sr (Table\,\ref{tab:stars}), is from \citet{2016AJ....151...13G}.

\subsubsection{TIC\,158275114}

TIC\,158275114 (KIC\,8677585) was classified as an roAp star by \citet{2011MNRAS.410..517B} from 10\,d of commissioning data from {\it Kepler}. It was further studied by \citet{2013MNRAS.432.2808B}, with \citet{2021FrASS...8...31H} analysing all the {\it Kepler} data. The {\it Kepler} data showed two low frequency modes with disputed origins, and many modes in the typical range for roAp stars, all of which show frequency variability. The {\it TESS} data analysed here, from sectors 14 and 26, do not show the same low frequency variability as the {\it Kepler} data did, although the signal in the {\it Kepler} data is significantly below the noise of the {\it TESS} data at the same frequencies. There is a new `signal' in the sector 14 data at a frequency of $0.012$\,mHz ($1.00$\,\cd), however this is not present in the SPOC reduced 30-min cadence data for the same sector, nor the subsequent observations of this star, so we class this as an instrumental (or pipeline) artefact. 

We detected two pulsation modes in the combined data for this star, significantly fewer than that seen in the {\it Kepler} data (Fig.\,\ref{fig:158275114}). The two modes were detected at $1.5853813\pm0.0000009$\,mHz ($136.97694\pm0.00008$\,\cd) and $1.6217146\pm0.0000019$\,mHz ($140.11614\pm0.00016$\,\cd), where the separation of the two modes is likely to be the large frequency separation, as previously reported. In Table\,\ref{tab:stars}, the A5\,p spectral classification is from \citet{1952ApJ...116..592M}.

\subsubsection{TIC\,159392323}

TIC\,159392323 (TYC\,3547-2692-1) was classified as an roAp star by  \citet{2015PhDT.......227H} through a survey of SuperWASP photometry, who also classified the star as F3p\,SrEuCr. There have been no further studies of this star in the literature. The {\it TESS} sector 15 data confirm the previous pulsation in this star at a frequency of $1.49021\pm0.00002$\,mHz ($128.754\pm0.001$\,\cd), and allowed for the detection of two new modes at $1.47658\pm0.00005$\,mHz ($127.576\pm0.004$\,\cd) and $1.49610\pm0.00003$\,mHz ($129.263\pm0.003$\,\cd). As with the previous literature, we did not detect a rotation period for this star.

\begin{figure}
\centering
\includegraphics[width=\columnwidth]{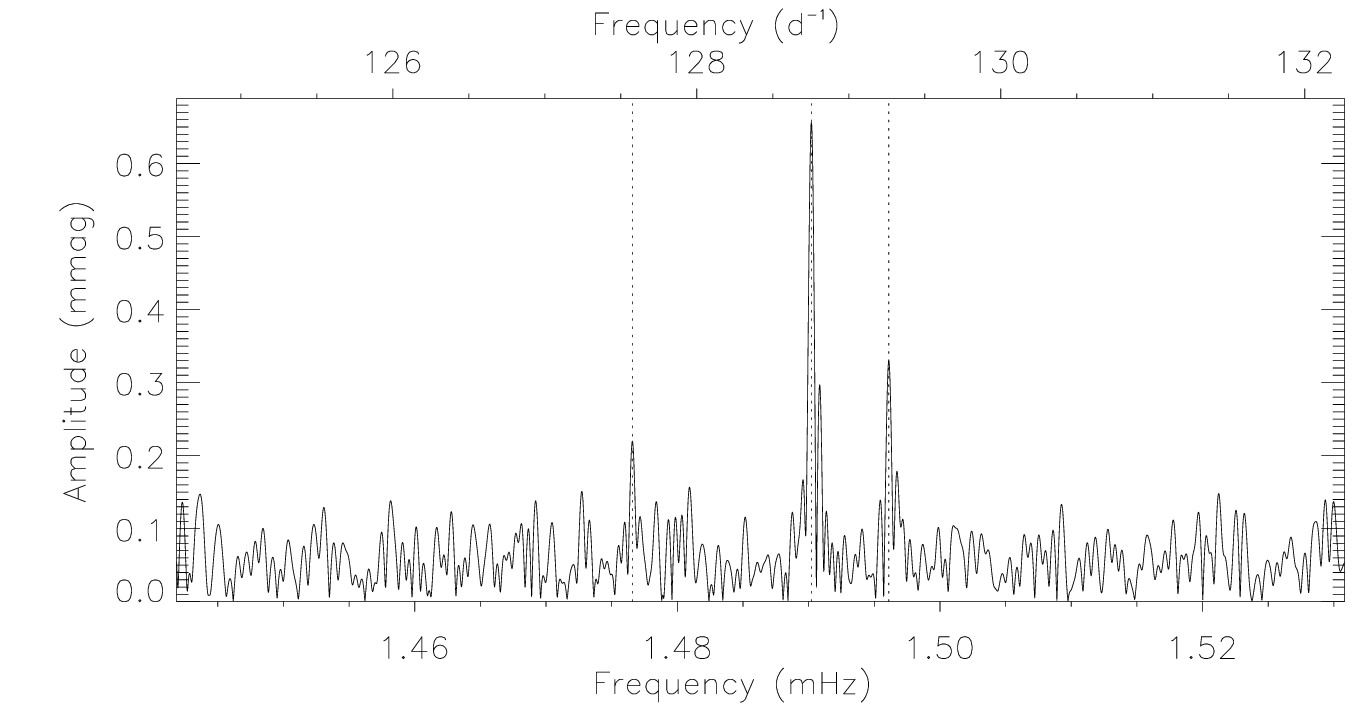}
\caption{Amplitude spectrum for TIC\,159392323. The vertical dotted lines identify the pulsation modes detected in the {\it TESS} data. The two lower amplitude modes are newly detected.}
\label{fig:159392323}
\end{figure}

\subsubsection{TIC\,169078762}

TIC\,169078762 (HD\,225914) was listed in the Henry Draper Catalogue and Extension \citep{1923AnHar..98....1C} as an A7 spectral type star, but found to be both chemically peculiar, A5Vp\,SrCr(Eu), and pulsating (in {\it Kepler} data) by \citet{2015MNRAS.452.3334S}. They performed a detailed analysis of this object and reported the pulsation mode at 0.7112\,mHz (61.45\,\cd) with an amplitude of 62.6\,$\muup$mag using the 4-yr long {\it Kepler} photometry, and measured a precise rotation period of $P_{\rm rot}=5.209358\pm0.000009$\,d. 

The analysis of high-resolution spectra revealed the following stellar parameters: $T_{\rm eff}=8100 \pm 200$\,K, $\log{g}=4.0 \pm 0.2$\,\cms, [Fe/H]=+0.31$\pm$0.24 and $v\sin i=14.6 \pm 1.6$\,\kms \citep{2015MNRAS.452.3334S}. Line profile variations caused by rotation are also evident. Lines of Sr, Cr, Eu, Mg and Si are strongest when the star is brightest (in the {\it Kepler} bandpass), while Y and Ba vary in antiphase with the other elements. The abundances of rare earth elements are only modestly enhanced compared to other roAp stars of similar $T_{\rm eff}$ and $\log{g}$ \citep{2015MNRAS.452.3334S}. 

{\it TESS} observed the star in sector 14 at 2-min cadence. Both the previously determined rotation period and pulsation frequency (Fig.\,\ref{fig:169078762}) are confirmed in these data.

\subsubsection{TIC\,171988782}

TIC\,171988782 (HD\,258048) was first classified as an F8 star \citep{2001KFNT...17..409K} but \citet{2014MNRAS.439.2078H} showed it to be F4p\,EuCr(Sr) using low resolution spectroscopy. They determined an effective temperate of $6200 \pm 280$\,K using the infrared flux method, and 6600\,K from Balmer line fitting. \citet{2014MNRAS.439.2078H} also classified the star to be an roAp star given the presence of a significant peak in the amplitude spectrum of SuperWASP photometry at $1.962$\,mHz (169.54\,\cd) with an amplitude of 1.49\,mmag. \citet{2018MNRAS.477.3145J} identified TIC\,171988782 as a W\,UMa type binary based on a period found in the ASAS-SN data. However, this period was not found in the SuperWASP data, nor in the present {\it TESS} data.

\begin{figure}
\centering
\includegraphics[width=\columnwidth]{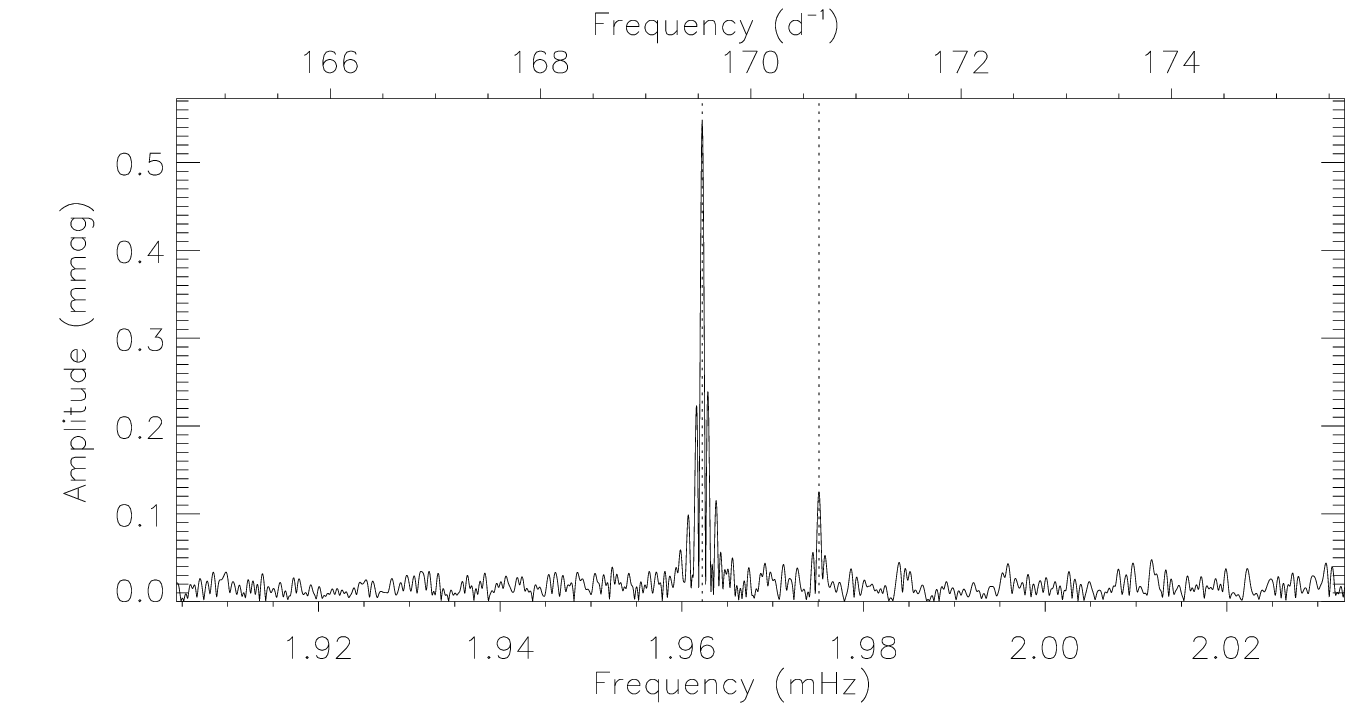}
\caption{Amplitude spectrum for TIC\,171988782. The vertical dotted lines identify the pulsation modes detected in the star.}
\label{fig:171988782}
\end{figure}

{\it TESS} observed the star in sector 20 of Cycle\,2. The principal mode identified in the SuperWASP data is confirmed at $1.962238\pm0.000005$\,mHz ($169.5374\pm0.0005$\,\cd) with a new secondary mode found at $1.975095\pm0.000023$\,mHz ($170.6482\pm0.0020$\,\cd; Fig.\,\ref{fig:171988782}). There is no indication of the rotation period in this star.

\subsubsection{TIC\,264509538}

TIC\,264509538 (KIC\,10685175) was discovered as an roAp star using super-Nyquist analysis on {\it Kepler}  Long Cadence (LC) data \citep{2019MNRAS.488...18H}. That work provided the first estimate for the principle pulsation frequency of $2.783$\,mHz ($240.45189$\,\cd) as well as the rotation period of $P_{\rm rot}=3.1020\pm0.0001$\,d. These values were reviewed by a new analysis based on {\it TESS} data by \citet{2020ApJ...901...15S}, where the fundamental pulsation mode and rotational period were found to be 2.217\,mHz ($191.5151$\,\cd) and $3.1028\pm0.0048$\,d, respectively. In that work the star was found to be an oblique quintuplet pulsator with sidelobes split by exactly the rotation frequency. The discrepancy between the two pulsation frequencies is a result of the incorrect selection of a Nyquist alias by \citet{2019MNRAS.488...18H}, an issue that is discussed in their work.

\citet{2019MNRAS.488...18H} obtained the effective temperature ($T_{\rm eff} = 8000 \pm 300$\,K) from LAMOST spectra, and a luminosity (scaled with the solar luminosity) of $\log L/{\rm L_\odot}=0.896\pm0.022$ from {\it Gaia} DR2. A mass of $M = 1.65\pm0.25$\,M$_{\odot}$ was also estimated through an interpolation over stellar tracks based on the models of \citet{2013MNRAS.436.1639C}.

Since the {\it TESS} data, from sectors 14 and 15 (Fig.\,\ref{fig:264509538}), have already  been analysed, we do not repeat the process here, but refer the reader to \citet{2020ApJ...901...15S}.

\subsubsection{TIC\,272598185}

TIC\,272598185 (KIC\,10483436) was discovered to be an roAp star by \citet{2011MNRAS.413.2651B} through the analysis of {\it Kepler} Short Cadence data. In that work, it was concluded that the star pulsates with a single mode at a frequency of $1.35$\,mHz (116.91\,\cd) with an amplitude of $69\pm2\,\muup$mag in the {\it Kepler} passband, however the {\it TESS} data analysed here, from sectors 14 and 15, showed no signs of variability with this frequency (Fig.\,\ref{fig:272598185}) which is most likely the result of the unfavourable longer wavelength observations of {\it TESS}. The rotation period presented in Table\,\ref{tab:stars} is more precise than that presented by  \citet{2011MNRAS.413.2651B} since the combined {\it TESS} observations analysed here have a longer baseline than the 27-d {\it Kepler} data set previously used.

\subsubsection{TIC\,273777265}

TIC\,273777265 (KIC\,6631188) was discovered to be an roAp star by \citet{2019MNRAS.488...18H}  through the analysis of {\it Kepler} LC data. The sector 14 and 15 {\it TESS} data showed low frequency variability which is attributed to rotation of the star. However, the derived rotation period is half of the value recorded in the literature, suggesting we detected only a harmonic of the rotation period of the star. The pulsation frequency found in the {\it TESS} data agreed with that in the literature within 2.5\,$\sigma$ ($1.49349\pm0.00001$\,mHz; $129.038\pm0.001$\,\cd; Fig.\,\ref{fig:273777265}). Interestingly for this object, the amplitude presented in Table\,\ref{tab:stars} is larger than the predicted intrinsic amplitude in the {\it Kepler} bandpass which is unexpected due to the redder wavelength coverage of {\it TESS}. The spectral type of this star, F0\,Sr (Table\,\ref{tab:stars}), is from \citet{2019MNRAS.488...18H}.

\subsubsection{TIC\,286992225}

TIC\,286992225 (TYC\, 2553-480-1) was discovered by \citet{2014MNRAS.439.2078H}  in SuperWASP data to be an roAp star. In that paper a single pulsation frequency of $2.726$\,mHz ($235.54$\,\cd)  was found, which is one of the highest frequencies found for roAp stars.  The authors classified the spectrum as A9p\,SrEu, and they found an effective temperature ranging from $7100\pm200$\,K (from Balmer lines) to $7500\pm400$\,K (from SED analysis). Moreover analysis of the H$\alpha$ profile seemed to indicate the presence of a core-wing anomaly \citep[e.g.,][]{2001A&A...367..939C}.

An analysis of {\it TESS} data for this star was presented by \citet{2022A&A...660A..70M}. The data show two modes separated by $35\,\muup$Hz, which the authors deduced that to be half of the large separation. TIC\,286992225 is also classified as a super-slowly rotating Ap (ssrAp) star, i.e., an Ap star with a rotation period greater than 50\,d.

We add no further analysis for this star with the {\it TESS} data (sector\,23; Fig.\,\ref{fig:286992225}), so refer the reader to \citet{2022A&A...660A..70M} for full details.

\subsubsection{TIC\,302602874}

TIC\,302602874 (TYC\,2488-1241-1) was, upon its discovery as a variable star, classified as A6p\,SrEu \citep{2014MNRAS.439.2078H}. The authors found $T_{\rm eff}$ to be 7800\,$\pm$\,200\,K (from Balmer lines) and 8300\,$\pm$\,500\,K (from SED fitting). The latter estimate is in line with $T_{\rm eff} = 8200$\,K provided in the TIC. 

This star was the subject of its own study in \citet{2018MNRAS.480.2405H}, in which the pulsation mode was concluded to be a distorted quadrupole mode. The {\it TESS} data analysed here (from sector 21; Fig\,\ref{fig:302602874}) add no further information for this star;  we refer the reader to \citet{2018MNRAS.480.2405H} for full details.

\section{Candidate roAp stars}
\label{sec:Cand_roAp}

In this section, we provide information on a number of stars that were flagged by the search teams as possible roAp stars. These stars are classed as candidate roAp stars since a combination of their spectral classification not indicating chemical peculiarity, their light curve does not show $\alpha^2$\,CVn rotational variability, their pulsation signature does not show signs of multiplets split by the rotation frequency, their presumed pulsation signal is close to a signal to noise limit of 4.0, or there is indiction they are a member of a binary system. The stars listed in this section would benefit from high-resolution spectroscopy to search for peculiarities and a magnetic field.


\subsection{TIC\,9171107}

TIC\,9171107 (TYC\,2479-429-1) has limited information available in the literature. The TIC gives an effective temperature of 6800\,K and $\log g$ of 4.2\,\cms\, while {\it Gaia} EDR3 gives $T_{\rm eff} = 6800\,K$ and $\log g$ of 3.7\,\cms\, \citep{2020yCat.1350....0G}. No pulsation detections have been reported.

\begin{figure}
\centering
\includegraphics[width=\columnwidth]{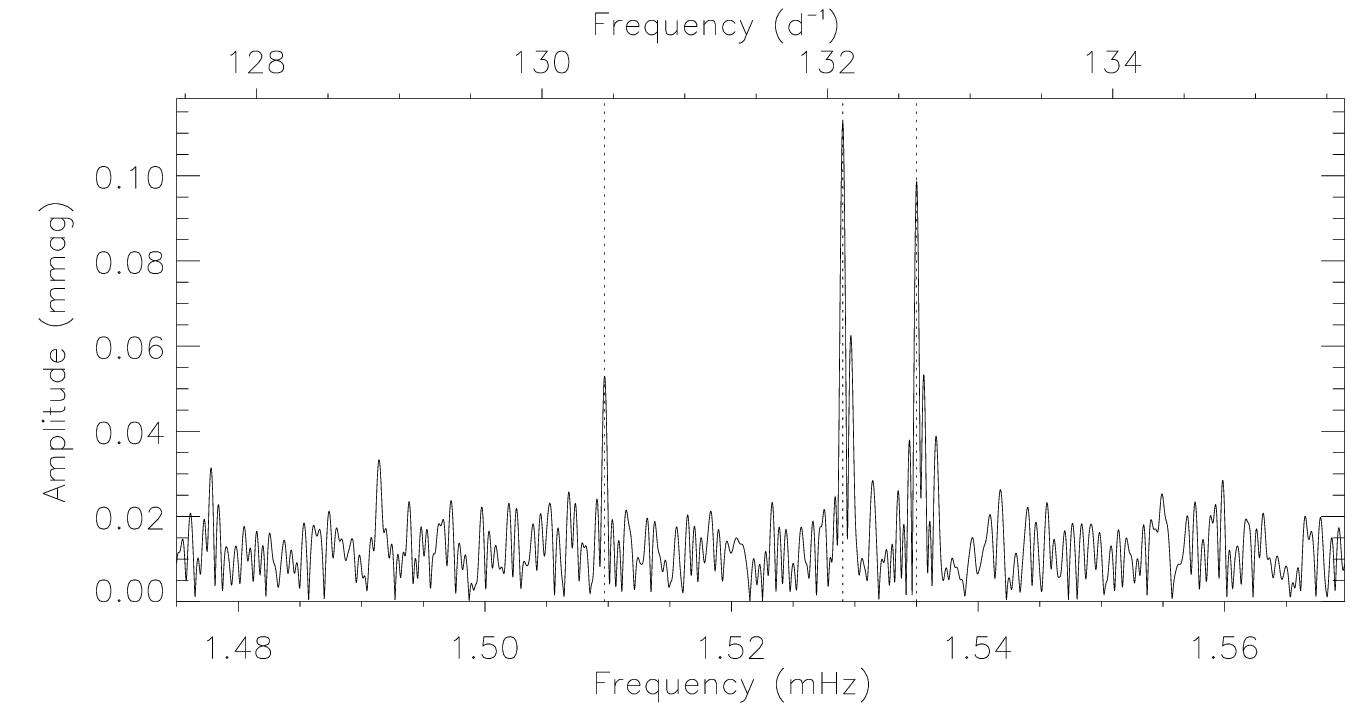}
\caption{Amplitude spectrum of TIC\,9171107, with the pulsation modes indicated with the vertical dotted lines.}
\label{fig:9171107_ft}
\end{figure}

TIC\,9171107 was observed in {\it TESS} sector 20 and showed no sign of rotation in its light curve. However, there are three significant short period signals in the amplitude spectrum of the light curve that we attribute to pulsation: $1.50971\pm0.00004$\,mHz ($130.439\pm0.003$\,\cd), $1.52902\pm0.00002$\,mHz ($132.107\pm0.002$\,\cd) and $1.53502\pm0.00002$\,mHz ($132.625\pm0.002$\,\cd; Fig.\,\ref{fig:9171107_ft}). The two lowest frequency modes are separated by 19.3\,$\muup$Hz, while the two highest frequency modes are separated by 6.0\,$\muup$Hz. The former separation may be half of the large frequency separation, while the latter separation may be the small frequency separation\footnote{defined as the difference between two odd or even degree modes that have a radial overtone difference of one}. 

There are no spectroscopic data of this object in the literature, and no Str\"omgren-Crawford indices that are available to classify the star as chemically peculiar. Therefore, we placed this star as a candidate roAp star until such data are available. 

\subsection{TIC\,275642037}

TIC\,275642037 (TYC\,4330-716-1), has no published spectral type. However, from photometry \citet{2008PASP..120.1128O} and \citet{2010PASP..122.1437P} both infer F0\,V. The TIC gives $T_{\rm eff} = 7700 \pm 400$\,K, while 7200\,K was obtained by \citet{2017MNRAS.471..770M} and 7550\,K by \citet{2006ApJ...638.1004A}. The {\it Gaia} DR3 flags this star as photometrically variable as "DSCT|GDOR|SXPHE", with a $g$-band amplitude of 5.9\,mmag at a single frequency of $8.1\,\muup$Hz (0.6965\,\cd, 1.436\,d). {\it Gaia} DR3 data also showed a variable radial velocity of $4.56\pm 0.84$\,\kms, with an amplitude of 10.67\,\kms and a spectral broadening of $44\pm10$\,\kms, which is a proxy for $v\sin i$ \citep{2023A&A...674A...8F}. There is no mention of any chemical peculiarities, rotational variability or pulsations in the literature. 

\begin{figure}
\centering
\includegraphics[width=\columnwidth]{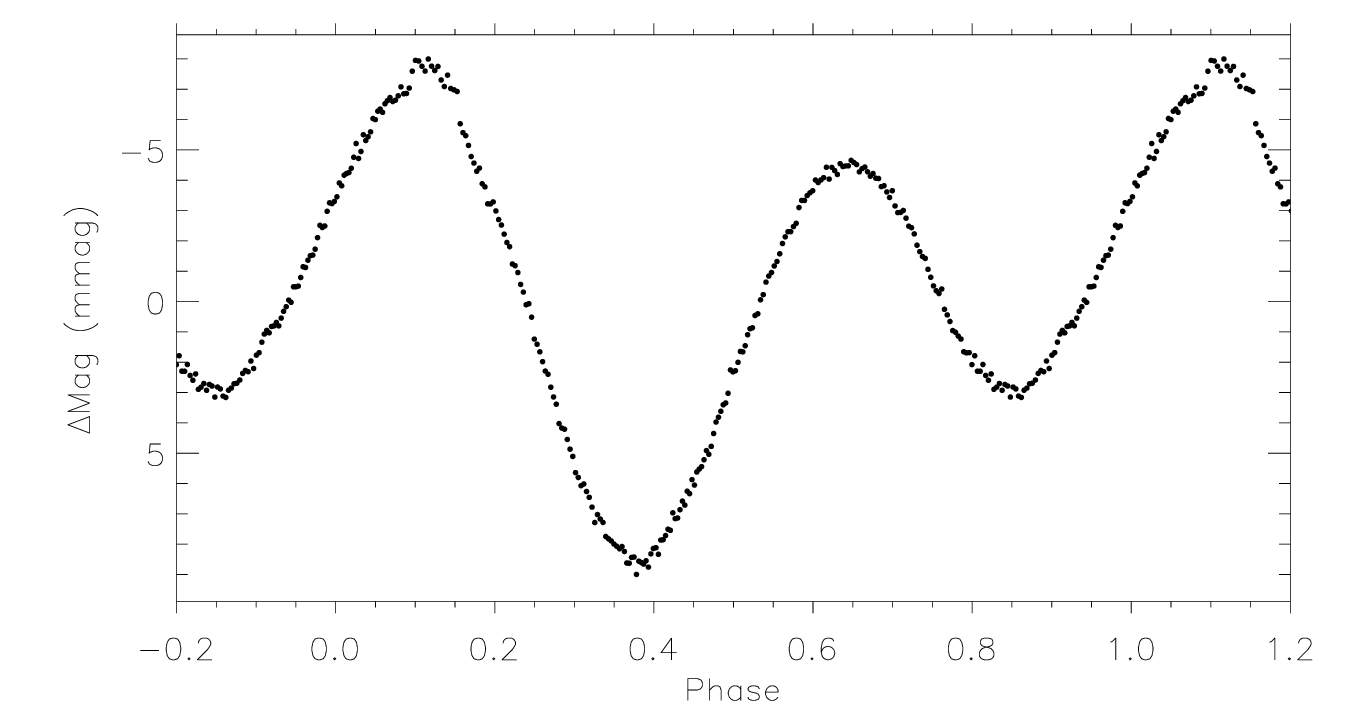}
\caption{Phase folded light curve of TIC\,275642037, phased on a period of $2.8745\pm0.0003$\,d. The data are binned 50:1}
\label{fig:275642037_rot}
\end{figure}

\begin{figure}
\centering
\includegraphics[width=\columnwidth]{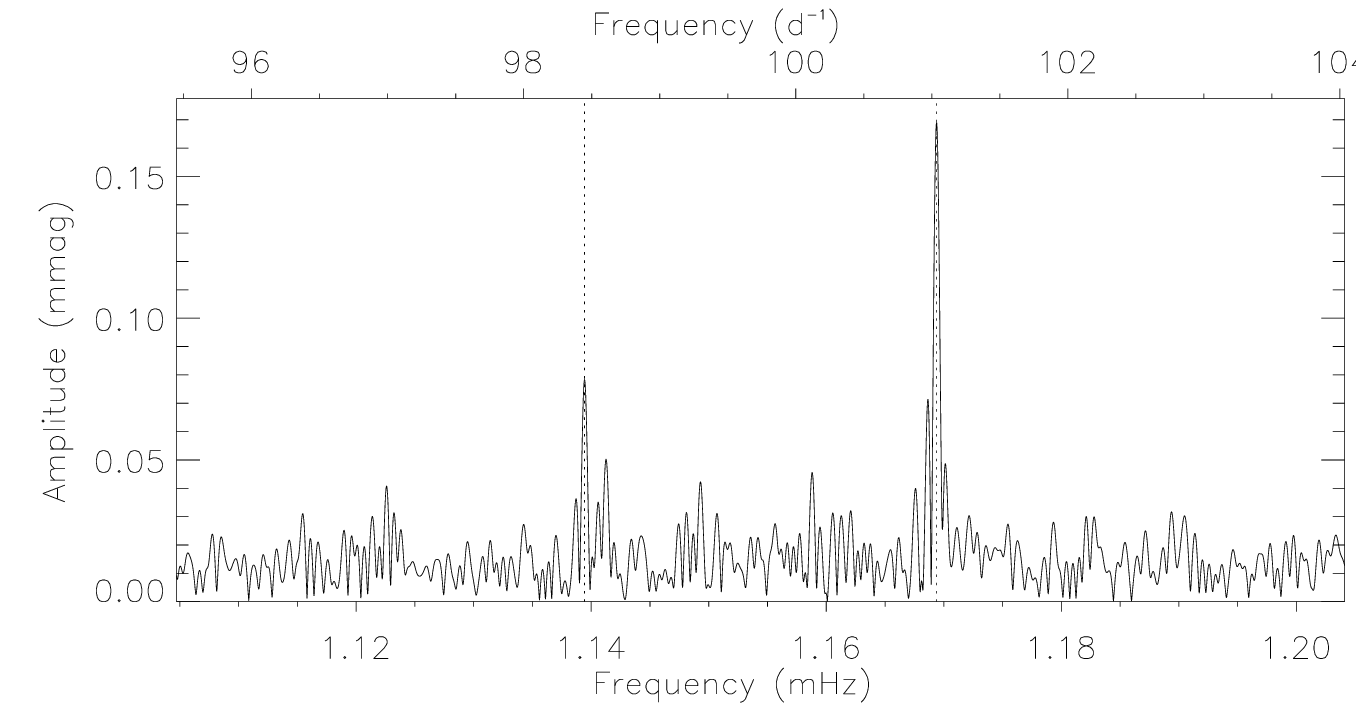}
\caption{Amplitude spectrum of TIC\,275642037 showing the pulsation modes (marked by the vertical dotted lines). }
\label{fig:275642037_ft}
\end{figure}

{\it TESS} observed this star during sector 18. There is clear low-frequency variability in the {\it TESS} light curve of this star that could either be attributed to rotation in the presence of stable spots, or an orbiting companion. The {\it Gaia} DR3 low-frequency result is half of the period we detected here ($P_{\rm rot}=2.8745\pm0.0003$\,d; Fig.\ref{fig:275642037_rot}), which is the dominant signal in an amplitude spectrum. The pulsations we detected (Fig.\,\ref{fig:275642037_ft}) at frequencies of $1.13943\pm0.00003$\,mHz ($98.447\pm0.003$\,\cd) and  $1.16939\pm0.00002$\,mHz ($101.035\pm0.001$\,\cd) are stable in both amplitude and phase over the observation period. They are also separated by 30\,$\umu$Hz, which could plausibly be half the large separation. Given the lack of rotational sidelobes to the pulsations, and confirmation that this is an Ap star, it remains a candidate roAp star until it is spectroscopically observed to determine any chemical peculiarity. 

\subsection{TIC\,298052991}

TIC\,298052991 (TYC\,4554-625-1) has been classified as an A2 star by \citet{2019A&A...623A..72K}. The same authors also performed an analysis to characterise the presence of a stellar/substellar mass around the star and found a physical companion orbiting it. They gave the mass, assuming a semi-major axis of 1\,au as $M_2=12_{-5}^{+6}$\,M$_{\rm{Jup}}$, using a primary mass and radius of $M_1=1.60\pm0.08$\,M$_{\odot}$ \citep{2000A&AS..141..371G} and $R_1=1.50\pm0.08$\,R$_{\odot}$ \citep{2004A&A...426..297K}. \citet{2012MNRAS.427..343M}  derived the stellar temperature and luminosity to be $T_{\rm{eff}}=9550$\,K and $L=9.6$\,L$_{\odot}$, respectively, from the comparison of the $\chi^2$ statistic between observed photometric data and a set of synthetic stellar spectra.  \citet{2022MNRAS.510.5743B} presented an analysis of the {\it TESS} data from sectors 14, 20, 21, 26, 40 and 41 and found a rotation period of $10.638$\,d and a pulsation frequency of $2.291$\,mHz (197.951\,\cd).

\begin{figure}
\centering
\includegraphics[width=\columnwidth]{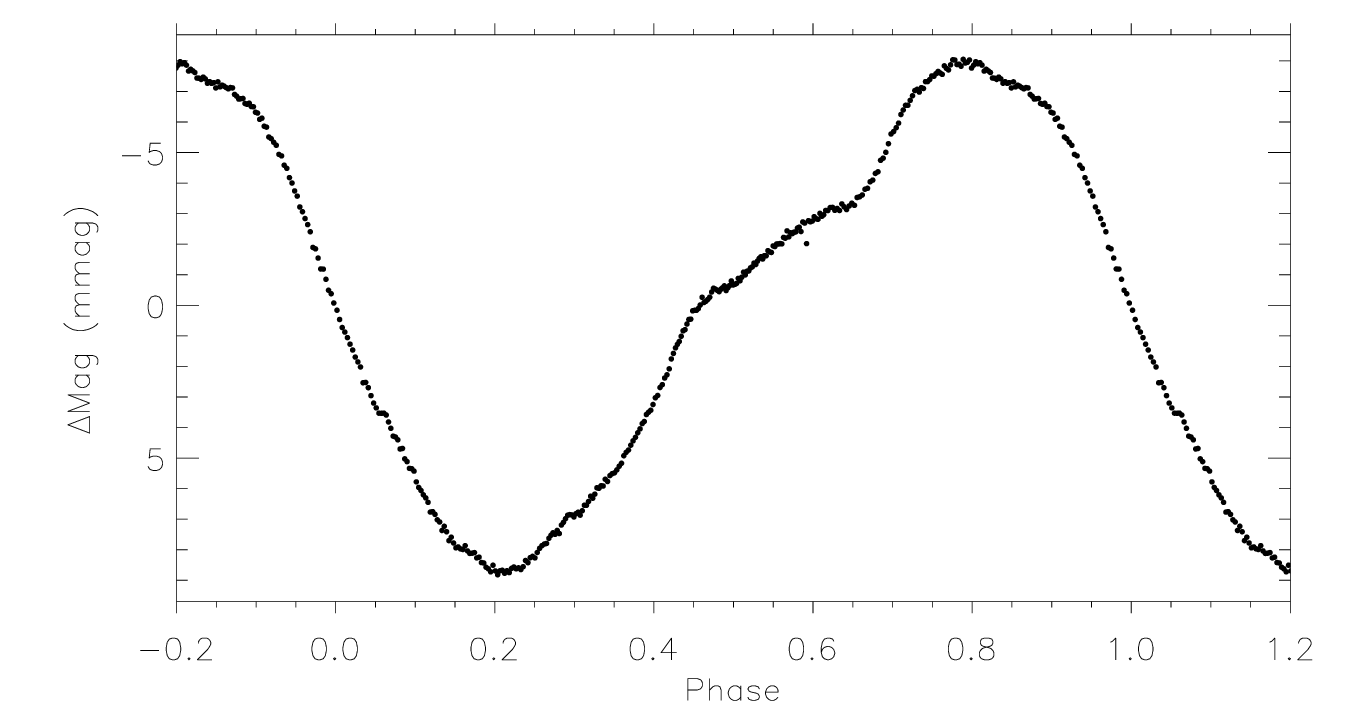}
\caption{Phase folded SAP light curve of TIC\,298052991, phased on a period of $10.5852\pm0.0001$\,d. The data are binned 200:1}
\label{fig:298052991_rot}
\end{figure}

\begin{figure}
\centering
\includegraphics[width=\columnwidth]{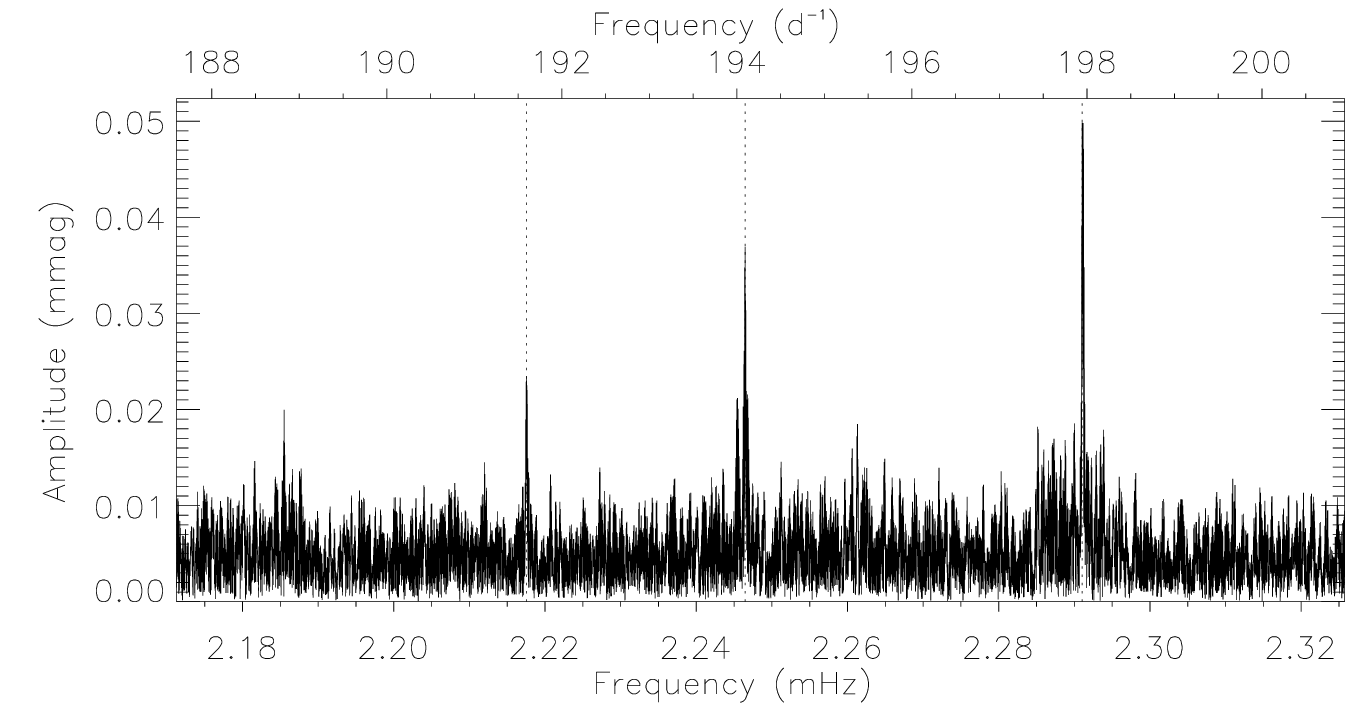}
\caption{Amplitude spectrum of TIC\,298052991 showing the pulsation modes (marked by the vertical dotted lines). }
\label{fig:298052991_ft}
\end{figure}

During Cycle\,2, TIC\,298052991 was observed during sectors 14, 20, 21 and 26. For this star, we used the SAP data to derive the rotation period as the signal in the PDC data had been altered by the processing. We derived a period of  $10.5852\pm0.0001$\,d (Fig.\,\ref{fig:298052991_rot}), which significantly differs from that previously determined given the use of the SAP data here. To investigate the pulsations, we pre-whitened the individual sectors to remove peaks below $0.116$\,mHz (10\,\cd) to the noise level of the high-frequency range. This allowed us to detect three pulsation frequencies in this star (Fig.\,\ref{fig:298052991_ft}; $2.217569\pm0.000003$\,mHz, $191.5979\pm0.0002$\,\cd; $2.246466\pm0.000002$\,mHz, $194.0947\pm0.0001$\,\cd; $2.291038\pm0.000001$\,mHz, $197.9457\pm0.0001$\,\cd), with a tentative detection of a fourth signal ($2.185497\pm0.000003$\,mHz, $188.8270\pm0.0003$\,\cd).

Since there is no evidence of chemical peculiarities or rotational splitting of modes, we are unable to confirm if this star is a bona fide roAp star, thus it remains a candidate for now.


\subsection{TIC\,405892692}

TIC\,405892692 (BD+49 3179) is a little-studied star. \citet{1952CoRut..32....1H} classified this $V=9.28$ mag star as F0. The TIC catalogue \citep{2019AJ....158..138S} lists the atmospheric parameters $T_{\rm eff}=7250$\,K and $\log g=4.24$\,\cms, however there are no indications of chemical peculiarity for this star. \citet{2022MNRAS.510.5743B} claimed the star to be a rotational variable with a pulsation at high frequency, based on {\it TESS} observations.

\begin{figure}
\centering
\includegraphics[width=\columnwidth]{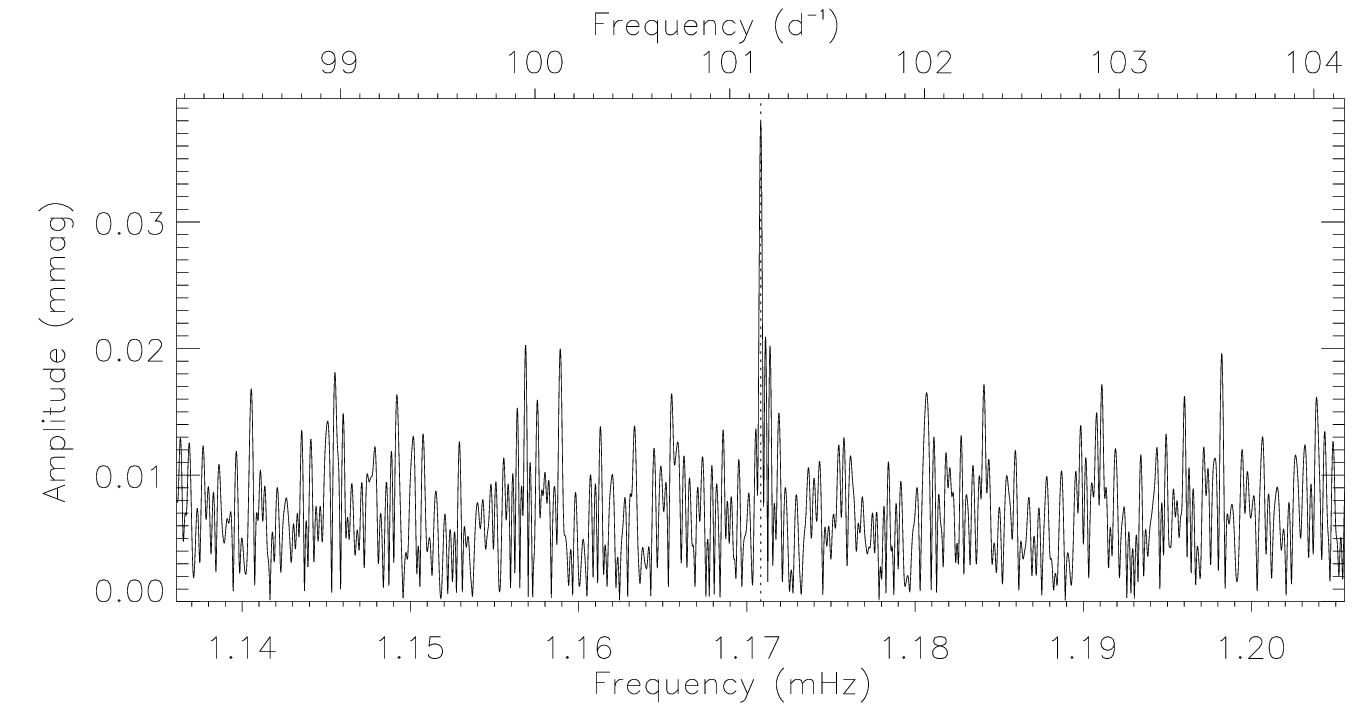}
\caption{Amplitude spectrum of TIC\,405892692 showing the detection of a pulsation mode (marked by the vertical dotted line). }
\label{fig:405892692_ft}
\end{figure}

This star was observed in sectors 14 and 15. We detected a low frequency peak of uncertain origin in the amplitude spectrum, corresponding to a period of $P=0.3786\pm0.0001$\,d. This is a short rotation period for an F0 star, so we postulated that this is in fact a signature of a binary, or higher order, system. This is tentatively supported by the {\it Gaia} DR3 binary probability flag being essentially unity \citep{2022yCat.1355....0G}.  The pulsation signature we detected is at a frequency of $1.17082\pm0.00002$\,mHz ($101.159\pm0.001$\,\cd; Fig.\,\ref{fig:405892692_ft}), which agrees with that previously reported.

Since there is little information available on this star, and the lack of any convincing evidence that it is chemically peculiar, this remains a candidate roAp star.


\subsection{TIC\,429251527}

TIC\,429251527 (TYC\,2185-478-1) is another star with very little information in the literature. This $V=10.5$\,mag star has no spectral classification or any other dedicated studies available. The atmospheric parameters, $T_{\rm eff}=7900$\,K and $\log g=4.66$\,\cms, are provided in the TIC \citep{2019AJ....158..138S}, while the {\it Gaia} DR3 data suggested a compatible value of $T_{\rm eff}=7950$\,K. The TIC $\log g$ value is anomalously high, with {\it Gaia} D3 providing a more likely value of $\log g=3.99$\,\cms.

\begin{figure}
\centering
\includegraphics[width=\columnwidth]{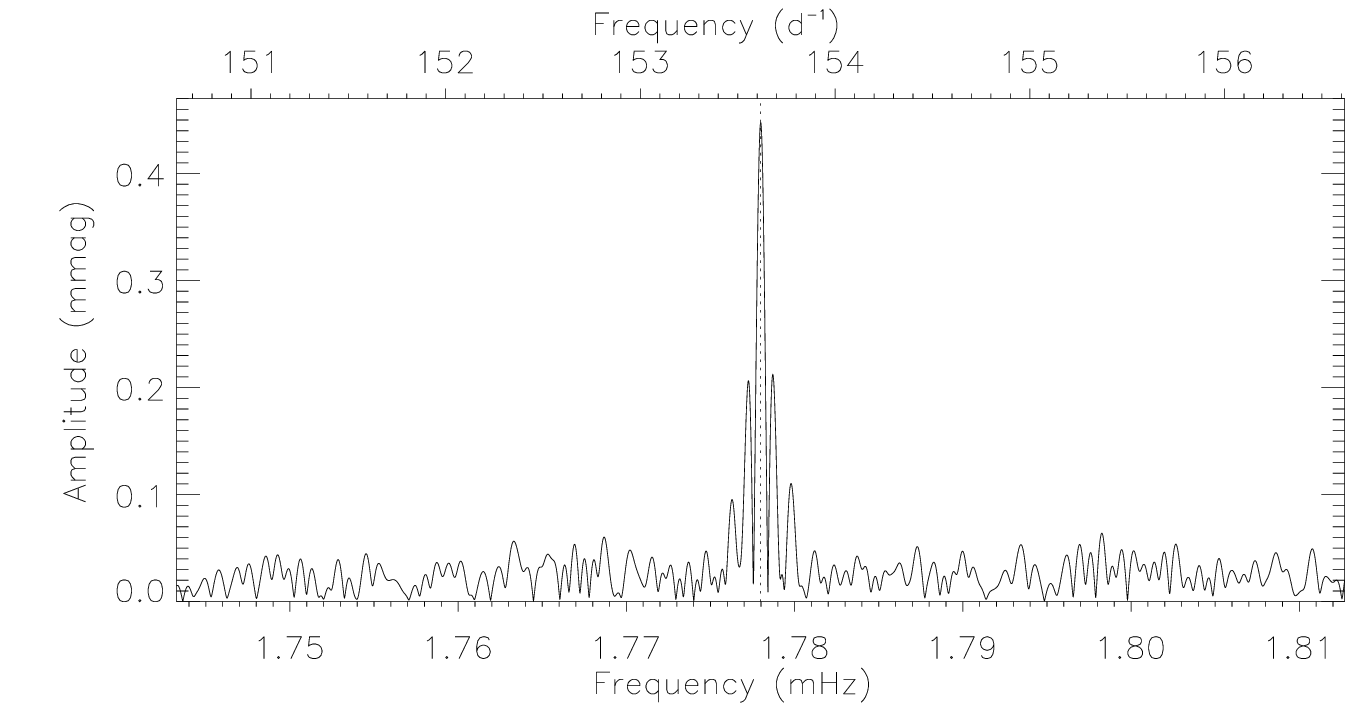}
\caption{Amplitude spectrum of TIC\,429251527 showing the pulsation mode marked by the vertical dotted line. }
\label{fig:429251527_ft}
\end{figure}

{\it TESS} observed the star in sector 15. The data did not show any rotation signature, but did show a pulsation with significant amplitude at $1.77798\pm0.00001$\,mHz ($153.6178\pm0.0008$\,\cd; Fig.\,\ref{fig:429251527_ft}). This is the first report of the pulsation in this star. Given the presence of only one mode, which does not show rotational splitting, and the lack of information in the literature, we cannot confirm this star to be an roAp star, and thus leave this as a candidate for now.

\subsection{TIC\,445796153}

TIC\,445796153 (HD\,34740) has several spectral classifications in the literature: Am (A3-A7) \citep{1967JO.....50..425B}; A0p or A2V \citep{1975AJ.....80..698P};  A0p \citep{1918AnHar..92....1C}. However, the peculiarity was marked as doubtful by \citet{2009A&A...498..961R}. The estimated effective temperature from {\it Gaia} DR3 is $T_{\rm eff}=10500$\,K \citep{2022yCat.1355....0G}, this, coupled with the spectral classifications, places this star at the hot end of the roAp range. \citet{1998A&AS..129..431H} measured Str\"omgren-Crawford indices of this star to be: $b-y=0.073$, $m_1=0.201$, $c_1=0.937$ and H$\beta =2.817$, giving $\delta m_1=0.005$ and $\delta c_1=0.127$. However, there is disparity between the measured H$\beta$ and $b-y$ values, with the former giving an equivalent $b-y$ of 0.129, redder than the observed value. The star is listed as a multiple system \citep{2001AJ....122.3466M} with a $\delta m_v=2.96$\,mag, which, based on the {\it Gaia} $T_{\rm eff}$ suggests a late B star with an early F companion. This may explain the discrepancies in the indices. The {\it TESS} data for TIC\,445796153 were analysed by \citet{2022MNRAS.510.5743B} to label the star as a possible $\alpha^2$\,CVn star (with a period of $P_{\rm rot}=5.576$\,d) with high-frequency $\delta$\,Sct modes. 

\begin{figure}
\centering
\includegraphics[width=\columnwidth]{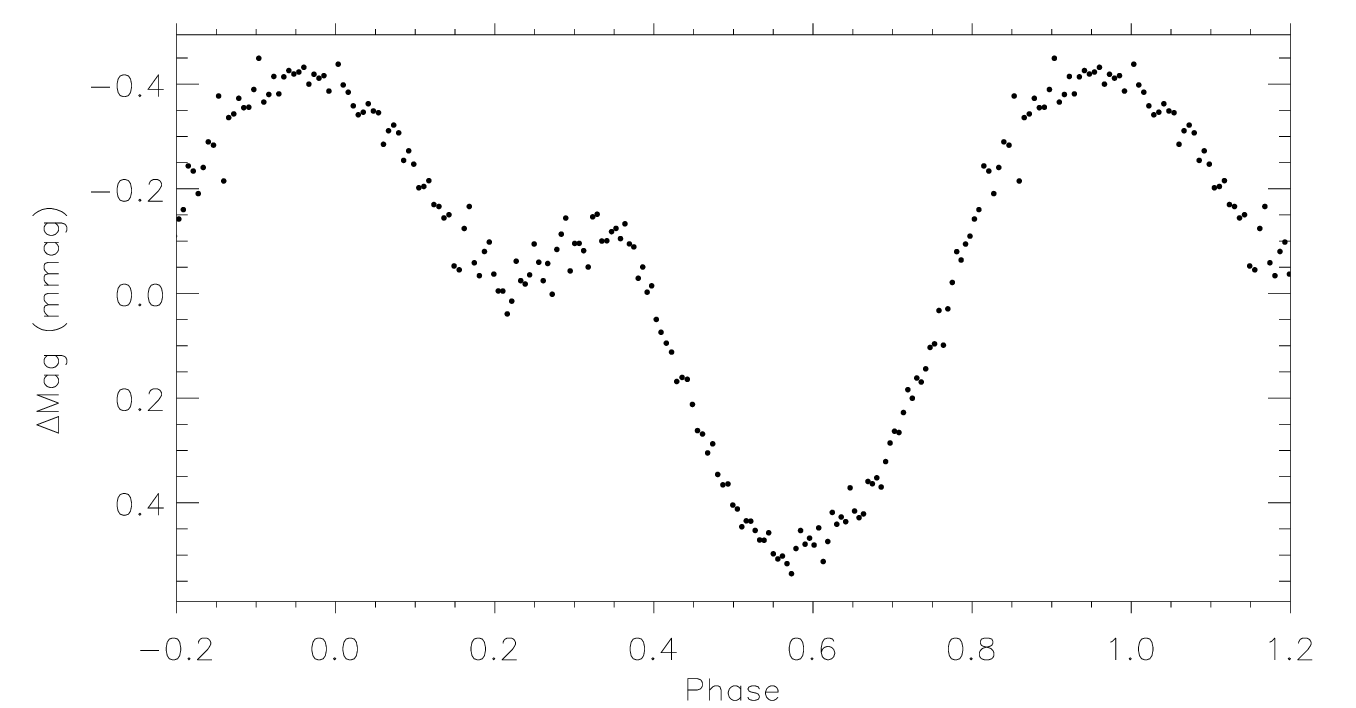}
\caption{Phase folded light curve of TIC\,445796153, phased on a period of $5.5721\pm0.0004$\,d. The data are binned 200:1}
\label{fig:445796153_rot}
\end{figure}

\begin{figure}
\centering
\includegraphics[width=\columnwidth]{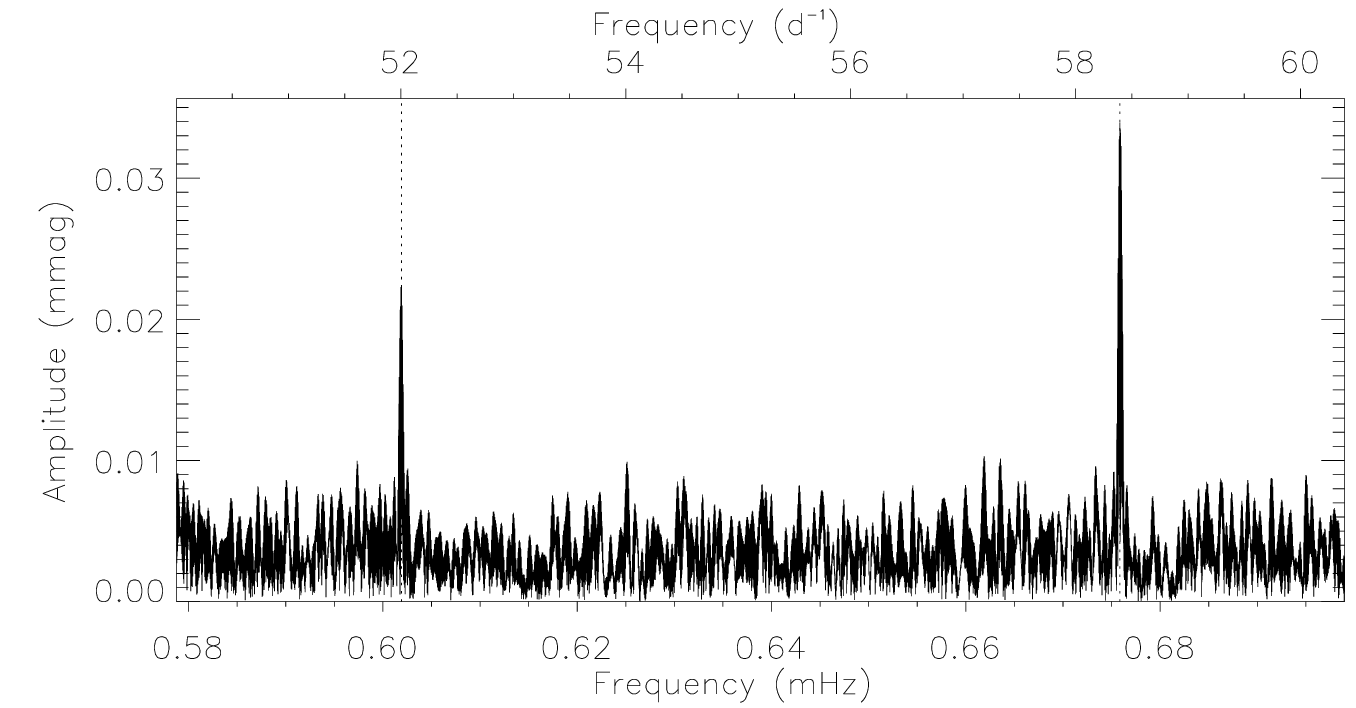}
\caption{Amplitude spectrum of TIC\,445796153. The two pulsation modes marked with vertical dotted lines.}
\label{fig:445796153_ft}
\end{figure}

Here, we used the {\it TESS} sector 19 and 26 data to confirm the rotation period of $P_{\rm rot}=5.5721\pm0.0004$\,d (Fig.\,\ref{fig:445796153_rot}). We also detected two pulsation modes (Fig.\,\ref{fig:445796153_ft}) that are in the frequency range where both $\delta$\,Sct and roAp pulsation modes can be found (e.g., TIC\,356088697, $0.646584\pm0.000008$\,mHz; \citetalias{2021MNRAS.506.1073H}): $0.601908\pm0.000002$\,mHz ($52.0049\pm0.0002$\,\cd) and $0.675849\,0.000002$\,mHz ($58.3934\,0.0001$\,\cd). It is unusual for a $\delta$\,Sct star to exhibit so few modes, especially at this amplitude. However, we cannot rule out the possibility that these modes are indeed $\delta$\,Sct in nature, as that requires an investigation into the pulsation driving which differs between the two classes and is beyond the scope of this work. We therefore class TIC\,445796153 as a candidate roAp star, especially given the lack of convincing spectral classification of the star being chemically peculiar. Spectroscopic monitoring to provide a binary solution, coupled with time-resolved spectra will help identify the origin of the pulsations in this binary system. Efforts should also be made to exclude the possibility of a contaminating $\delta$\,Sct star.

\subsection{TIC\,91224991}

TIC\,91224991 (HD\,191380) has a spectral classification in the literature of F8 \citep{1923AnHar..98....1C}. \citet{2019AJ....158..138S} estimate the effective temperature and $\log g$ to be 6700\,K and 3.97\,\cms, respectively. The effective temperature is determined from the dereddened Bp-Rp colour, while the $\log g$ value is derived using the estimated mass and radius.  Another estimation of $T_{\rm eff}$ is given by \citet{2006ApJ...638.1004A} as 6650\,K, while \citet{2019A&A...628A..94A} derived $T_{\rm eff}$ = 6960\,K based on narrow-band photometry. 

 \citet{2022MNRAS.510.5743B} categorised this star as an A star with high frequency variability, based on the {\it TESS} data. He estimated a luminosity $\log L/{\rm L}_{\odot} = 0.96$ using {\it Gaia} EDR3 parallaxes \citep{2016A&A...595A...1G, 2021A&A...649A...1G} in conjunction with interstellar reddening corrections from \citet{2017AstL...43..472G} and a bolometric correction calibration by \citet{2013ApJS..208....9P}.  \citet{2022MNRAS.510.5743B} found a characteristic frequency $1.160$\,mHz (100.213\,\cd) and also one harmonic, but does not give a rotation frequency. This star was also observed using long baseline interferometry by \citet{2019MNRAS.490.3158C}, who found an angular diameter corresponding to a radius of $R=2.13$\,R$_\odot$. 

\begin{figure}
\centering
\includegraphics[width=\columnwidth]{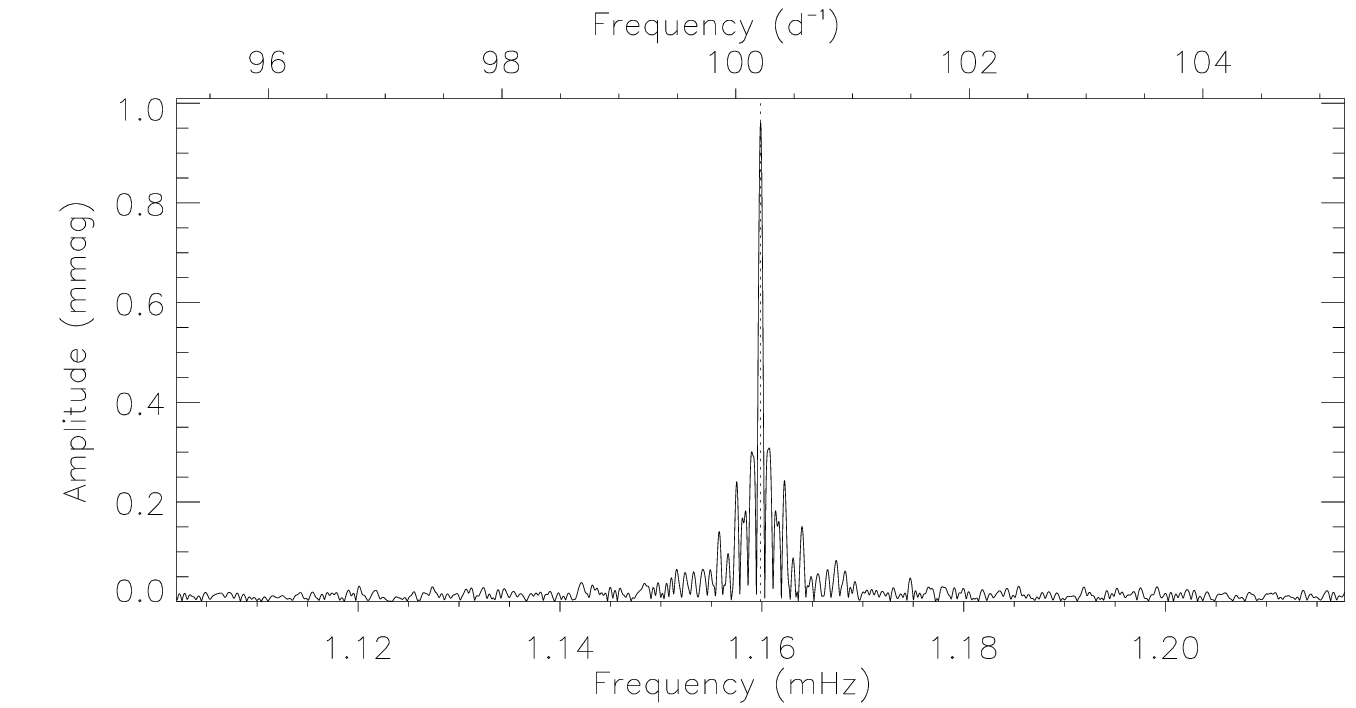}
\caption{Amplitude spectrum of TIC\,91224991. The vertical dotted line marks the pulsation mode.}
\label{fig:91224991_ft}
\end{figure}

This star was added to the {\it TESS} 2-min target list as a result of a tentative detection of variability in KELT data \citep{2007PASP..119..923P} that was made available to the TASOC WG4 members. The {\it TESS} data, from sector 14, clearly showed the presence of high frequency variability. We determined the frequency to be $1.159876\pm0.000002$\,mHz ($100.2133\pm0.0002$\,\cd; Fig.\,\ref{fig:91224991_ft}), and note the presence of the second harmonic at twice this frequency. Our results are consistent with those of \citet{2022MNRAS.510.5743B}. There are no rotational sidelobes to the pulsation peak in the amplitude spectrum, which is expected given the lack of a rotation signal in the {\it TESS} data. 

\section{The remaining roAp stars}

To provide a complete sample of roAp stars, we must include those that have not been observed by {\it TESS} and those that have been observed or discovered beyond the Cycle\,1 and Cycle\,2 observations. Therefore, in Table\,\ref{tab:further_roAp} we present these stars. Where there are now {\it TESS} data available, we follow the procedures outlined above to provide the rotation and pulsation properties. For each star, we present their amplitude spectrum in the range of the pulsations in Appendix\,\ref{sec:app_remaining}, and for stars that have yet to be observed, we provide the literature information.

\begin{table*}
    \centering
    \caption{Details of the roAp stars not observed in Cycles\,1 or 2. The columns provide the TIC identifier and star name, the TIC\,v8.1 {\it TESS} magnitude, the spectral type, the effective temperature, as provided in the TIC, the sectors in which {\it TESS} observed the target, where applicable. The final three columns provide the stellar rotation period derived in this work, the pulsation frequency(ies) (note that sidelobes that arise from oblique pulsation are not listed), and the pulsation amplitude(s) seen in the {\it TESS} data. A `*' indicates excess power remains in the amplitude spectrum after extracting the modes listed. Frequencies marked with a $^\dagger$ are derived from sidelobes. A $\ddag$ denotes that the SAP data were used to derive the value. For the stars not observed by {\it TESS} the superscript after the TIC ID give the data reference, and we neglect providing the amplitude since various different techniques and/or filters were used, thus making them non-comparable to the {\it TESS} results.}
    \label{tab:further_roAp}
    \begin{tabular}{lcccccrcc}
        \hline
         \multicolumn{1}{c}{TIC} & {HD/TYC}    	& {\it TESS} & {Spectral} 	 & $T_{\rm eff}^{\rm TIC}$   & {Sectors} &  \multicolumn{1}{c}{{$P_{\rm rot}$}}  	& \multicolumn{1}{c}{Pulsation frequency}	&  \multicolumn{1}{c}{Pulsation amplitude}     \\
                       		             &   {name}          	&  mag & {type}            &   {(K)}                               & {} &  \multicolumn{1}{c}{(d)}         		&   \multicolumn{1}{c}{(mHz)}			& \multicolumn{1}{c}{(mmag)}  \\
        \hline
\multicolumn{6}{l}{\textit{roAp stars discovered in TESS data}}\\

33604636 	& 42605		& 8.78	& Ap\,SrEuCr	& 8280	& 33	& $2.7793\pm0.0011$ &$1.25733\pm0.00002$* &$0.068\pm0.007$	\\

36576010		& 216018		& 7.45	& A9Vp\,SrCrEu	&7750	& 42	& No signature$^a$ &$1.92919\pm0.00003$ & $0.045\pm0.006$ 	\\

198781841 	& 2134-154-1	& 10.58	&	--		&7730	& 40,53,54	& No signature		&$1.8913785\pm0.0000005$ &$0.25\pm0.01$\\
											& 	& 	&			&	& 		& 		&$1.9271666\pm0.0000003$ &$0.40\pm0.01$\\
											& 	& 	&			&	& 		& 		&$1.9629854\pm0.0000009$ &$0.15\pm0.01$\\

229960986	& 4221-240-1	&10.38	&	--	& 8150	& 40,41,47-55 		& $2.093996\pm0.000003$& $2.2179825\pm0.0000009$ & $0.082\pm0.006$ 	\\

453826702	& 22032		& 8.81	& Ap\,SrEuCr	& 8100	& 31	&  $4.8525\pm0.0007$ 	&$1.57518\pm0.00003$* & $0.051\pm0.007$	\\
			& 			& 		&			&		& 	&					&$1.60998\pm0.00001$ & $0.131\pm0.007$\\
			& 			& 		&			&		& 	& 					&$1.64107\pm0.00003$ & $0.061\pm0.007$\\

456673854	& 59702		& 8.56	& A8V		& 7770	& 34	& $5.438\pm0.001$		&$3.13157\pm0.00004^\dagger$	&<0.03		\\
			& 			& 		&			&		& 	& 					&$3.16928\pm0.00002$			&$0.113\pm0.008$\\
			& 			& 		&			&		& 	& 					&$3.20693\pm0.00004^\dagger$	&<0.02	\\

630844439 & 9348-1029-1& 8.44	&	A8m		&		& 27,28 &$1.6618\pm0.0003$	& $3.36021\pm0.00001$ &$0.13\pm0.01$\\ \\

\multicolumn{6}{l}{\textit{Known roAp stars prior to TESS launch}}\\
        
96855460 	& 185256		& 9.55	& Fp\,SrEu	& 7100	&27	&		$\sim24$	&$1.62766\pm0.00003$ & $0.13\pm0.01$\\

100196783	& 193756		&8.98	& Ap\,SrCrEu	&7590	& 67	& $3.0096\pm0.0005$& $1.28405\pm0.00001$ & $0.199\pm0.008$	\\

163587609	& 101065		& 7.53	& F8/G0p		& 6125	& 37 & No signature$^a$			&$1.315020\pm0.000017$ & $0.065\pm0.005$\\
			& 			& 		&			&		& 	& 					&$1.369901\pm0.000010$ & $0.116\pm0.005$\\
			& 			& 		&			&		& 	& 					&$1.372881\pm0.000001$* & $0.958\pm0.005$\\
			& 			& 		&			&		& 	& 					&$1.393128\pm0.000010$ & $0.112\pm0.005$\\
			& 			& 		&			&		& 	& 					&$1.427743\pm0.000039$ & $0.028\pm0.005$\\

293265536	& 4-562-1		& 9.93	& A9p\,SrEu(Cr)& 7640	& 42	& No signature		&$1.73910\pm0.00001$&$0.32\pm0.01$	\\

299000970	& 176232		& 5.69	& A7p\,ArEu	& 7520	& 53,54	& No signature$^a$		&$1.336785\pm0.000013$&$0.012\pm0.001$	\\
			& 			& 		&			&		& 	& 					&$1.396832\pm0.000003$&$0.045\pm0.001$	\\
			& 			& 		&			&		& 	& 					&$1.417638\pm0.000018$&$0.009\pm0.001$	\\
			& 			& 		&			&		& 	& 					&$1.427100\pm0.000005$&$0.029\pm0.001$	\\
			& 			& 		&			&		& 	& 					&$1.447892\pm0.000006$&$0.027\pm0.001$	\\
			& 			& 		&			&		& 	& 					&$1.508482\pm0.000024$&$0.006\pm0.001$	\\

318007796	& 190290		& 9.61	& Ap\,EuSr	& 7560	& 27,39	& $4.09070\pm0.00002$	& $2.2307867\pm0.0000015$ & $0.049\pm0.007$	\\
			& 			& 		&			&		& 		& 					& $2.2723808\pm0.0000002$ & $0.307\pm0.007$\\
			
354619745$^\ddag$	& 201601		& 4.48	& A9Vp\,SrCrEu	& 7490	& 55		& No signature$^a$	&$1.311230\pm0.000009$ & $0.044\pm0.002$ \\
			& 			& 		&			&		& 		& 					& $1.338619\pm0.000001*$ & $0.317\pm0.002$ \\
			& 			& 		&			&		& 		& 					& $1.366022\pm0.000003$ & $0.124\pm0.002$ \\

383521659$\ddag$	& 137909		& 3.40	& F2Vp\,SrEuSi		& 		& 51			&	$\sim18.3$ & No signal in TESS data	& < 0.005\\

407929868$\ddag$	& 24355		&9.43	& A7/F0		& 7830	& 42,43,44		& $27.724\pm0.008$& $2.5961335\pm0.0000003$ & $2.056\pm0.006$\\	

420687462	& 122970		& 7.93	& F0			& 6800	& 50			& No signature & $1.476848\pm0.000035$ & $0.053\pm0.007$ \\
			& 			& 		&			&		& 		& 				& $1.477817\pm0.000015$ & $0.120\pm0.007$ \\
			& 			& 		&			&		& 		& 				& $1.502519\pm0.000005$ & $0.290\pm0.006$ \\

445543326	& 12098		& 7.91	& 	F0\,Eu		& 7760	& 58		&$5.4816\pm0.0008$ &$2.12206\pm0.00003$ & $0.062\pm0.007$ \\
			& 			& 		&			&		& 		& 				& $2.12636\pm0.00004$ & $0.040\pm0.007$ \\
			& 			& 		&			&		& 		& 				& $2.30602\pm0.00003$ & $0.054\pm0.007$ \\

 \multicolumn{6}{l}{\textit{Known roAp stars not observed by {\it TESS}}}\\
 
153101639$^b$	& --			& 11.87	& A7Vp\,SrEu(Cr)	&7510	& & $3.6747\pm0.0005$ & $1.7585027\pm0.0000002$\\

335303863$^c$	& 137949		& 6.45	& F0Vp\,EuGdSr	&7400 & &>10\,000$^d$ &$2.01475317\pm0.00000007$\\

411247704$^e$	& 196470		& 9.49	& Fp\,SrEu	&7450 & & None published& $1.544$	\\

413938178$^f$	& 148593 		& 8.71	& Ap\,SrEuCr	& 7390  & &None published & $1.56\pm0.01$\\

432223926$^g$	& 134214		& 7.17	& F2Vp\,SrCrEu	& 7040 & &None published & $2.9496\pm0.0004$\\

465996299$^h$	& 177765		& 8.87	& Ap\,SrEuCr		&7420	& &>13\,500 & $0.702580\pm0.000006$\\
			&			&		&				&		& & & $0.713382\pm0.000015$\\
			&			&		&				&		& & & $0.717106\pm0.000010$\\

\hline
\multicolumn{9}{l}{$^a$See text for rotation period discussion. $^b$\citet{2018MNRAS.476..601H}; $^c$\citet{2018MNRAS.480.2976H}; $^d$\citet{2022MNRAS.514.3485G}; $^e$\citet{1990IBVS.3506....1M}}\\
\multicolumn{9}{l}{$^f$\citet{2013MNRAS.431.2808K}; $^g$\citet{1994MNRAS.270..115K}; $^h$\citet{2016IBVS.6185....1H}}\\
            \end{tabular}
\end{table*}

After accounting for the stars already presented as roAp stars in \citetalias{2021MNRAS.506.1073H}, this work, and that of \citet{2021RNAAS...5..268J} who reported TIC\,198781841 and TIC\,229960986 to be roAp stars, there are 27 additional stars listed in \citet{2022MNRAS.510.5743B} as roAp stars, or candidate roAp stars. Of these 27, 11 appeared in \citet{2014MNRAS.439.2078H} as either Am or marginal Am (Am:) stars. Four of these 11 stars have no {\it TESS} 2-min cadence data so it is unclear how \citet{2022MNRAS.510.5743B} decided a different variability type than the original classification, four further stars showed clear $\delta$\,Sct pulsations and one is a known binary with a compact pulsator. Of the two final Am(:) stars, TIC\,22113439 is listed as a double lined spectroscopic binary in LAMOST DR9 data \citep{2023ApJS..266...18Z} which explained the low frequency variability combined with the $\delta$\,Sct pulsations, and careful examination of the light curve of TIC\,22132451 showed that the low-frequency variability changed phase over the observations (sectors 40, 52 and 53), which is not expected of an $\alpha^2$\,CVn variable. This, coupled with the pulsations being of a $\delta$\,Sct nature, means we rejected this as an roAp star.

Beyond the Am(:) stars, we considered TIC\,427400331. This star is listed as a Be star by \citet{2022AJ....163..226L} who commented on the presence of Slowly Pulsating B (SPB) star g-modes, and noted the unusual presence of high-frequency variability. We checked the high-frequency modes and found they did not show oblique pulsation, despite a confirmed rotation period. Two low resolution spectra, in the Be Star Spectra (BeSS) database did not show strong absorption lines typical of Ap stars, and the {\it Gaia} DR3 Re-normalised Unit Weight Error (RUWE) value is >25, strongly indicating a binary star. We therefore rejected this star as an roAp star, pending an investigation to the source of the high-frequency variability. We rejected a further eight stars on the grounds of either them being obvious $\delta$\,Sct stars or having no signal at the claimed frequency. Finally, two stars of the 27 roAp stars in \citet{2022MNRAS.510.5743B} are listed as candidates in our work (TIC\,410163387 in \citetalias{2021MNRAS.506.1073H}, and TIC\,405892692 here). Given there are no additional data products available, we did not include these stars as roAp in Table\,\ref{tab:further_roAp}. The five remaining stars are included as roAp stars in Table\,\ref{tab:further_roAp}.

In other work, \citet{2020A&A...639A..31M} and \citet{2022A&A...660A..70M} noted 10 candidate roAp stars that were detected as part of a systematic search of the {\it TESS} data for ssrAp stars. Nine stars, one of which is the same as a candidate in \citet{2022MNRAS.510.5743B}, are rejected as roAp stars here due to the lack of a detection of an appropriate peak in the amplitude spectra. One star, TIC\,72802368 from \citet{2020A&A...639A..31M}, remains a candidate as a signal at the reported frequency ($1.207677\pm0.000001$\,mHz) is detected in two (sectors 36 and 62) of the three available sectors.

Below, we provide comments on all stars in Table\,\ref{tab:further_roAp} with {\it TESS} data:

\begin{itemize}
\item {\it TIC\,33604636 (HD\,42605)} -- There are clear signs of residual power in the amplitude spectrum after the removal of the pulsation frequency. It is not clear if there are further modes present, or amplitude/frequency variability at play (cf. Fig\,\ref{fig:33604636}).

\item {\it TIC\,36576010 (HD\,216018)} -- There may be additional modes around 1.79\,mHz that are not fully resolved (cf. Fig\,\ref{fig:36576010}). The rotation period is disputed in the literature, see Mathys et al., (submitted) for a discussion. 

\item {\it TIC\,198781841 (TYC\,2134-154-1)} -- The modes in this star are split by $\sim\,35.8\,\muup$Hz, which is plausibly half the large frequency separation \citep[see][]{2021RNAAS...5..268J} (cf. Fig\,\ref{fig:198781841}).

\item {\it TIC\,229960986 (TYC\,4221-240-1)} -- Application of the oblique pulsator model for this star resulted in a $\tan i\tan\beta$ value of $6.10\pm0.43$, and the conclusion that the mode is a slightly distorted dipole mode due to unequal phases between the pulsation mode and the sidelobes (cf. Fig\,\ref{fig:229960986}).

\item {\it TIC\,453826702 (HD\, 22032)} -- The three modes identified in this star are separated by an average of $\sim33\,\muup$Hz which is plausibly half of the large frequency separation for this star. After removing the three identified modes, there remains excess power of unknown origin in the same frequency range (cf. Fig\,\ref{fig:453826702}).

\item {\it TIC\,456673854 (HD\,59702)} -- The amplitude spectrum of this star showed, in increasing frequency, a dipole mode, a quadrupole mode and another dipole mode. The dipole modes form doublets with only the sidelobes visible while the quadrupole mode showed the pulsation and the $\pm2\nu_{\rm rot}$ sidelobes. With the application of the oblique pulsator model, we found that all three modes have $\tan i\tan\beta$ in agreement ($7.57\pm3.38$ for the first dipole, $10.91\pm3.84$ for the quadrupole, and $5.29\pm2.57$ for the second dipole), with the phase relations suggesting the first two modes are non-distorted, while the second dipole mode is suggested to be a distorted mode. This star will benefit from an in-depth study (cf. Fig\,\ref{fig:456673854}).

\item{\it TIC\,630844439 (TYC\,9348-1029-1)} -- The amplitude spectrum of this star showed the presence of both rotation and g-mode pulsations. It is noted in the literature as a binary star, so we postulate that the g-modes are from a $\gamma$\,Dor companion to an roAp star. The high-frequency peaks form a triplet, suggesting a single dipole mode in the roAp star. Applying the oblique pulsator model to the triplet resulted in $\tan i\tan\beta =2.48\pm0.27$, and the indication that the mode is slightly distorted (cf. Fig\,\ref{fig:630844439}).

\item{ \it TIC\,96855460 (HD\,185256)} -- The SAP data were used for this star to provide a lower limit on the rotation period. No new information is gleaned about the pulsation mode (cf. Fig\,\ref{fig:96855460}).

\item {\it TIC\,100196783 (HD\,193756)} -- This star showed strong g-mode pulsation superimposed on the rotation signature. This is the first time this is mentioned in the literature. While there is no confirmation that this star is a binary, the {\it Gaia} RUWE ($\approx3.6$) suggests otherwise. The {\it TESS} data provide no new information on the roAp pulsation (cf. Fig\,\ref{fig:100196783}).

\item{\it TIC\,163587609 (HD\,101065)} -- While there are five pulsation modes detected in the star, there is no clear indication of the large frequency separation. Three modes are separated by $\sim55\,\muup$Hz, but this value is in disagreement with $64\,\muup$Hz from \citet{2008A&A...490.1109M}. It is unclear which, if either, of these large separation values is correct. There is also excess power remaining in the amplitude spectrum around the principal peak after it was removed (cf. Fig\,\ref{fig:163587609}). This star has a lower limit on its rotation period of $\sim 43$\,yr \citep{2018MNRAS.477.3791H}.

\item{\it TIC\,293265536 (TYC\,4-562-1)} -- No new information is provided by the {\it TESS} data (cf. Fig\,\ref{fig:293265536}).

\item{\it TIC\,299000970 (HD\,176232)} -- With the identification of six pulsation modes, we were able to identify two recurrent separations of $\sim60\,\muup$Hz and $\sim30\,\muup$Hz, which we interpreted as the large frequency separation, and half of it. This is different from the value adopted by \citet{2008A&A...483..239H} of $50\,\muup$Hz, based on MOST photometry (cf. Fig\,\ref{fig:299000970}). \citet{2019MNRAS.483.3127S} propose a rotation period much greater than 12\,yr.

\item{\it TIC\,318007796 (HD\,190290)} -- The {\it TESS} data showed both modes in this star to be multiplets thus allowing for the determination that the lowest frequency mode is a dipole mode, with the other being a quadrupole mode, which suggests a large frequency separation of $\sim83\,\muup$Hz. Applying the oblique pulsator model resulted in $\tan i\tan\beta$ values for the dipole and quadrupole of $4.01\pm0.61$ and $3.51\pm0.27$, respectively, with the phase variations relations for each mode suggesting they are both distorted (cf. Fig\,\ref{fig:318007796}).

\item{\it TIC\,354619745 (HD\,201601)} -- We list three pulsation modes in this star, but note there is residual power remaining in the amplitude spectrum after they are removed. The three modes are each separated by $27.4\,\muup$Hz, which is presumably half of the large frequency separation. The principal mode showed a harmonic at twice its frequency, which is joined by a second peak whose frequency is the sum of the two highest amplitude modes. This non-linear interaction is only the third occurrence noted in the literature after 33\,Lib \citep{2018MNRAS.480.2976H} and $\alpha$\,Cir \citetalias{2021MNRAS.506.1073H} (cf. Fig\,\ref{fig:354619745}). The rotation period of this star is expected to be of order $100$\,yr \citep{2016MNRAS.455.2567B}.

\item{\it TIC\,383521659 (HD\,137909)} -- We did not detect the pulsation mode in this star in the {\it TESS} data. The pulsation was detected spectroscopically with an amplitude of about $30$\,m\,s$^{-1}$ \citep{2007MNRAS.380..741K}, and is suspected to be below the {\it TESS} noise level (cf. Fig\,\ref{fig:383521659}).

\item{\it TIC\,407929868 (HD\,24355)} -- The {\it TESS} data of this star showed a highly distorted quadrupole mode, but did not add additional information over the literature (cf. Fig\,\ref{fig:407929868}).

\item{\it TIC\,420687462 (HD\,122970)} -- The {\it TESS} data do not add new information to the literature for this star (cf. Fig\,\ref{fig:420687462}).

\item{\it TIC\,445543326 (HD\,12098)} -- This star showed a dipole triplet with a second mode close in frequency, and a third mode at higher frequency. Applying the oblique pulsator model to the dipole triplet resulted in a $\tan i\tan\beta$ value of $2.28\pm0.30$, with the phase relations showing this mode to be very distorted. A more detailed analysis of this star is required (cf. Fig\,\ref{fig:445543326}).

\end{itemize}


\section{Conclusions}
\label{sec:conc}

This work constitutes the second phase of analysing the {\it TESS} data in the search for high-frequency pulsations in Ap stars. We have used the 2-min cadence Cycle\,2 data of all hot ($T_{\rm eff}>6000$\,K) stars in the Northern Ecliptic hemisphere, amounting to over 36\,000 stars. This has enabled us to: identify seven new roAp stars unreported as such in the literature, present a detailed analysis of a further 9 roAp stars previously identified as new in the {\it TESS} data, show that four of 16 well established roAp stars do not show variability in the analysed {\it TESS} data, and present 7 candidate roAp stars where insufficient spectral data exist to admit the star to the roAp class. 

Three of the seven new roAp stars are multiperiodic pulsators, thus offering the opportunity for a more complete asteroseismic analysis. Furthermore, of these seven stars, five were previously not identified as chemically peculiar stars.This has important ramifications for the continued search for roAp stars in {\it TESS} data, and indeed other data sets. We have shown that catalogues of CP stars are, understandably, incomplete and as such, the search for high-frequency pulsations should include stars previously not identified as CP. 

All of these new roAp stars are rotationally variable stars, and of the 32 roAp stars presented here, almost 70\,per\,cent (22 stars) show variability due to spots. When we combined the results from this paper, \citetalias{2021MNRAS.506.1073H}, and all roAp stars in the literature, we found 63\,per\,cent (71 of 112 roAp stars) showed rotation signatures in their light curves. While some periods are still to be determined, many of the stars showed periods less than 10\,d. A future investigation will explore the potential correlations between rotation rates and pulsation properties (Cunha et al., in prep.), with the super slowly rotating Ap stars showing a higher incidence of roAp stars than is found for all Ap stars \citep{2020A&A...639A..31M,2022A&A...660A..70M}.

We have identified a few examples of short timescale mode variability, which while not as extreme as in some cases, show that the modes in the roAp stars can be unstable on the order of days to weeks. One example here is TIC\,26833276 where we see changes in mode amplitude between {\it TESS} orbits (cf., Fig.\,\ref{fig:26833276_ft}). More significant is TIC\,435263600, which is complex in both its pulsation spectrum (Fig\,\ref{fig:435263600_ft}) and its stellar spectrum (Fig.\,\ref{fig:435263600_spec}). With a presumed close companion, this star provides a prime opportunity to investigate whether interactions between roAp stars and companions affect their pulsation behaviour.

As with \citetalias{2021MNRAS.506.1073H}, we note the lack of detection of roAp pulsations in the hottest of the Ap stars. The theoretical blue edge of the roAp instability strip reaches to temperatures of about 10\,000\,K \citep{2002MNRAS.333...47C} whereas the hottest known pulsator based on the TIC temperature lies at $T_{\rm eff_{\rm TIC}}\sim8540$\,K \citep[TIC\,119327278;][]{2021MNRAS.506.1073H}. This presents a significant challenge for modelling the driving of pulsation in the hottest Ap stars, showing the need to further develop theoretical understanding of these stars.

Our results show that the incidence of roAp stars among the Ap stars is low. Before the launch of {\it TESS}, the TASOC WG4 proposed to observe all stars listed as Ap in the \citet{2009A&A...498..961R} catalogue with 2-min cadence during the nominal mission. This amounted to 1381 stars proposed, of which 1285 have been observed to date. Of the 112 stars presented across our two papers, 73 appear in that catalogue. This implies that the incidence of roAp stars in the known Ap star population is just 5.5\,per\,cent. This raises the fundamental question of: why do only some Ap stars pulsate? We must consider that, as we have shown, some pulsation amplitudes are below the {\it TESS} detection limit, but that is only a fraction of the known roAp stars. Thanks to our work, we now have a statistically significant sample of roAp stars with which we can investigate this conundrum, and drive forward our theoretical understanding of these complex stars.

These works, \citetalias{2021MNRAS.506.1073H} and the current paper, represent the most comprehensive and collaborative observational study of the roAp stars observed during the nominal mission of {\it TESS}. In this collective work, we have identified 19 new roAp stars, presented an analysis of 87 more, and included a further six which {\it TESS} has not yet observed. Our approach has resulted in a homogeneous sample of roAp stars in both the detection methods used and their analysis techniques and software. This resource will stand as the definitive collection of roAp stars observed during Cycles\,1 \& 2 of the {\it TESS} mission, with other roAp stars noted in the literature from later {\it TESS} Cycles 4, 5 and 6. 

\section*{Data availability}
All photometric data used in this work are available to download from the MAST server\footnote{\url{https://mast.stsci.edu/portal/Mashup/Clients/Mast/Portal.html}}. LAMOST data are available to download from the LAMOST data archive\footnote{\url{http://www.lamost.org/dr9/}}. HERMES data are available on reasonable request from the authors. 

\section*{Acknowledgements}

DLH acknowledges financial support from the Science and Technology Facilities Council (STFC) via grant ST/M000877/1. 
MSC acknowledges the support of Funda\c c\~ao para a Ci\^encia e Tecnologia FCT/MCTES, Portugal, through national funds by these grants UIDB/04434/2020, UIDP/04434/2020.FCT, 2022.03993.PTDC and CEECIND/02619/2017. 
MLM, AA, JPG and ARB  acknowledge financial support from project PID2019-107061GB-C63 from the `Programas Estatales de Generaci\'on de Conocimiento y Fortalecimiento Cient\'ifico y Tecnol\'ogico del Sistema de I+D+i y de I+D+i Orientada a los Retos de la Sociedad' and from the Severo Ochoa grant CEX2021-001131-S funded by MCIN/AEI/10.13039/501100011033. MLM also acknowledges financial support from Spanish public funds for research under project ESP2017-87676-C5-5-R.
VA was supported by a research grant (00028173) from VILLUM FONDEN.
SBF received financial support from the Spanish State Research Agency (AEI) Projects No. PID2019-107061GB-C64: ``Contribution of the UGR to the PLATO2.0 space mission. Phases C/D-1''. SBF also thanks the resources received from the PLATO project collaboration with Centro de Astrobiolog\'ia (PID2019-107061GB-C61).
This work was supported by the PRODEX Experiment Agreement No. 4000137122 between the ELTE E\"otv\"os Lor\'and University and the European Space Agency (ESA-D/SCI-LE-2021-0025). This project has received funding from the HUN-REN Hungarian Research Network.
MS acknowledges the support by Inter-transfer grant no LTT-20015.
DMB gratefully acknowledges a senior postdoctoral fellowship from Research Foundation Flanders (FWO; grant number: 1286521N), and the Engineering and Physical Sciences Research Council (EPSRC) of UK Research and Innovation (UKRI) in the form of a Frontier Research grant under the UK government's Horizon Europe funding guarantee (grant number: [EP/Y031059/1]).
OK and VK acknowledge financial support from the Natural Sciences and Engineering Research Council of Canada (NSERC), and from Mitacs.
ZsB and \`AS acknowledge the financial support of the KKP-137523 `SeismoLab' \`Elvonal grant of the Hungarian Research, Development and Innovation Office (NKFIH). ZsB also acknowledges the support by the J\`anos Bolyai Research Scholarship of the Hungarian Academy of Sciences.
DLB gratefully acknowledges support from NASA (NNX16AB76G, 80NSSC22K0622) and the Whitaker Endowed Fund at Florida Gulf Coast University
LFM acknowledges the financial support from the UNAM under grant PAPIIT IN117323.
AGH acknowledges funding support from ``FEDER/Junta de Andaluc\`ia-Consejer\`ia de Econom\'a y Conocimiento'' under project E-FQM-041-UGR18 by Universidad de Granada and from the Spanish State Research Agency (AEI) project PID2019-107061GB-064.
GH acknowledges financial support from the Polish National Science Center (NCN), grant no. 2021/43/B/ST9/02972.
AH acknowledges support from the Science and Technology Facilities Council (STFC) grant No. ST/X000915/1.
OK acknowledge support by the Swedish Research Council (project 2019-03548).
CCL acknowledges Discovery Grant support by the Natural Science and Engineering Research Council (NSERC) of Canada.
PM acknowledges the support from the European Union's Horizon 2020 research and innovation program under grant agreements No. 730890 and 101004719.
ARB acknowledges funding support from Spanish public funds for research from projects ESP 2017-87676C5-5-R and PRE 2018-084322 from the `Programa Estatal de Promoci\'on del Talento y su Empleabilidad del Plan Estatal de Investigaci\'on Cient\'ifica y T\'ecnica y de Innovaci\'on 2013-2016', from project PID 2019-107061GBC63 from the `Programas Estatales de Generaci\'on de Conocimiento y Fortalecimiento Cient\'ifico y Tecnol\'ogico del Sistema de I+D+i y de I+D+i Orientada a los Retos de la Sociedad', all from the Spanish Ministry of Science, Innovation and Universities (MCIU) and from the Severo Ochoa grant CEX2021-001131-S funded by MCIN/AEI/10.13039/501100011033.
SJM was supported by the Australian Research Council (ARC) through Future Fellowship FT210100485.
IS acknowledges a partial financial support from BFNI via grant K$\Piup$-06-H58/30.
JCS acknowledges support from the Spanish State Research Agency (AEI) Projects No. PID2019-107061GB-C64: ``Contribution of the UGR to the PLATO2.0 space mission. Phases C/D-1''.
This research was supported by the KKP-137523 `SeismoLab' \`Elvonal grant of the Hungarian Research, Development and Innovation Office (NKFIH) and by the MW-Gaia COST Action (CA18104).
TW thanks the supports from the National Key Research and Development Program of China (Grant No. 2021YFA1600402), the B-type Strategic Priority Program of Chinese Academy of Sciences (Grant No. XDB41000000), the National Natural Science Foundation of China (Grants No. 12133011 and 12273104), the Youth Innovation Promotion Association of Chinese Academy of Sciences, the Yunnan Ten Thousand Talents Plan Young \& Elite Talents Project, and from the International Centre of Supernovae, Yunnan Key Laboratory (No. 202302AN360001).
EZ acknowledges Iran National Science Foundation (INSF) for its total support under grant No. 4002562.
WZ acknowledge the support from the National Natural Science Foundation of China (NSFC) No. 12273002 and 12090040/2.

This publication was produced within the framework of institutional support for the development of the research organisation of Masaryk University.
This paper includes data collected by the {\it TESS} mission, which are publicly available from the Mikulski Archive for Space Telescopes (MAST). Funding for the {\it TESS} mission is provided by NASA's Science Mission directorate. Funding for the {\it TESS} Asteroseismic Science Operations Centre is provided by the Danish National Research Foundation (Grant agreement no.: DNRF106), ESA PRODEX (PEA 4000119301) and Stellar Astrophysics Centre (SAC) at Aarhus University. We thank the {\it TESS} team and staff and TASC/TASOC for their support of the present work. This research has made use of the SIMBAD database, operated at CDS, Strasbourg, France.

Guoshoujing Telescope (the Large Sky Area Multi-Object Fiber Spectroscopic Telescope LAMOST) is a National Major Scientific Project built by the Chinese Academy of Sciences. Funding for the project has been provided by the National Development and Reform Commission. LAMOST is operated and managed by the National Astronomical Observatories, Chinese Academy of Sciences.

Some of the data in this work are observations obtained with the HERMES spectrograph, which is supported by the Research Foundation - Flanders (FWO), Belgium, the Research Council of KU Leuven, Belgium, the Fonds National de la Recherche Scientifique (F.R.S.-FNRS), Belgium, the Royal Observatory of Belgium, the Observatoire de Gen{\`e}ve, Switzerland, and the Th{\"u}ringer Landessternwarte Tautenburg, Germany, and were assembled by the Mercator Telescope, operated on the island of La Palma by the Flemish Community, at the Spanish Observatorio del Roque de los Muchachos of the Instituto de Astrof{\'i}sica de Canarias.

This work has made use of the BeSS database, operated at LESIA, Observatoire de Meudon, France: \url{http://basebe.obspm.fr}.




\bibliographystyle{mnras}
\bibliography{references} 



\appendix

\section{Author Affiliations}
\label{app:affiliations}
$^{1}$Jeremiah Horrocks Institute, University of Central Lancashire, Preston PR1 2HE, UK\\
$^{2}$South African Astronomical Observatory, P.O. Box 9, Observatory 7935, Cape Town, South Africa\\
$^{3}$Instituto de Astrof\'isica e Ci\^encias do Espa\c co, Universidade do Porto CAUP, Rua das Estrelas, PT4150-762 Porto, Portugal\\
$^{4}$Stellar Variability Group, Institute of Astrophysics of Andalusia, Department of Stellar Physics, IAA-CSIC, Granada, Spain\\
$^{5}$Centre for Space Research, North-West University, Mahikeng 2745, South Africa\\
$^{6}$National Space Institute, Technical University of Denmark, Elektrovej, DK-2800 Kgs. Lyngby, Denmark\\
$^{7}$F\'isica Te\'orica y del Cosmos Dept., Universidad de Granada (UGR), 18071, Granada, Spain\\
$^{8}$Royal Observatory of Belgium, Ringlaan 3, B-1180 Brussels, Belgium\\
$^{9}$ELTE E{\"o}tv{\"o}s Lor\'and University, Gothard Astrophysical Observatory, Szombathely, Szent Imre h. u. 112., H-9700, Hungary\\
$^{10}$MTA-ELTE  Lend{\"u}let "Momentum" Milky Way Research Group, Hungary\\
$^{11}$HUN-REN?SZTE Stellar Astrophysics Research Group, H-6500 Baja, Szegedi \'ut, Kt. 766, Hungary
$^{12}$Department of Astronomy and Space Sciences, Science Faculty, Erciyes University, 38030 Melikgazi, Kayseri, T\"urkiye\\
$^{13}$Ankara University, Faculty of Science, Dept. of Astronomy and Space Sciences, 06100, Tandogan - Ankara, Turkey\\
$^{14}$Astronomical Institute of the Czech Academy of Sciences, Fri\v{c}ova 298, CZ-25165 Ond\v{r}ejov, Czech Republic\\
$^{15}$Department of Theoretical Physics and Astrophysics, Masaryk University, Kotl\'{a}\v{r}sk\'{a} 2, CZ-61137 Brno, Czech Republic\\
$^{16}$Institute for Astronomy, University of Hawaii, Honolulu, USA\\
$^{17}$Department of Astronomy, Peking University, Beijing 100871, China\\
$^{18}$The Kavli Institute for Astronomy and Astrophysics, Peking University, Beijing 100871, China\\
$^{19}$School of Mathematics, Statistics and Physics, Newcastle University, Newcastle upon Tyne, NE1 7RU, UK\\
$^{20}$Institute of Astronomy, KU Leuven, 3001 Leuven, Belgium\\
$^{21}$D\'epartement de Physique et d\'Astronomie, Universit\'e de Moncton, Moncton, N.B., Canada E1A 3E9\\
$^{22}$Virtualmech R\&D division. Parque Empresarial Nuevo Torneo. Arquitectura 1, 41015 Sevilla, Spain\\
$^{23}$Konkoly Observatory, HUN-REN Research Centre for Astronomy and Earth Sciences, MTA Centre of Excellence, H-1121 Budapest, Konkoly Thege Mikl\'os\'ut 15-17, Hungary\\
$^{24}$Department of Chemistry and Physics, Florida Gulf Coast University, 10501 FGCU Blvd S, Fort Myers, FL 33965, USA\\
$^{25}$Institute of Geophysics, University of Tehran, Tehran, Iran\\
$^{26}$Instituto de Astronom\'{\i}a--Universidad Nacional Aut\'onoma de M\'exico,  Ensenada, BC 22860, Mexico\\
$^{27}$Theoretical Physics and the Cosmos, University of Granada, E-18071, Granada, Spain\\
$^{28}$Department of Physics, Institute for Advanced Studies in Basic Sciences (IASBS), Zanjan 45137-66731, Iran\\
$^{29}$XTD-NTA, MS T-082, Los Alamos National Laboratory, Los Alamos, NM 87545 USA\\
$^{30}$Stellar Astrophysics Centre, Aarhus University, DK-8000 Aarhus C, Denmark\\
$^{31}$Nicolaus Copernicus Astronomical Center, Polish Academy of Sciences, ul. Bartycka 18, 00-716, Warszawa, Poland\\
$^{32}$Centre for Fusion, Space and Astrophysics, Department of Physics, University of Warwick, Coventry CV4 7AL, UK\\
$^{33}$Department of Physics and Kavli Institute for Astrophysics and Space Research, Massachusetts Institute of Technology, 77 Massachusetts Ave, Cambridge, MA 02139, USA\\
$^{34}$Department of Physics and Astronomy, Uppsala University, Box 516, SE 75120, Uppsala, Sweden\\
$^{35}$Mount Allison University, 69 York St, Sackville, NB, Canada\\
$^{36}$Instytut Astronnomiczny, Uniwersytet Wroc\l{}awski, Kopernika 11, 51-622 Wroc\l{}aw, Poland\\
$^{37}$Astronomical Observatory, University of Warsaw, Al.~Ujazdowskie~4, 00-478~Warszawa, Poland\\
$^{38}$National Astronomical Research Institute of Thailand, 260 Moo 4, T. Donkaew, A. Maerim, Chiang Mai 50180, Thailand\\
$^{39}$Centre for Astrophysics, University of Southern Queensland, Toowoomba, QLD 4350, Australia\\
$^{40}$Instituto de Astrof\'isica de Andaluc\'ia - CSIC, 18008 Granada, Spain\\
$^{41}$Department of Theoretical Physics and Astrophysics, Masaryk University, Kotl\'a\v rsk\'a 2, 611 37 Brno, Czech Republic\\
$^{42}$Instituto de Astrof\'isica de Andaluc\'ia (CSIC), Glorieta de la Astronom\'ia S/N, E18008 Granada, Spain \\
$^{43}$Department of Physics, Faculty of Science, University of Zanjan, Zanjan, Iran\\
$^{44}$Max-Planck-Institut f\"ur Sonnensystemforschung, 37077 G\"ottingen, Germany\\
$^{45}$Astrophysics Group, Keele University, Keele Rd, Keele ST5 5BG, UK\\
$^{46}$Institute of Astronomy and NAO, Bulgarian Academy of Sciences, blvd.Tsarigradsko chaussee 72, Sofia, Bulgaria\\
$^{47}$F\`isica Te\`orica y del Cosmos Dept. Universidad de Granada. Campus de Fuentenueva s/n. 18071. Granada, Spain\\
$^{48}$ELTE E\"otv\"os Lor\'and University, Institute of Physics, H-1117, P\'azm\'any P\'eter s\'et\'any 1/A, Budapest, Hungary\\
$^{49}$Yunnan Observatories, Chinese Academy of Sciences, 396 Yangfangwang, Guandu District, Kunming, 650216, People's Republic of China\\
$^{50}$Key Laboratory for the Structure and Evolution of Celestial Objects, Chinese Academy of Sciences, 396 Yangfangwang, Guandu District, Kunming, 650216, People's Republic of China\\
$^{51}$Center for Astronomical Mega-Science, Chinese Academy of Sciences, 20A Datun Road, Chaoyang District, Beijing, 100012, People's Republic of China\\
$^{52}$University of Chinese Academy of Sciences, Beijing 100049, People's Republic of China\\
$^{53}$International Centre of Supernovae, Yunnan Key Laboratory, Kunming 650216, P. R. China\\
$^{54}$Institute of Theoretical Physics, Shanxi University, Taiyuan 030006, China\\
$^{55}$Department of Physics, Faculty of Science, University of Zanjan, University Blvd., Zanjan, Postal Code: 45371-38791, Zanjan, Iran.\\
$^{56}$Institute for Frontiers in Astronomy and Astrophysics, Beijing Normal University, 102206, Beijing, P. R. China\\
$^{57}$Department of Astronomy, Beijing Normal University, 100875, Beijing, P. R. China\\
$^{58}$Department of Physics and Kavli Institute for Astrophysics and Space Research, Massachusetts Institute of Technology, Cambridge, MA 02139, USA\\
$^{59}$Department of Earth, Atmospheric and Planetary Sciences, Massachusetts Institute of Technology, Cambridge, MA 02139, USA\\
$^{60}$Department of Aeronautics and Astronautics, MIT, 77 Massachusetts Avenue, Cambridge, MA 02139, USA\\

\section{Spectral plots}
Here we provide the spectral  plots used to derive new or updated spectral classifications. Information of the sources can be found in Sec.\,\ref{sec:spec}.

\begin{figure}
\centering
\includegraphics[width=\columnwidth]{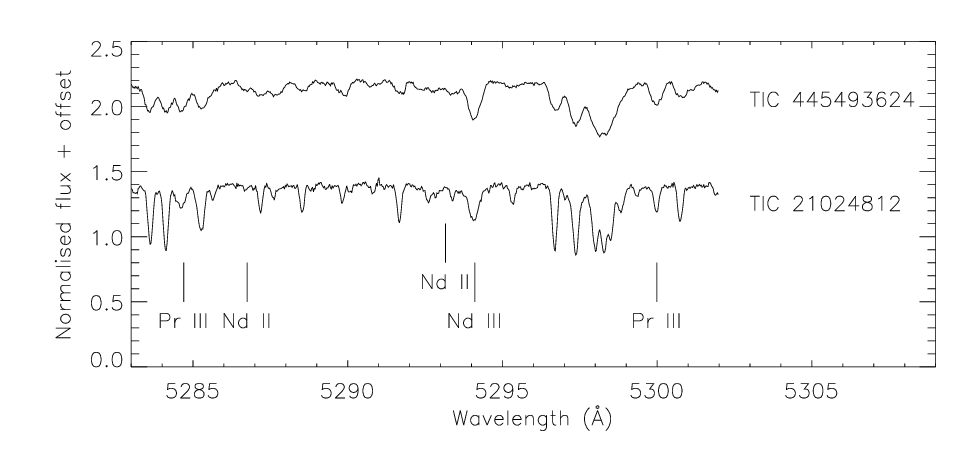}
\caption{HERMES spectra of TIC\,445493624 and TIC\,21024812 in the blue region.}
\label{fig:HERMES_BLUE}
\end{figure}

\begin{figure}
\centering
\includegraphics[width=\columnwidth]{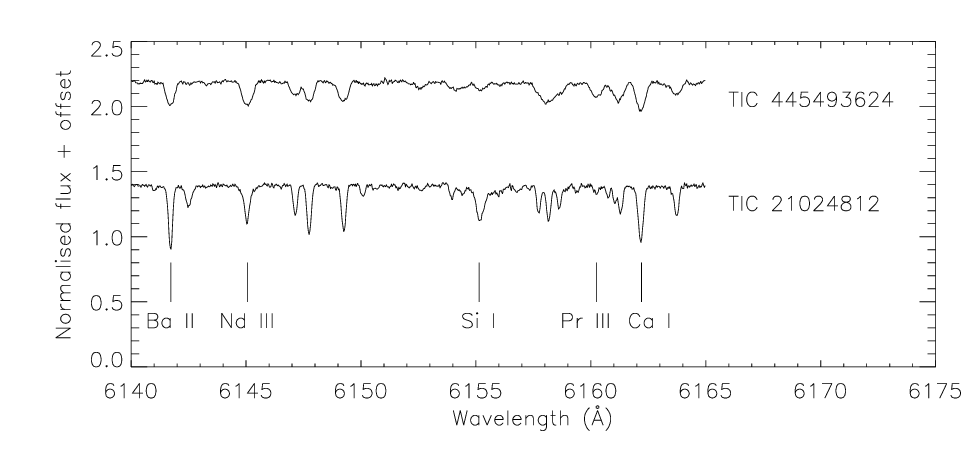}
\caption{HERMES spectra of TIC\,445493624 and TIC\,21024812 in the red region.}
\label{fig:HERMES_RED}
\end{figure}

\begin{figure}
\centering
\includegraphics[width=\columnwidth]{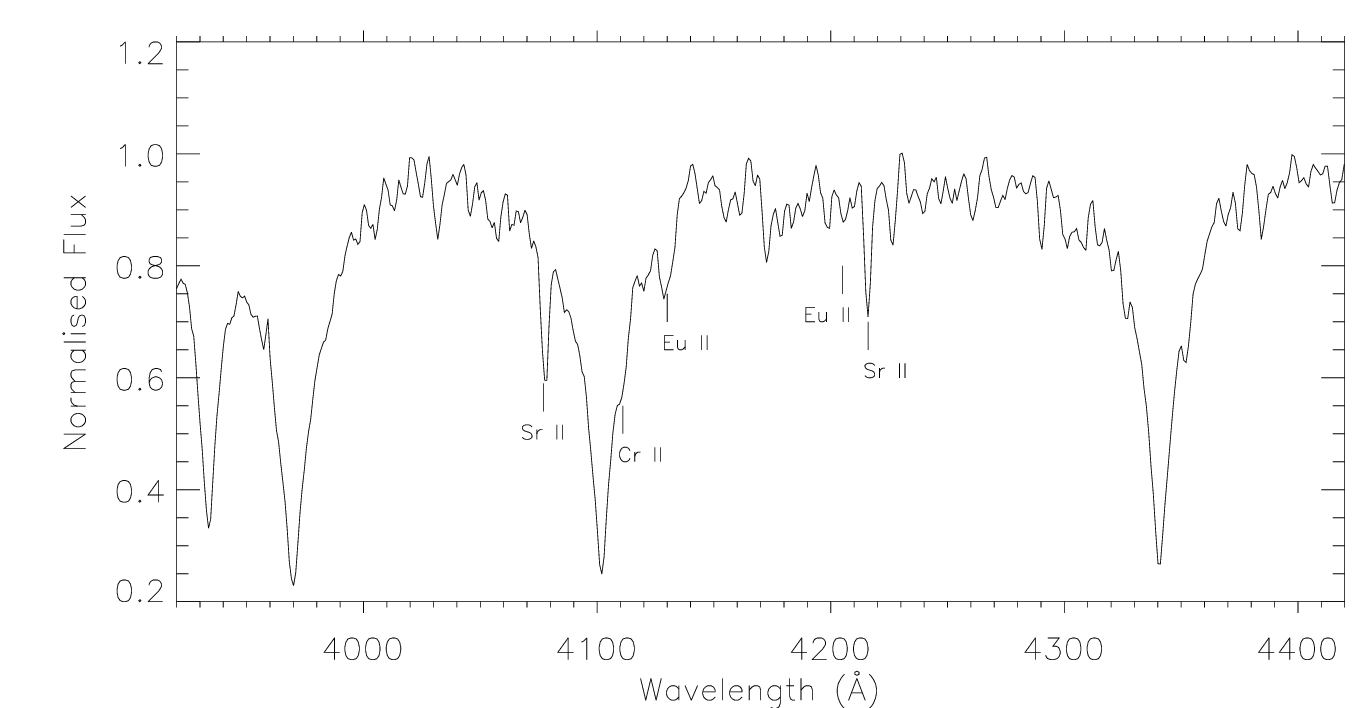}
\caption{LAMOST low resolution spectrum of TIC\,21024812.}
\label{fig:21024812_LAMOST}
\end{figure}

\begin{figure}
\centering
\includegraphics[width=\columnwidth]{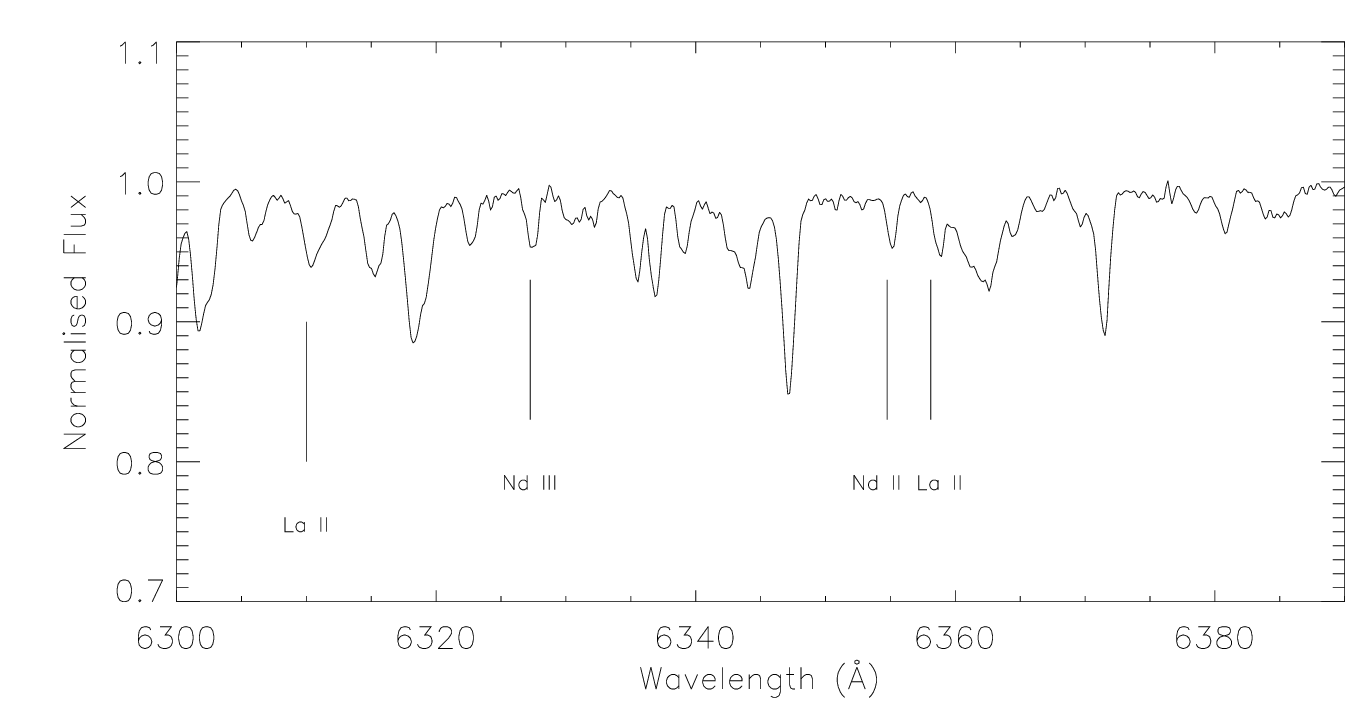}
\caption{Segment of the LAMOST medium resolution spectrum of TIC\,259017938.}
\label{fig:259017938_LAMOST}
\end{figure}

\begin{figure}
\centering
\includegraphics[width=\columnwidth]{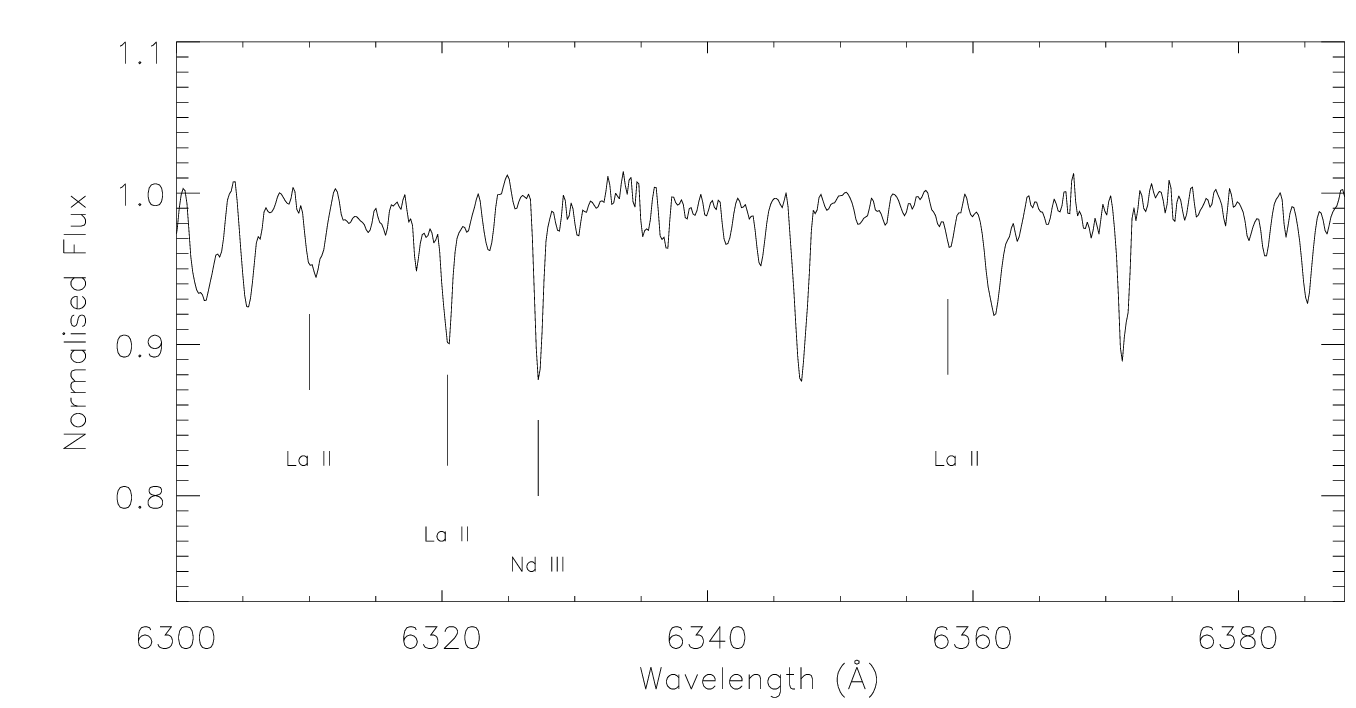}
\caption{Segment of the LAMOST medium resolution spectrum of TIC\,26833276.}
\label{fig:26833276_LAMOST}
\end{figure}

\begin{figure}
\centering
\includegraphics[width=\columnwidth]{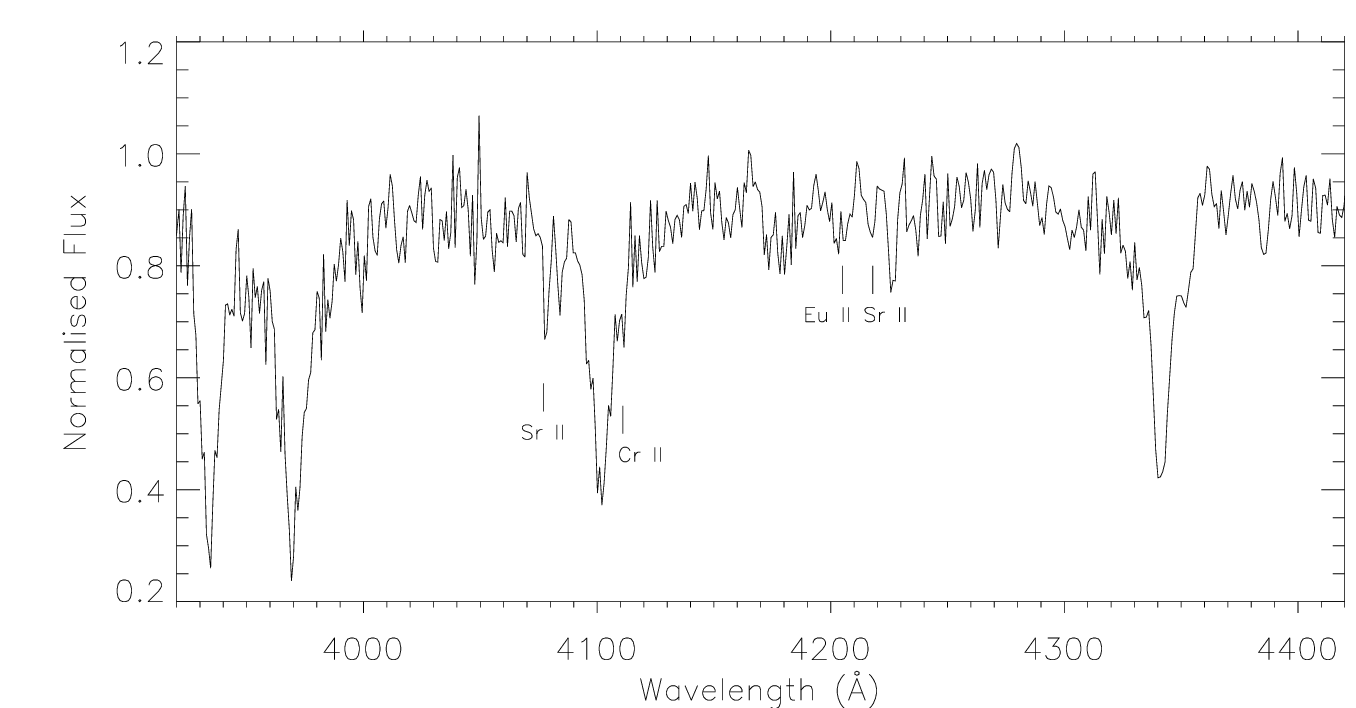}
\caption{LAMOST low resolution spectrum of TIC\,101624823 showing the clear presence of Cr\,{\sc{ii}}, with indications of Sr\,{\sc{ii}} and Eu\,{\sc{ii}} indicated. }
\label{fig:101624823_spec}
\end{figure}

\begin{figure}
\centering
\includegraphics[width=\columnwidth]{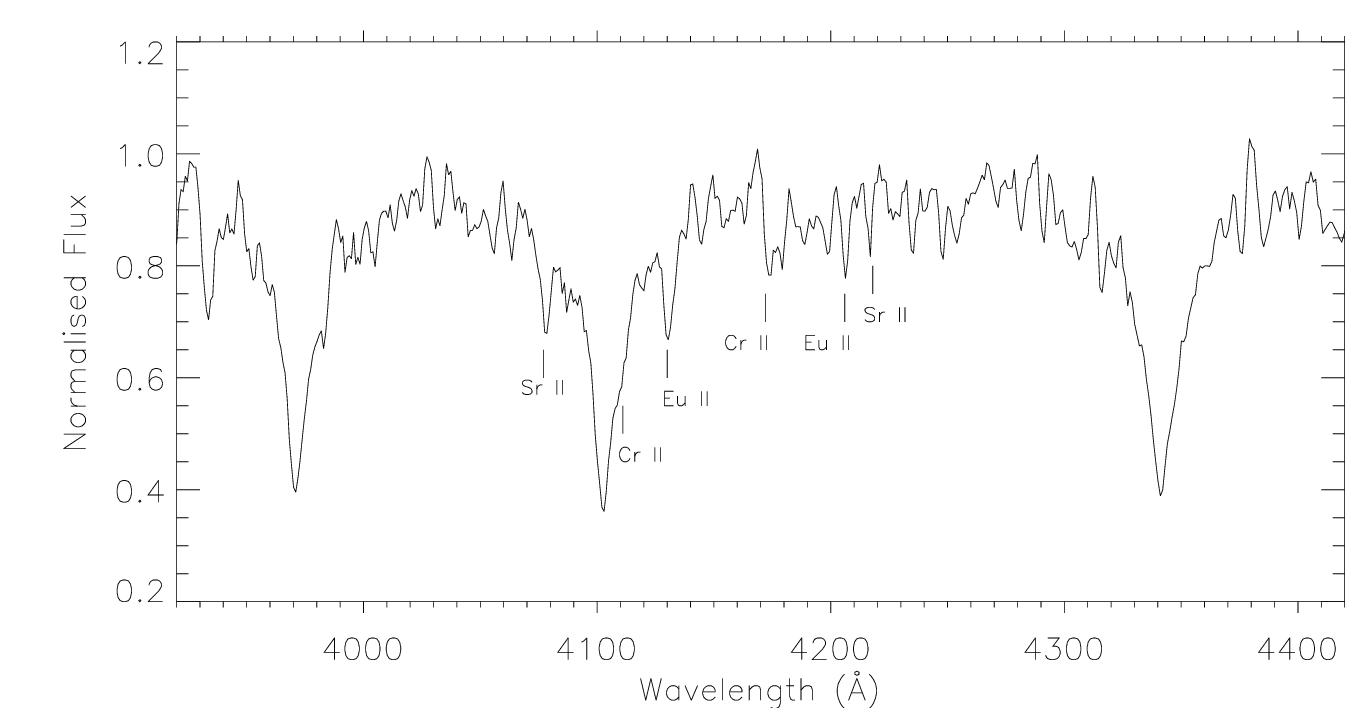}
\caption{LAMOST low resolution spectrum of TIC\,118247716 showing the clear presence of Sr\,{\sc{ii}}, Eu\,{\sc{ii}} and Cr\,{\sc{ii}}. }
\label{fig:118247716_spec}
\end{figure}

\begin{figure}
\centering
\includegraphics[width=\columnwidth]{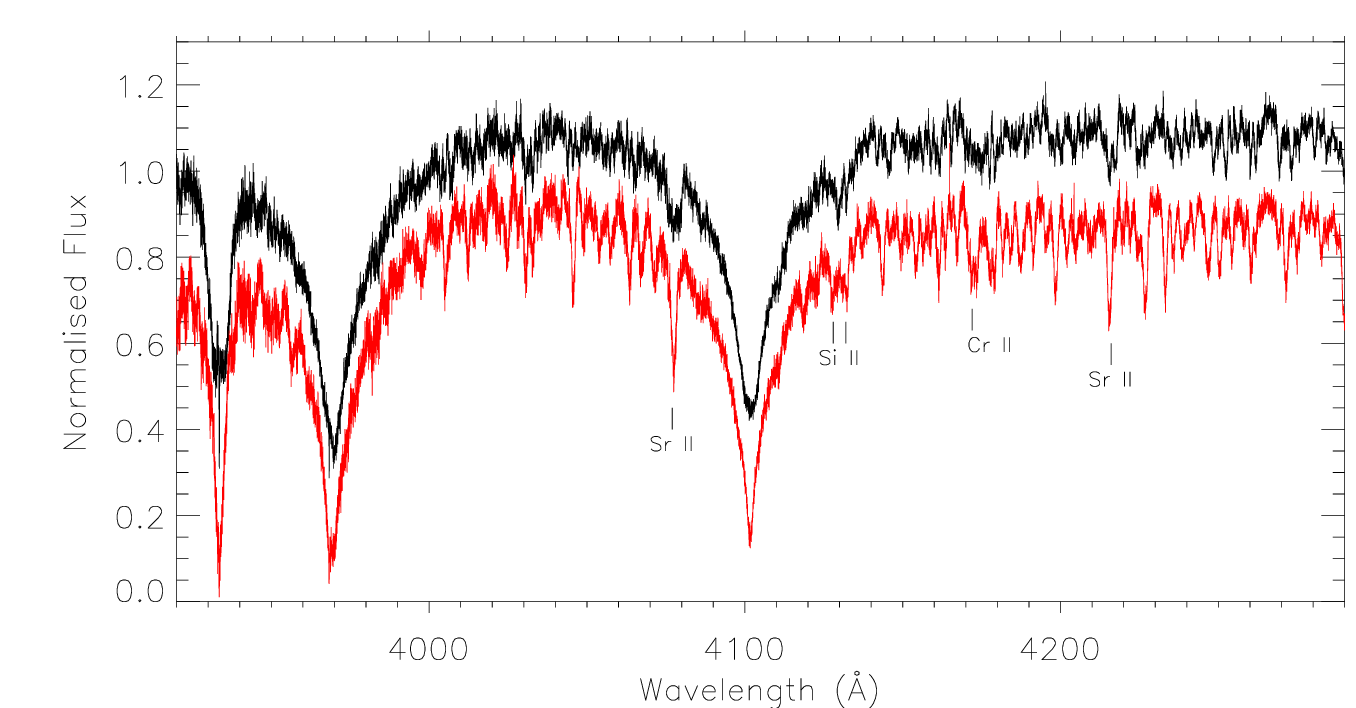}
\includegraphics[width=\columnwidth]{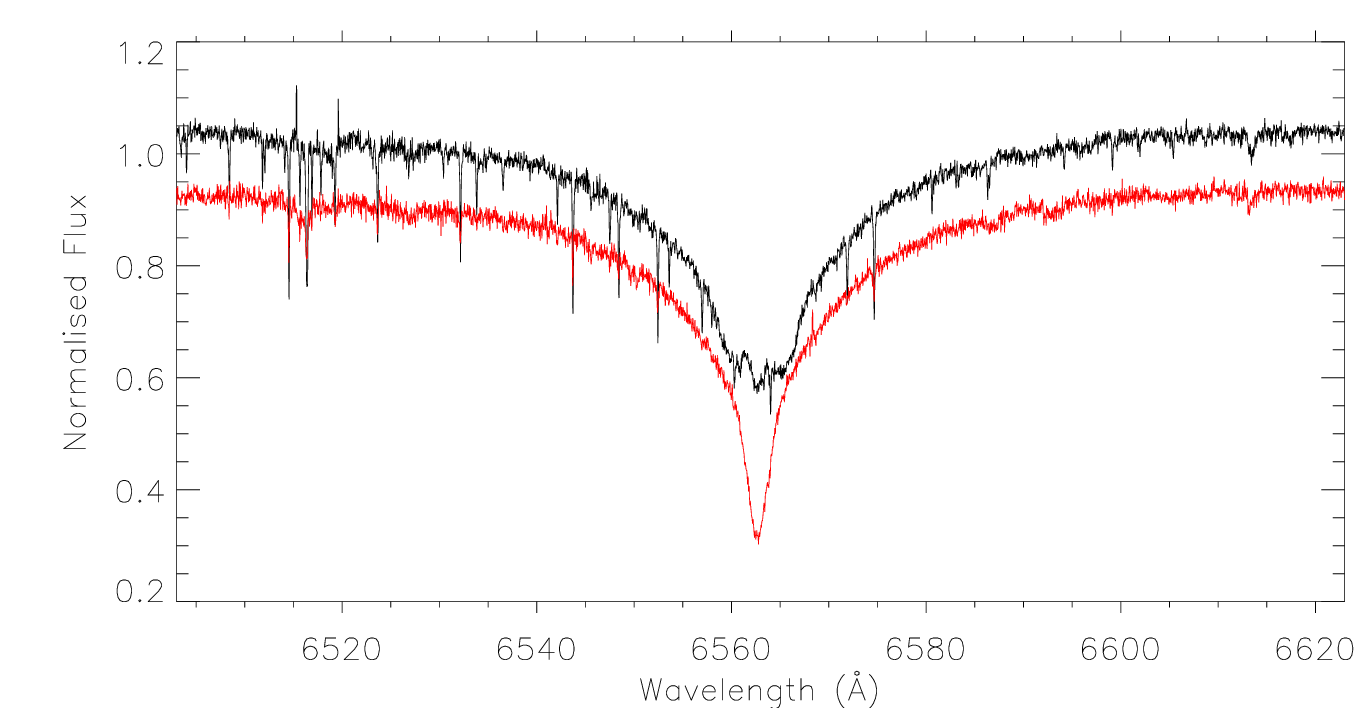}
\caption{HERMES spectra of TIC\,435263600 showing the binary nature of the system. The two epochs are plotted in black (obtained at BJD$=2458807.5357$) and red (obtained at BJD$=58812.5439$), offset in intensity for clarity. The top panel shows the region classically used for spectral classification (note the significant change in the Ca\,{\sc{ii}}\,K line profile), while the bottom panel shows the region around H$_\alpha$. It is easy to see how multiple different classifications can arise from observing the star at different epochs. The sharp absorption lines in the lower panel are telluric lines.}
\label{fig:435263600_spec}
\end{figure}

\begin{figure}
\centering
\includegraphics[width=\columnwidth]{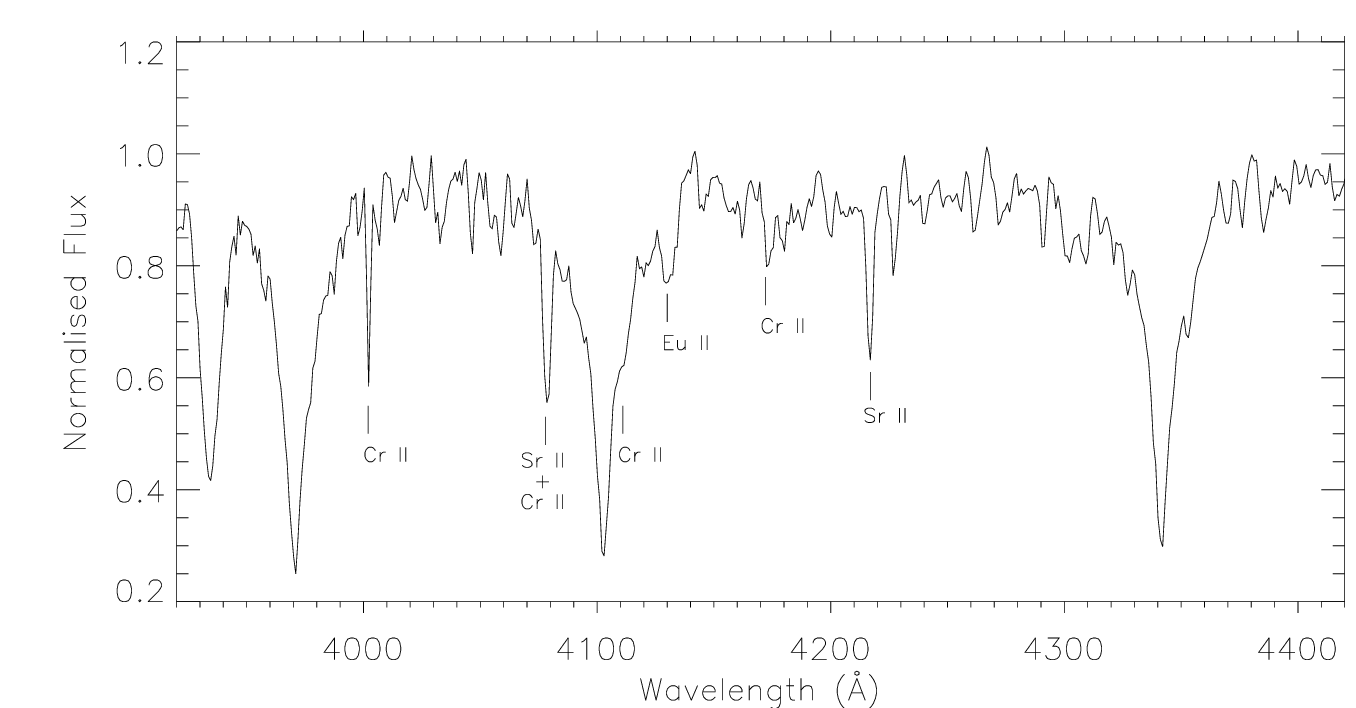}
\caption{LAMOST low resolution spectrum of TIC\,467074220 showing the clear presence of Sr\,{\sc{ii}} and Cr\,{\sc{ii}} with an indication of Eu\,{\sc{ii}}. }
\label{fig:467074220_spec}
\end{figure}

\section{Amplitude spectra of roAp stars known prior to the launch of {\it TESS}}
\label{app:known}
Here we provide plots of the amplitude spectra of {\it TESS} light curves of roAp stars known prior to the mission launch.

\begin{figure}
\centering
\includegraphics[width=\columnwidth]{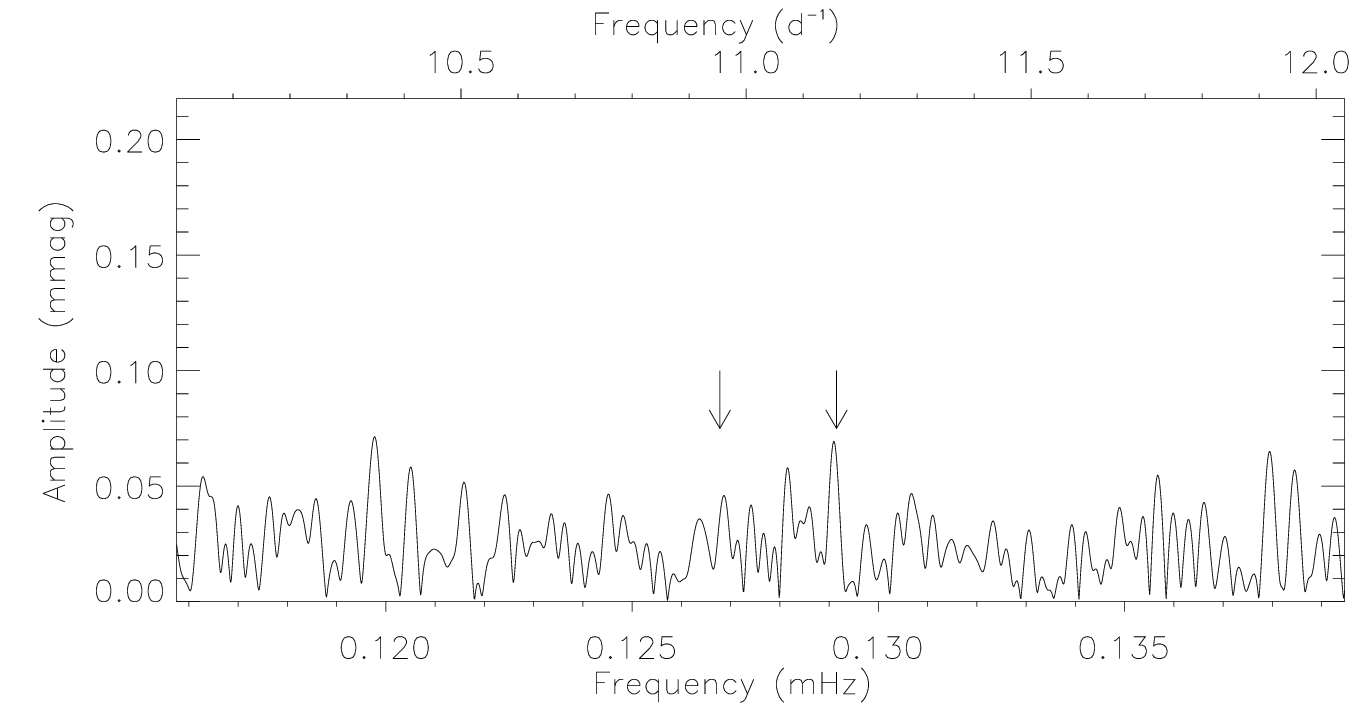}
\includegraphics[width=\columnwidth]{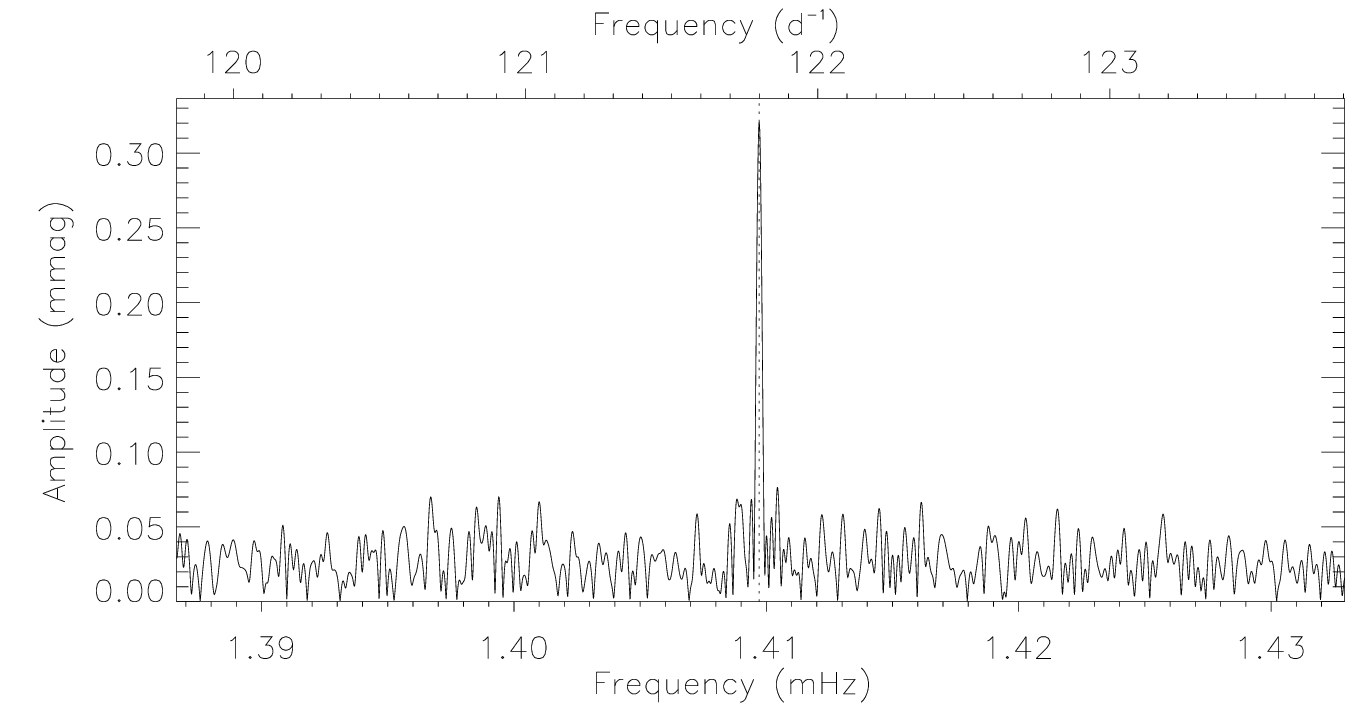}
\caption{Amplitude spectrum for TIC\,26418690 in the regions of previously reported variability. The top panel shows the low frequency range where $\delta$\,Sct modes have been reported (indicated by the arrows), but are below the {\it TESS} detection limit. The bottom panel shows the {\it TESS} detection of the roAp pulsation in this star.}
\label{fig:26418690}
\end{figure}

\begin{figure}
\centering
\includegraphics[width=\columnwidth]{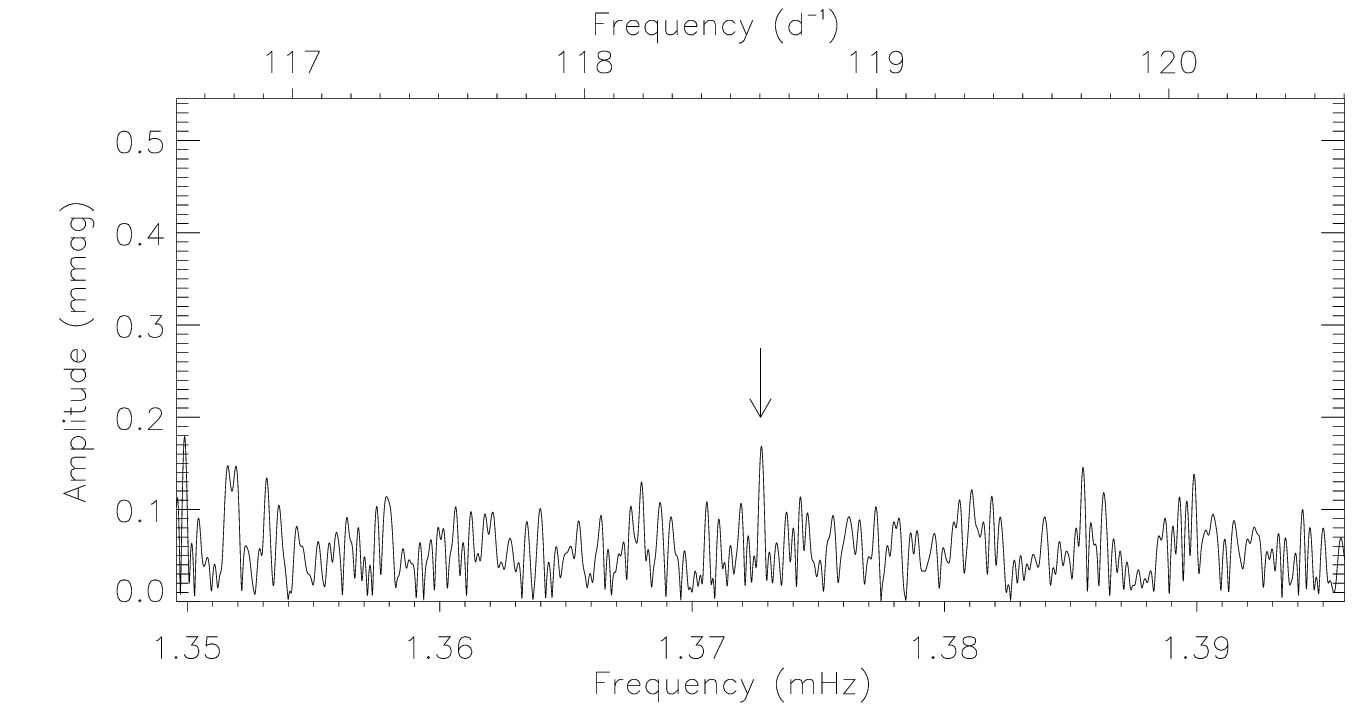}
\caption{Amplitude spectrum for TIC\,26749633 showing the tentative detection of the known pulsation mode in the {\it TESS} data.}
\label{fig:26749633}
\end{figure}

\begin{figure}
\centering
\includegraphics[width=\columnwidth]{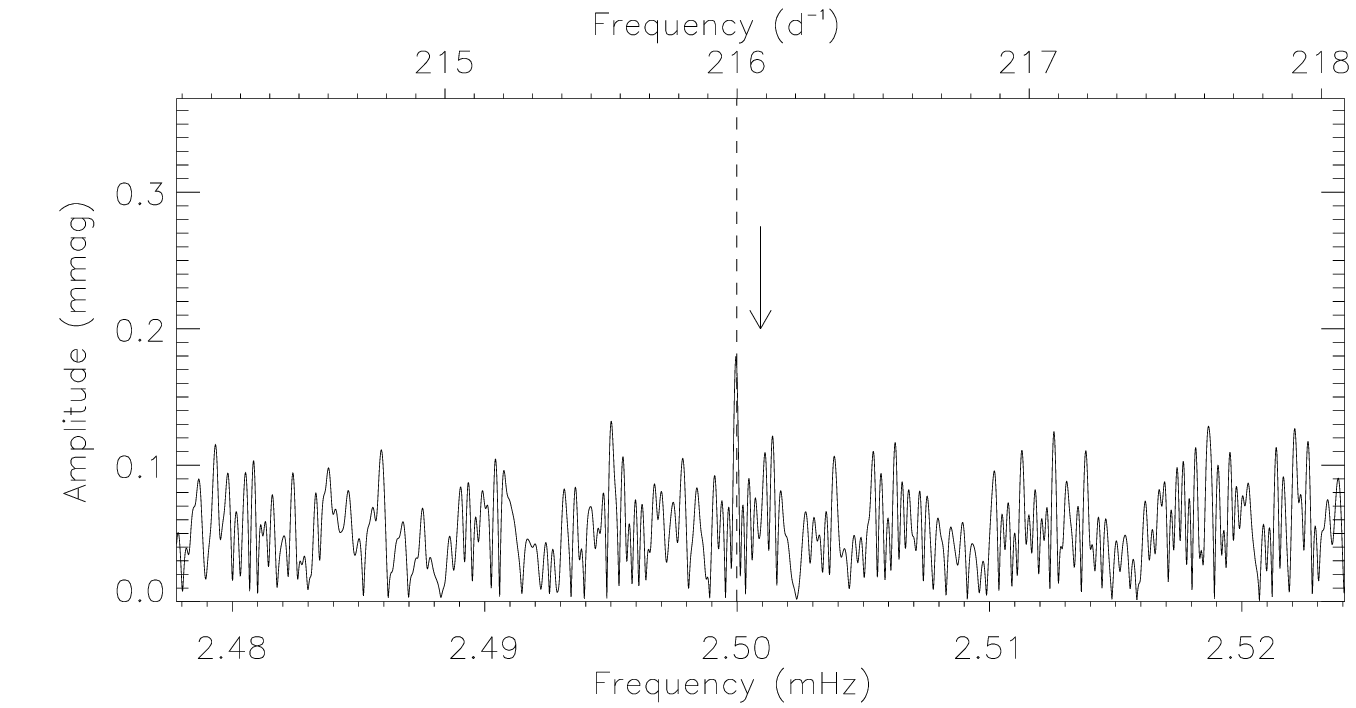}
\caption{Amplitude spectrum for TIC\,27395746 showing the detection of the $\nu-\nu_{\rm rot}$ sidelobe (indicated by the dashed vertical line) of the unseen pulsation mode (with a frequency indicated by the arrow).}
\label{fig:27395746}
\end{figure}

\begin{figure}
\centering
\includegraphics[width=\columnwidth]{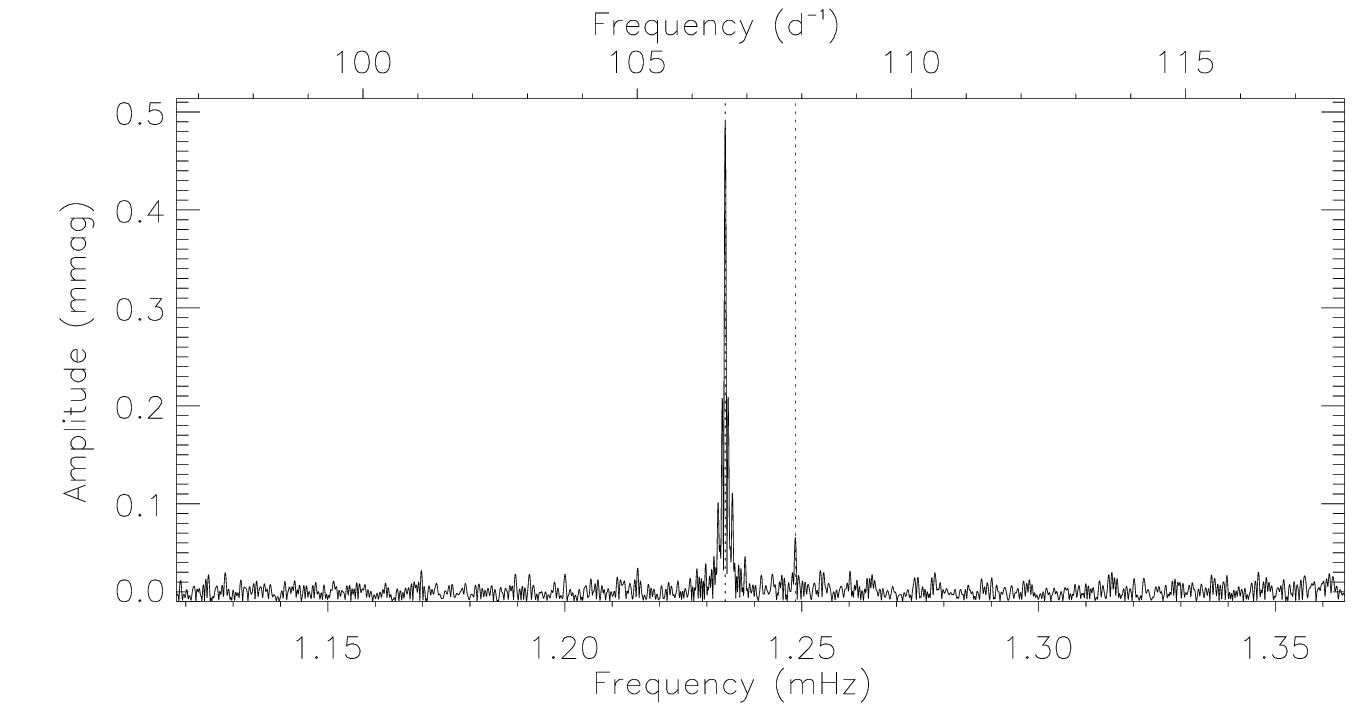}
\caption{Amplitude spectrum for TIC\,77128654. The vertical dotted lines identify the pulsation modes.}
\label{fig:77128654}
\end{figure}

\begin{figure}
\centering
\includegraphics[width=\columnwidth]{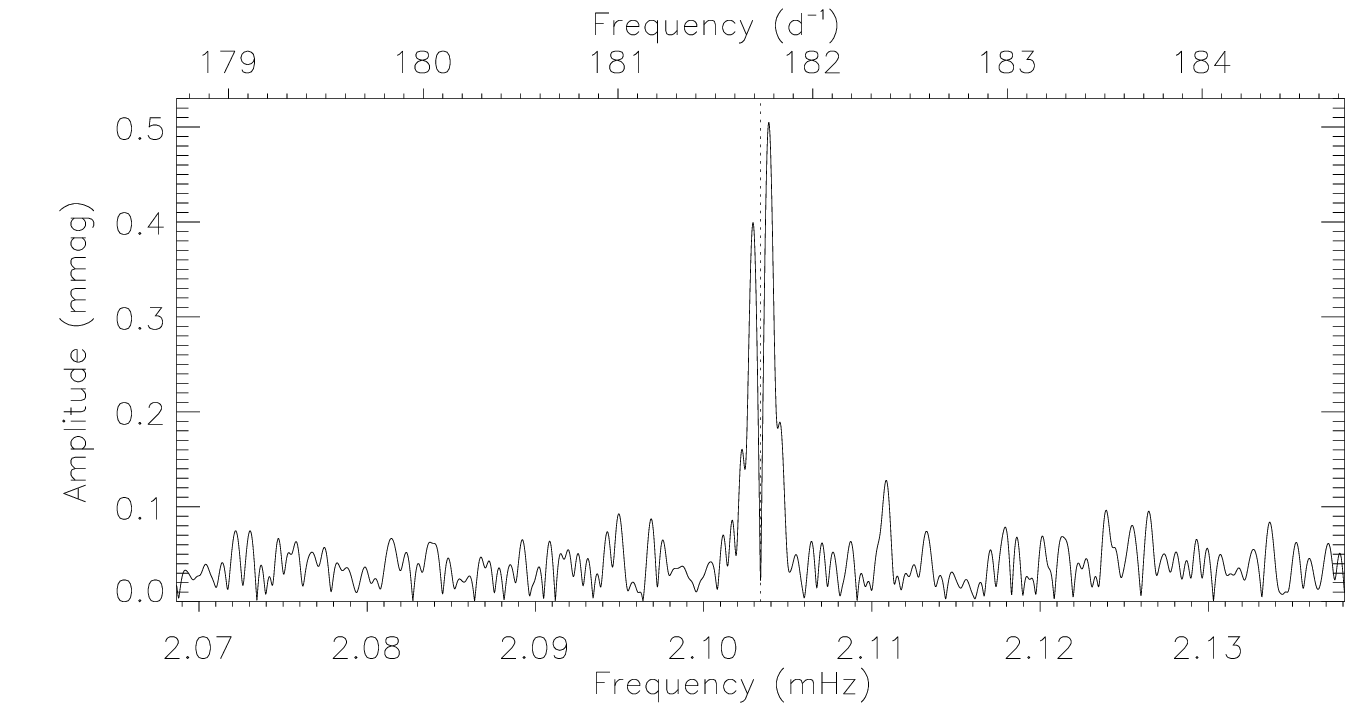}
\caption{Amplitude spectrum for TIC\,123231021. The vertical dotted lines identify the pulsation mode provided in the literature. The fact that the mode is split here demonstrates the significant frequency variability present in this star.}
\label{fig:123231021}
\end{figure}

\begin{figure}
\centering
\includegraphics[width=\columnwidth]{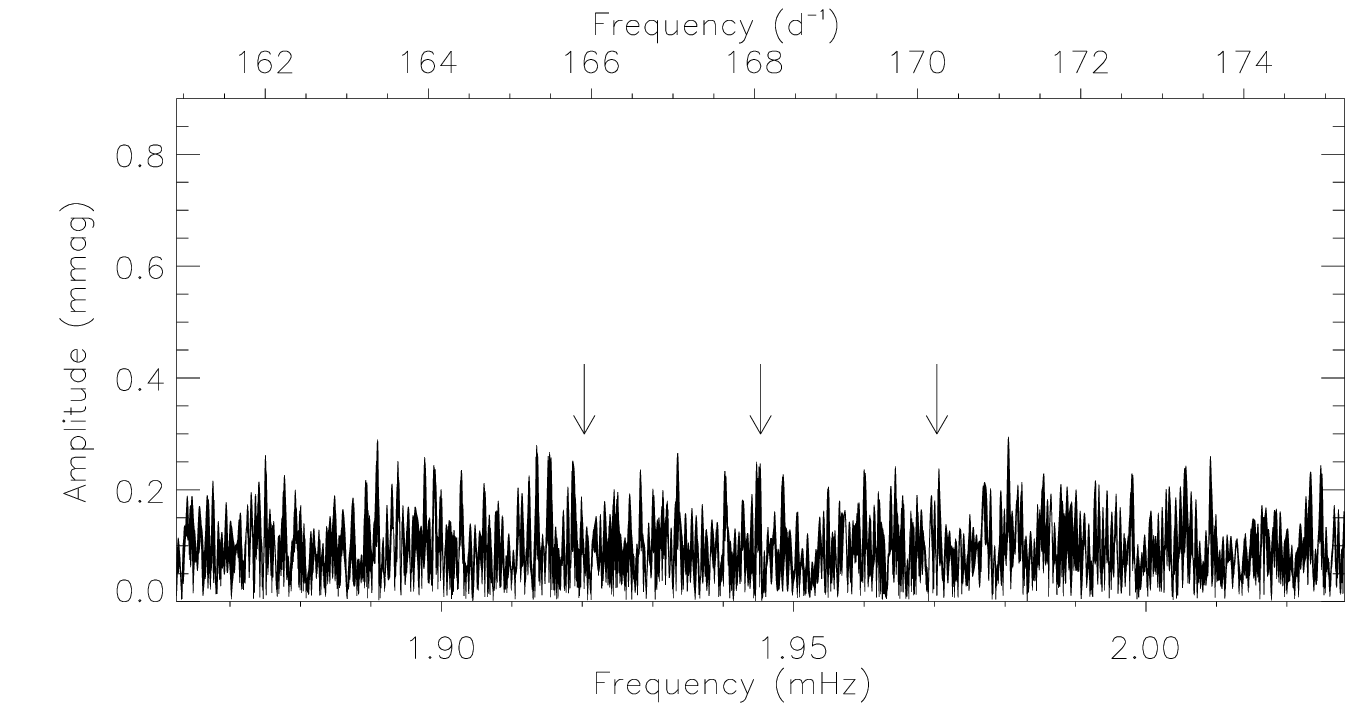}
\caption{Amplitude spectrum for TIC\,158216369. The arrows identify the locations of the pulsation modes listed in the literature. Their amplitudes are around or below the noise level in the {\it TESS} data.}
\label{fig:158216369}
\end{figure}

\begin{figure}
\centering
\includegraphics[width=\columnwidth]{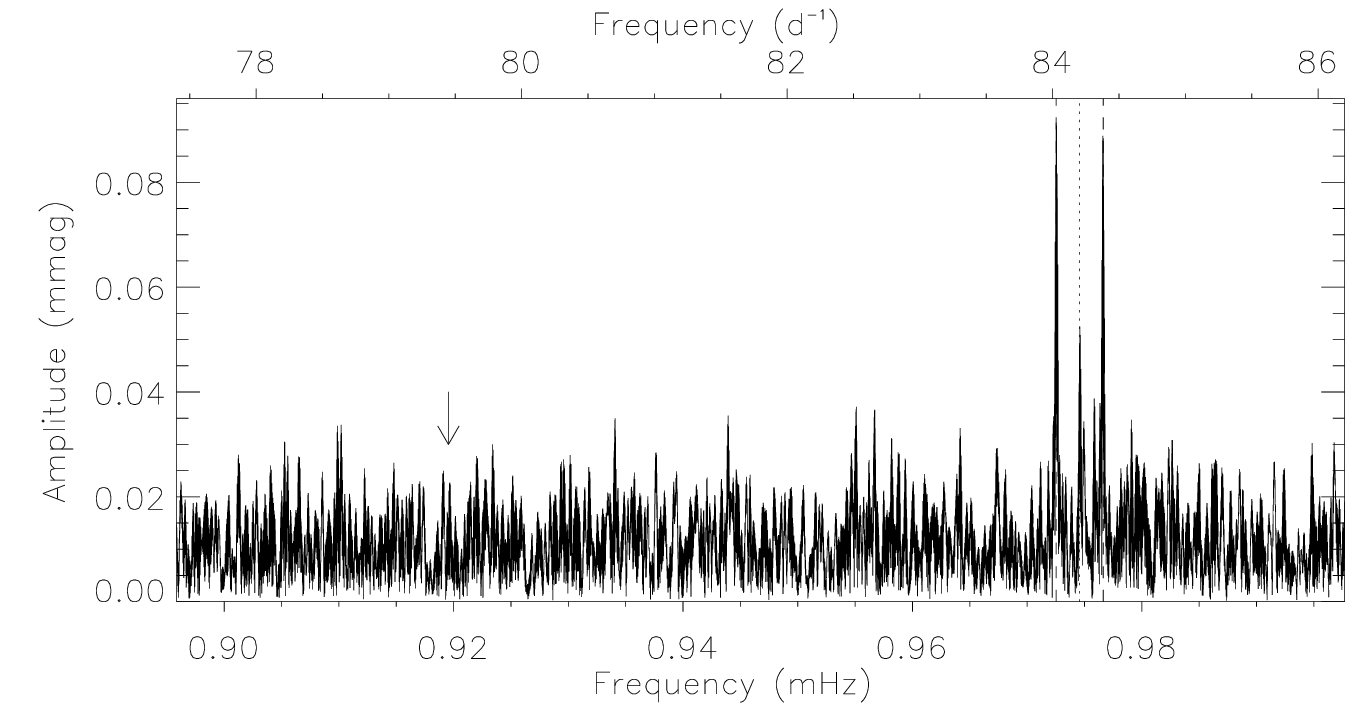}
\caption{Amplitude spectrum for TIC\,158271090. The vertical dotted line identifies the pulsation mode, with the dashed lines identifying the rotationally split sidelobes. The arrow identifies the location of a second mode in this star which is below the noise level in the {\it TESS} data.}
\label{fig:158271090}
\end{figure}

\begin{figure}
\centering
\includegraphics[width=\columnwidth]{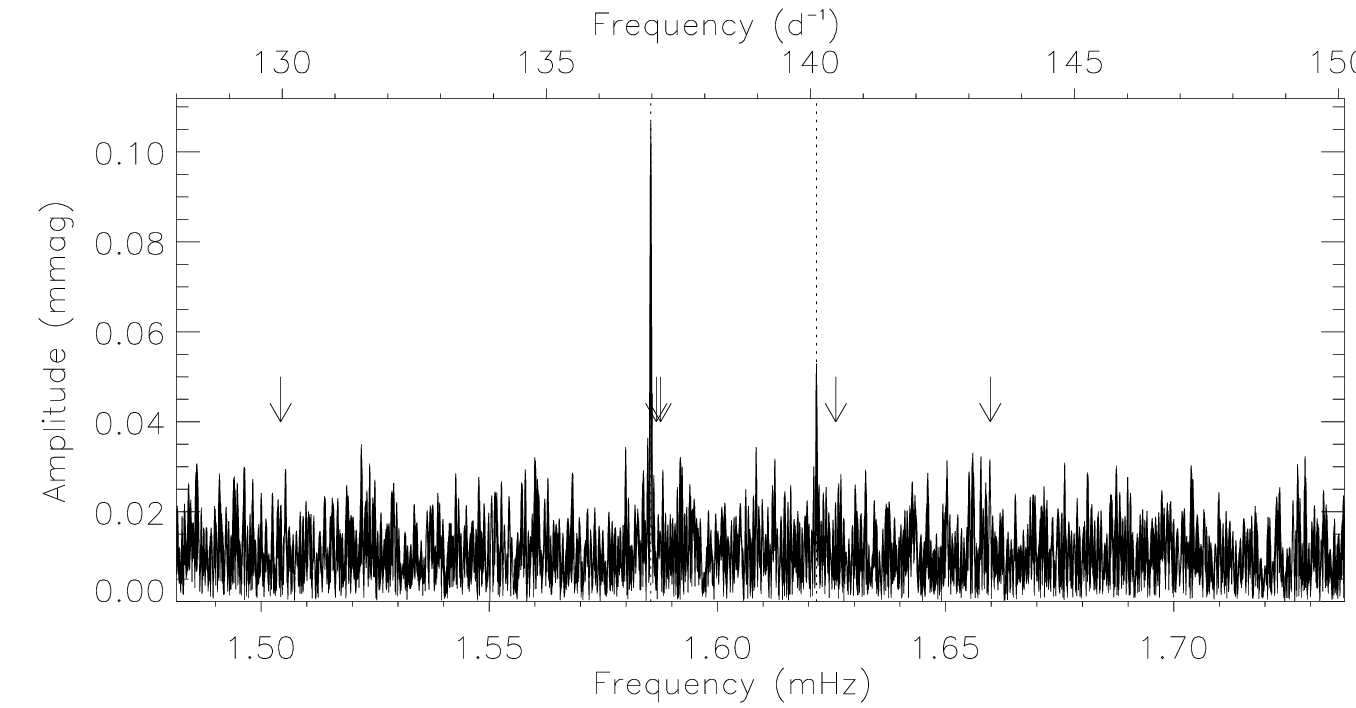}
\caption{Amplitude spectrum for TIC\,158275114. The vertical dotted lines identify the pulsation modes detected in the {\it TESS} data, while the arrows indicate previously known pulsation modes.}
\label{fig:158275114}
\end{figure}

\begin{figure}
\centering
\includegraphics[width=\columnwidth]{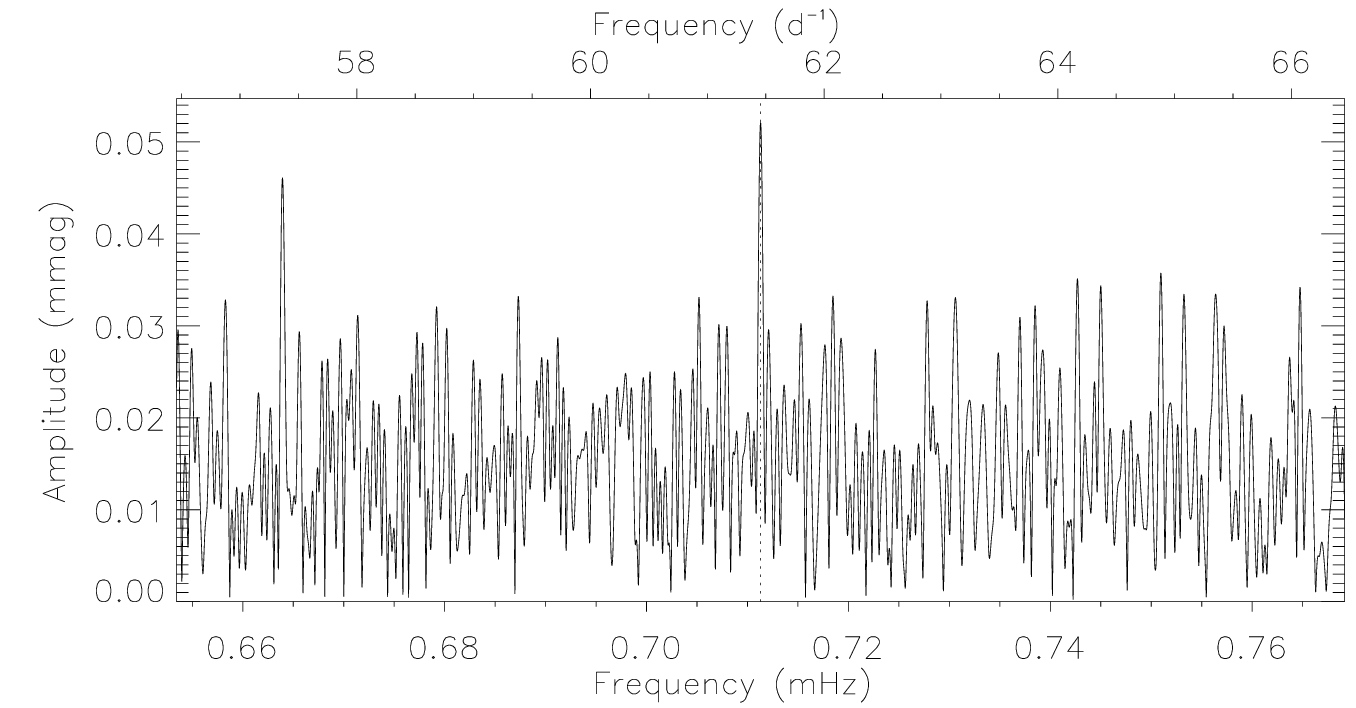}
\caption{Amplitude spectrum for TIC\,169078762. The vertical dotted line identifies the pulsation mode detected in the {\it TESS} data.}
\label{fig:169078762}
\end{figure}

\begin{figure}
\centering
\includegraphics[width=\columnwidth]{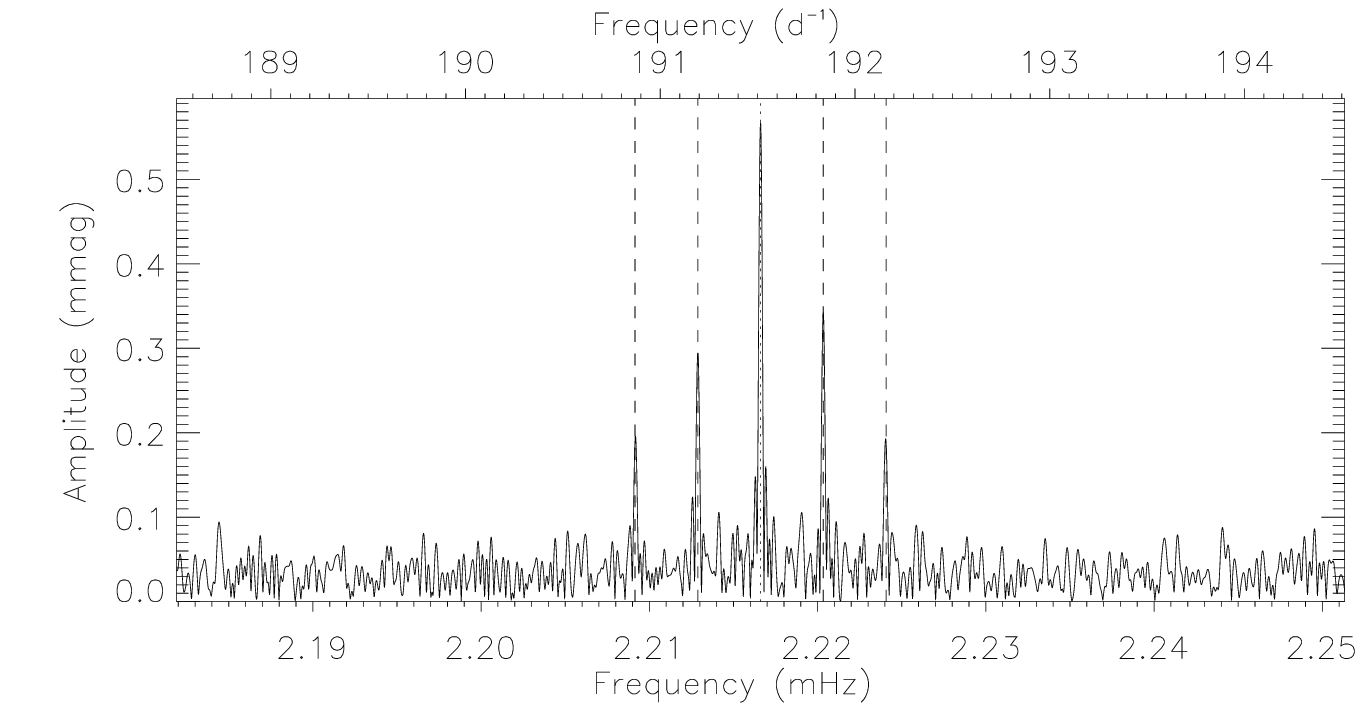}
\caption{Amplitude spectrum for TIC\,264509538. The vertical dotted line identifies the pulsation mode, with the dashed lines indicating the rotationally split sidelobes.}
\label{fig:264509538}
\end{figure}

\begin{figure}
\centering
\includegraphics[width=\columnwidth]{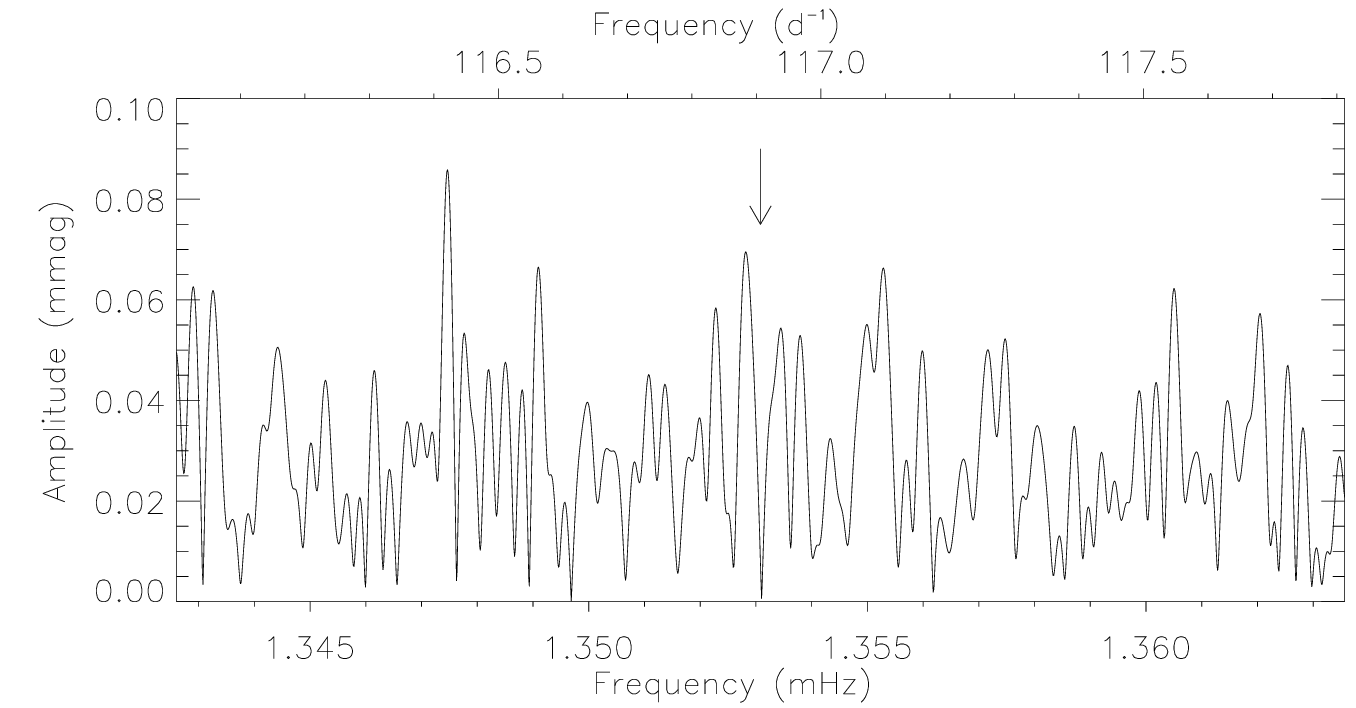}
\caption{Amplitude spectrum for TIC\,272598185. The arrow indicates the previously known pulsation mode, which is not detected in the {\it TESS} data.}
\label{fig:272598185}
\end{figure}

\begin{figure}
\centering
\includegraphics[width=\columnwidth]{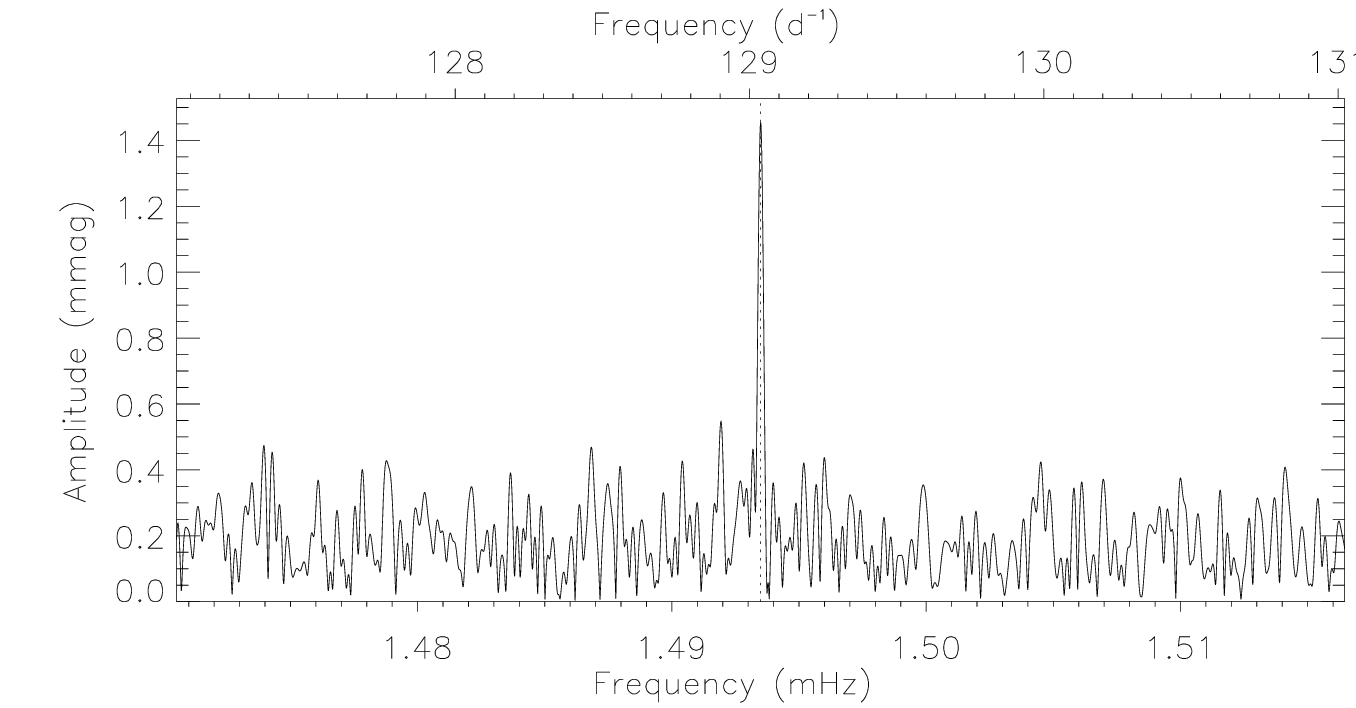}
\caption{Amplitude spectrum for TIC\,273777265. The vertical dotted line identifies the pulsation mode detected in the {\it TESS} data.}
\label{fig:273777265}
\end{figure}

\begin{figure}
\centering
\includegraphics[width=\columnwidth]{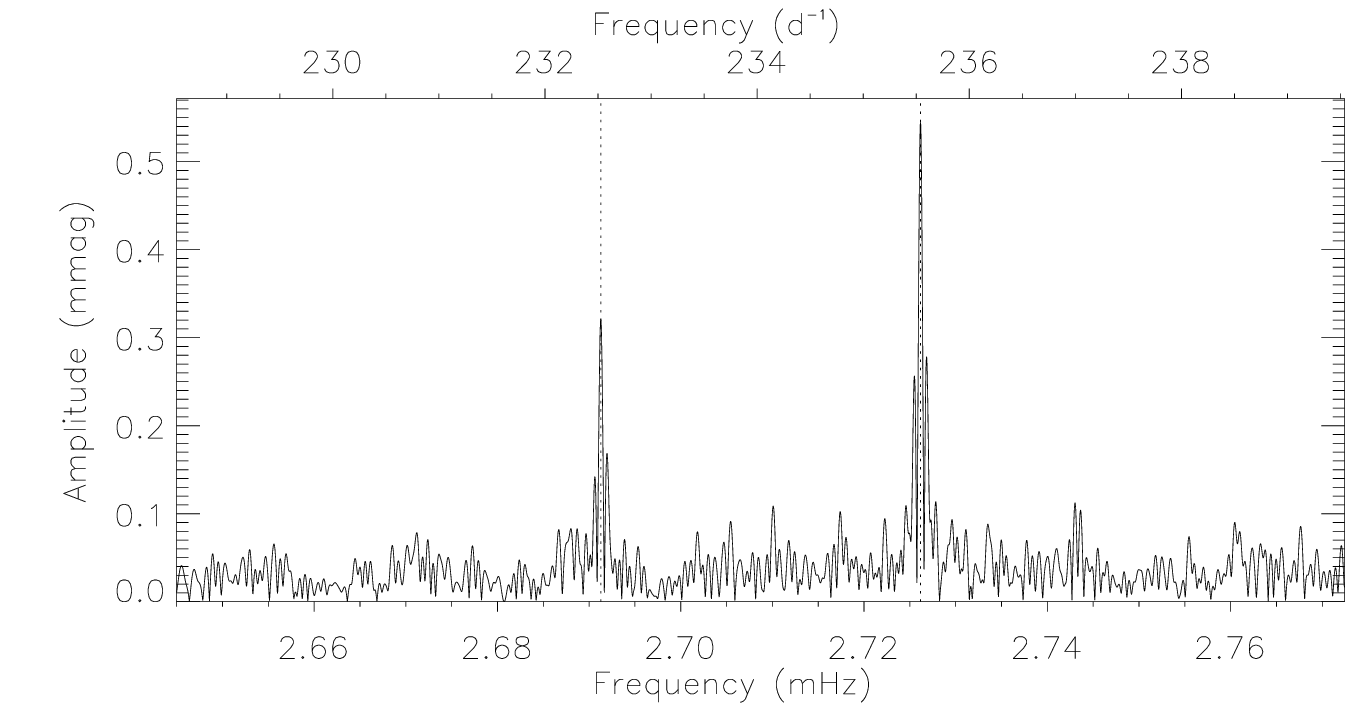}
\caption{Amplitude spectrum for TIC\,286992225. The vertical dotted lines identify the pulsation modes detected in the {\it TESS} data; the lower frequency mode is a new detection.}
\label{fig:286992225}
\end{figure}

\begin{figure}
\centering
\includegraphics[width=\columnwidth]{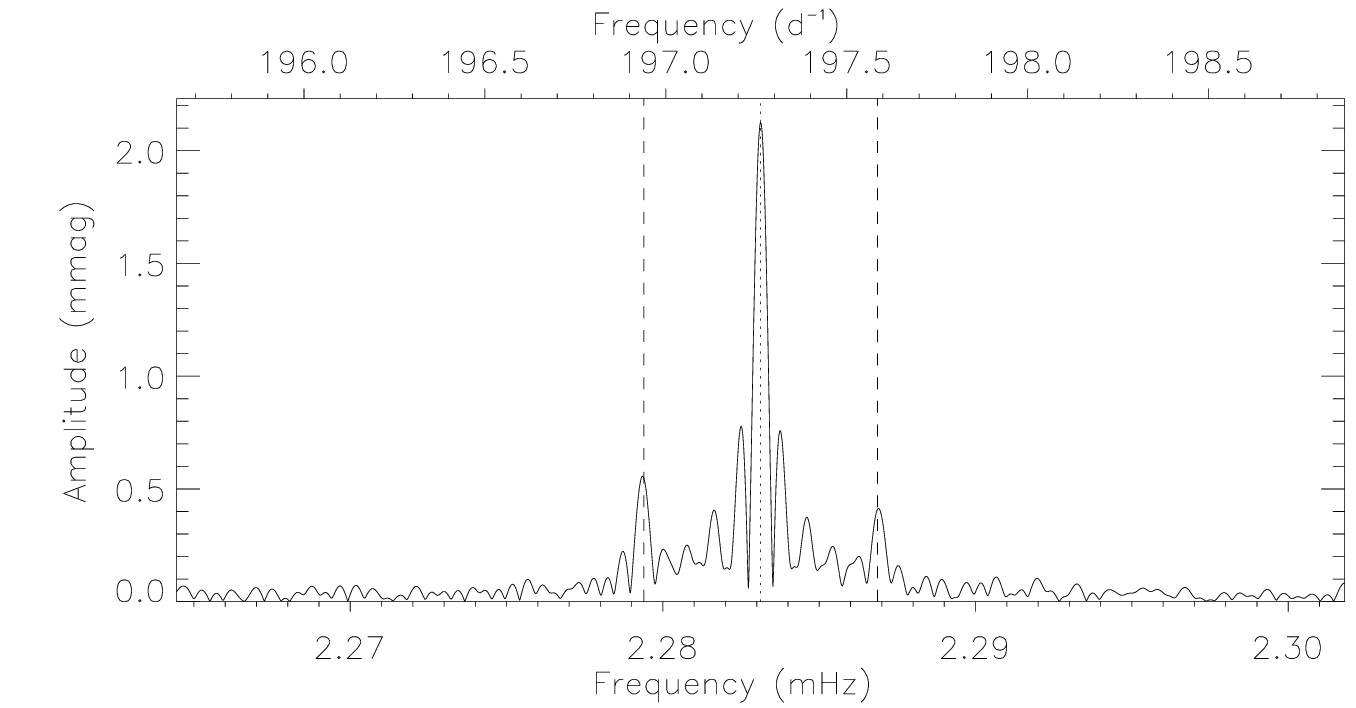}
\caption{Amplitude spectrum for TIC\,302602874. The vertical dotted line identifies the pulsation mode, while the dashed lines identify the rotationally split sidelobes.}
\label{fig:302602874}
\end{figure}

\section{Amplitude spectra of roAp stars observed beyond Cycle\,2}
\label{sec:app_remaining}
Here we provide plots of the amplitude spectra of {\it TESS} light curves of roAp observed beyond Cycle\,2.

\begin{figure}
\centering
\includegraphics[width=\columnwidth]{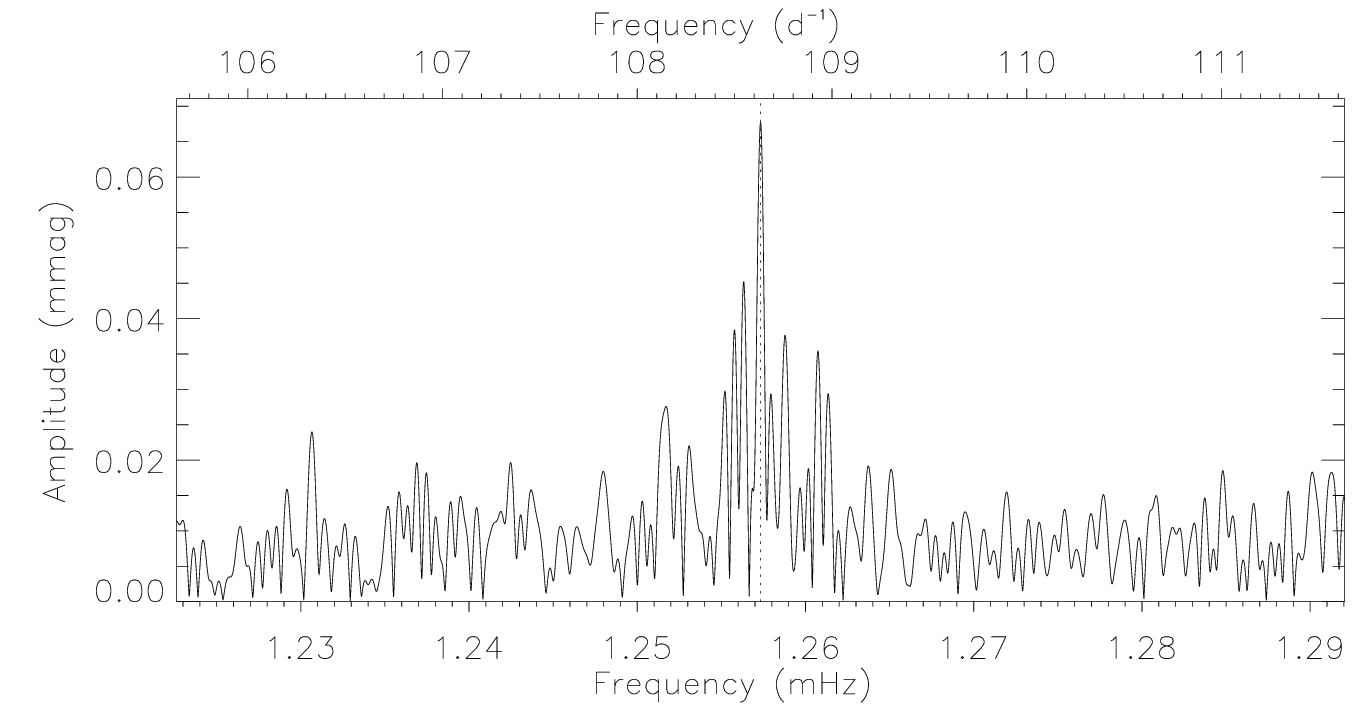}
\caption{Amplitude spectrum for TIC\,33604636. The vertical dotted line identifies the pulsation mode. After the removal of this peak, there is residual power remaining in this frequency range.}
\label{fig:33604636}
\end{figure}

\begin{figure}
\centering
\includegraphics[width=\columnwidth]{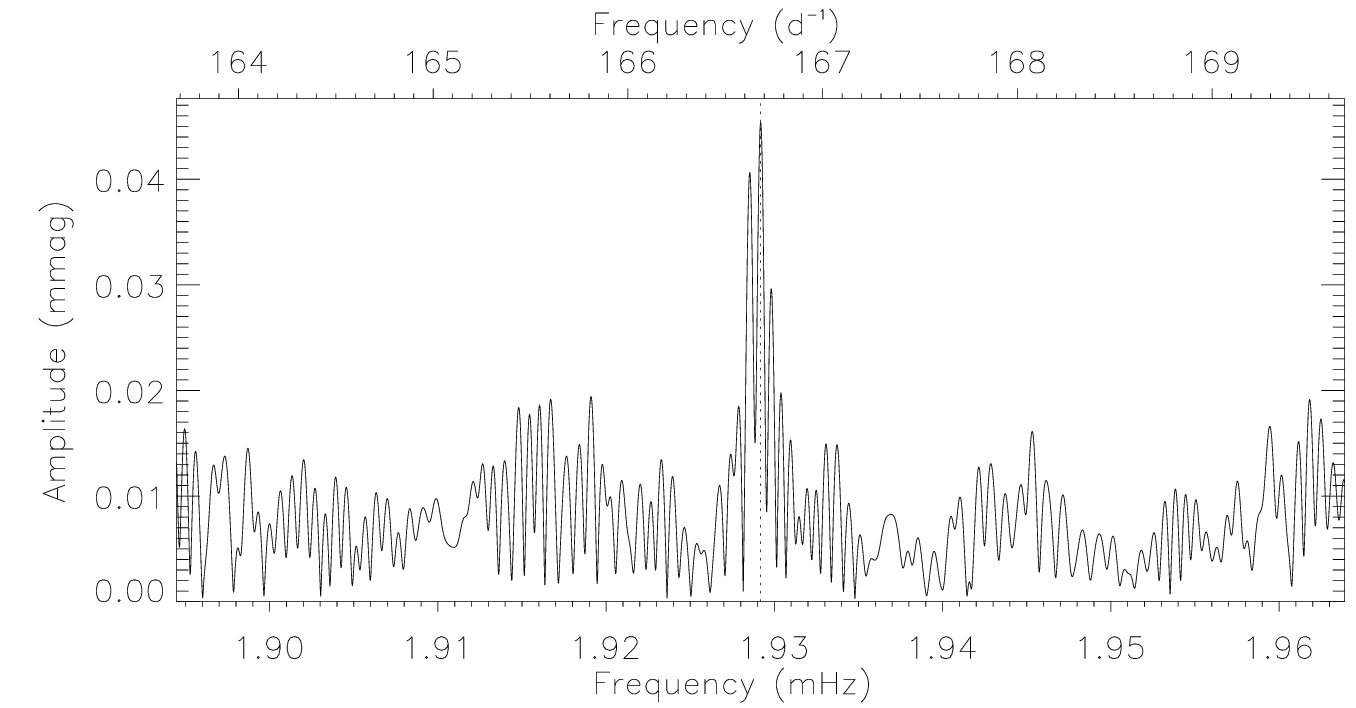}
\caption{Amplitude spectrum for TIC\,36576010. The vertical dotted line identifies the pulsation mode.}
\label{fig:36576010}
\end{figure}

\begin{figure}
\centering
\includegraphics[width=\columnwidth]{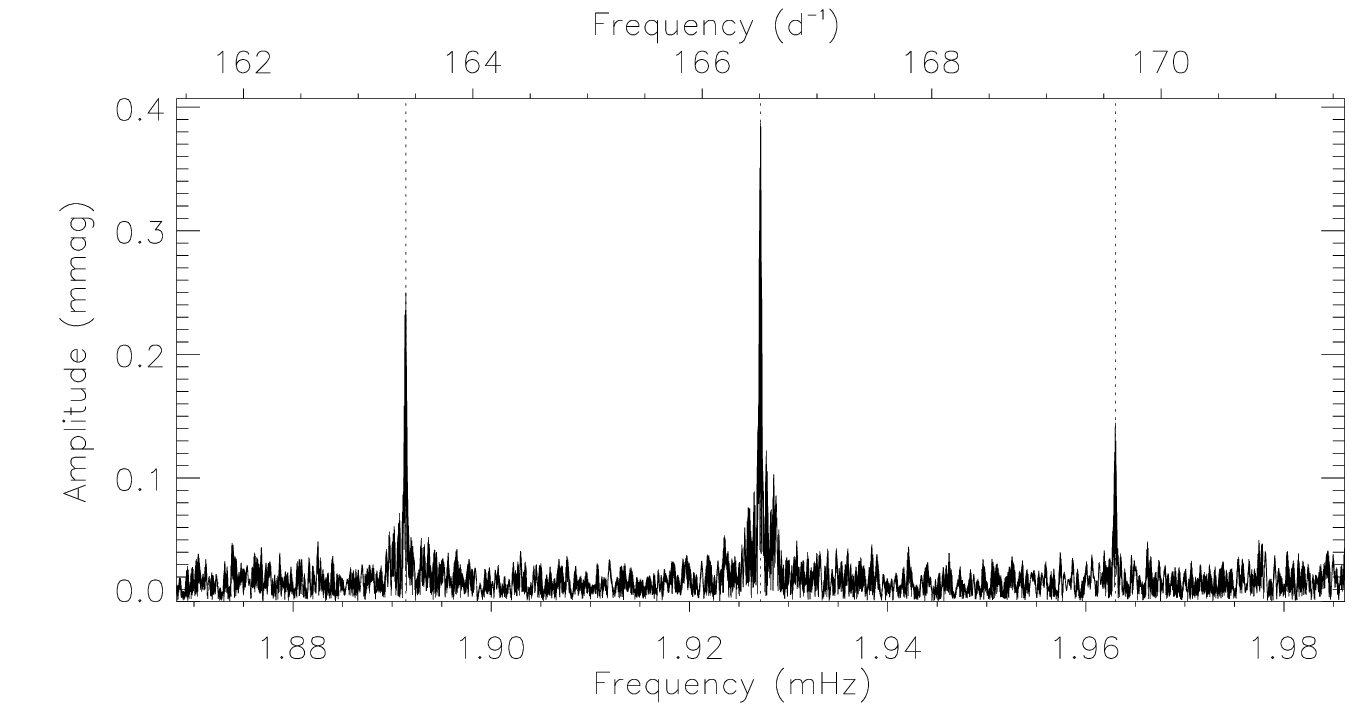}
\caption{Amplitude spectrum for TIC\,198781841. The vertical dotted lines identify the pulsation modes.}
\label{fig:198781841}
\end{figure}

\begin{figure}
\centering
\includegraphics[width=\columnwidth]{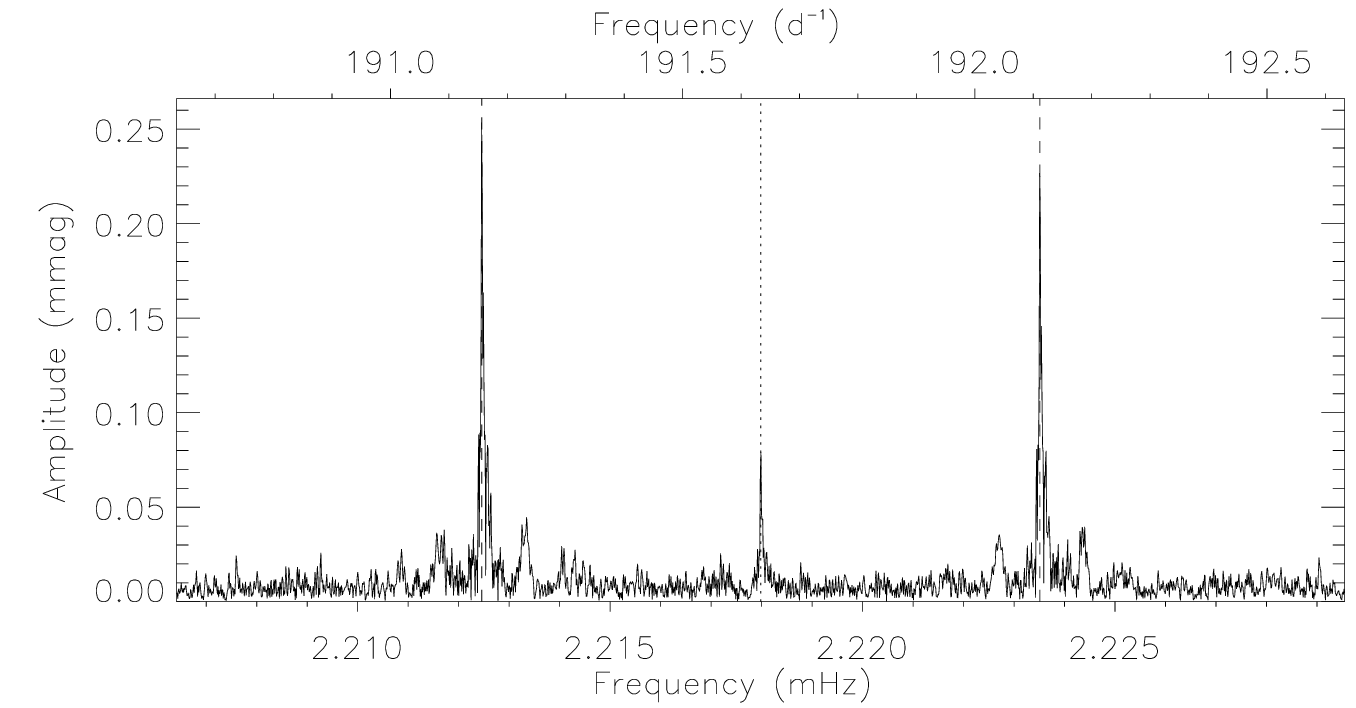}
\caption{Amplitude spectrum for TIC\,229960986. The vertical dotted line identifies the pulsation mode, with the dashed lines indicating the rotational sidelobes.}
\label{fig:229960986}
\end{figure}

\begin{figure}
\centering
\includegraphics[width=\columnwidth]{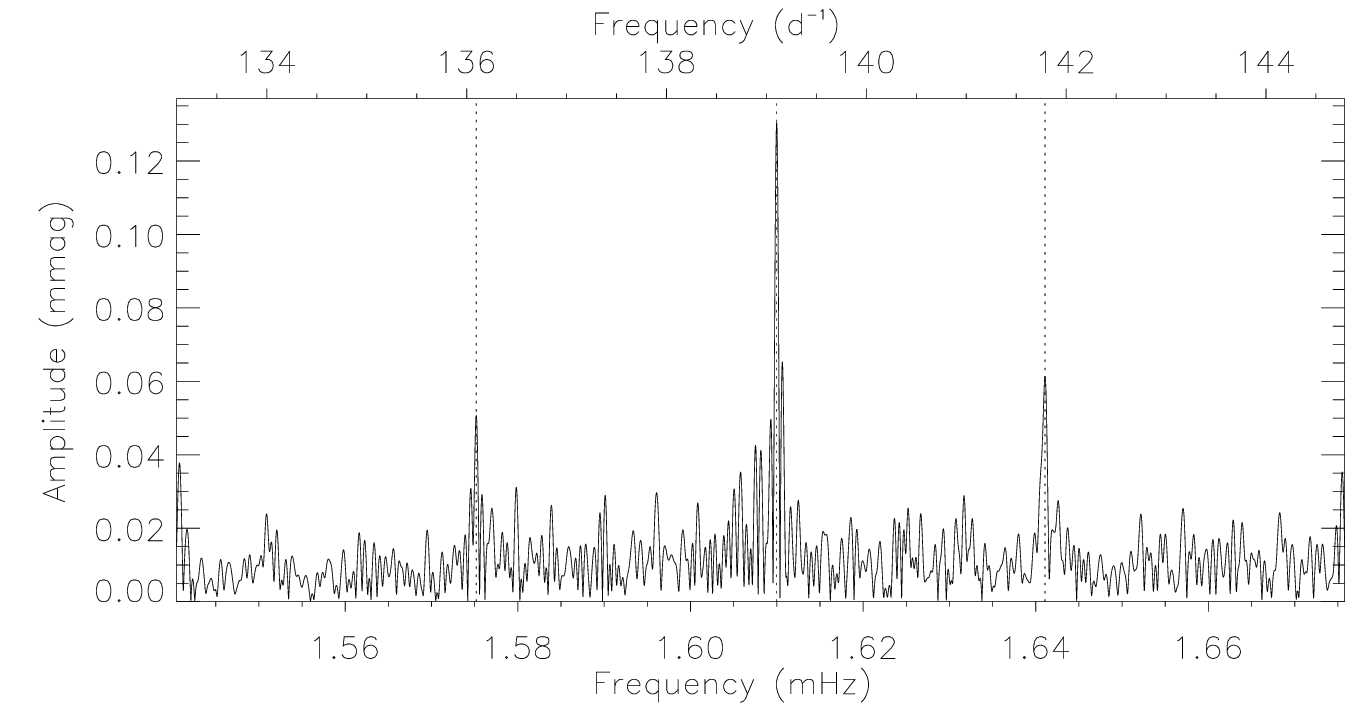}
\caption{Amplitude spectrum for TIC\,453826702. The vertical dotted lines identify the pulsation modes.}
\label{fig:453826702}
\end{figure}

\begin{figure}
\centering
\includegraphics[width=\columnwidth]{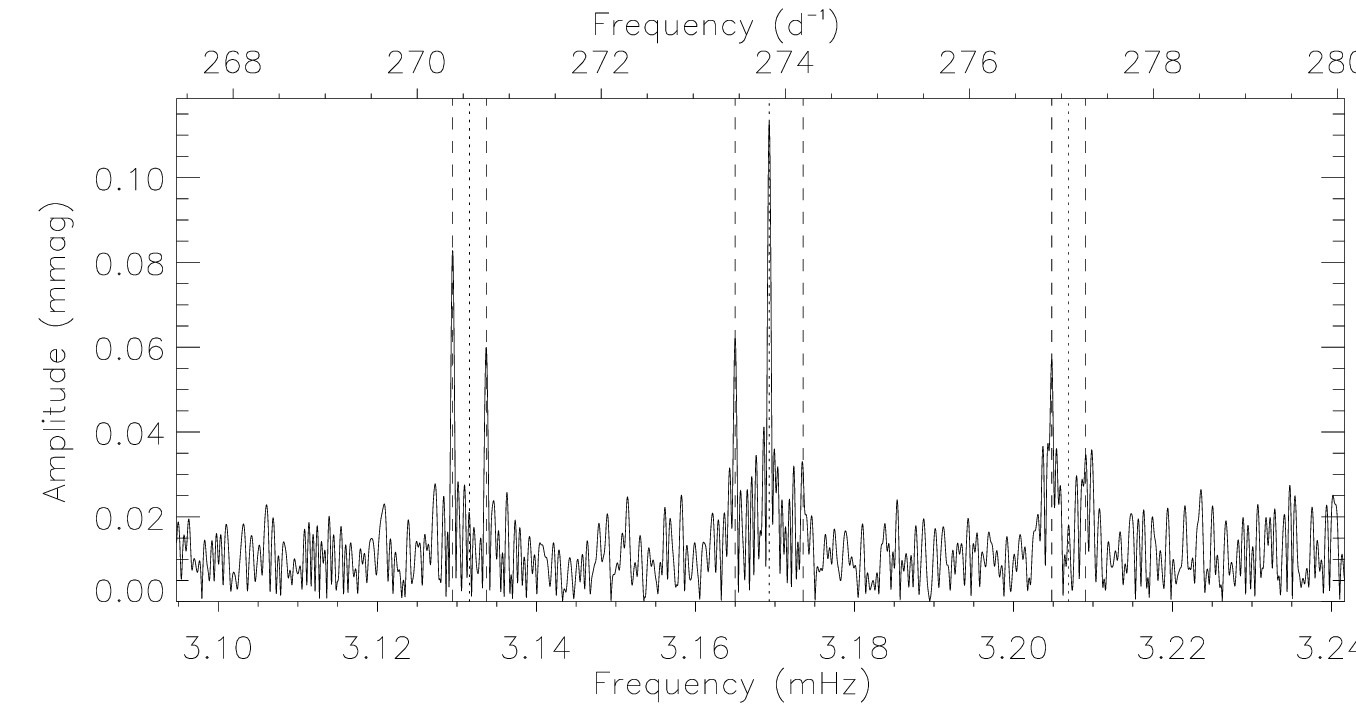}
\caption{Amplitude spectrum for TIC\,456673854. The vertical dotted line identifies the pulsation mode. We conclude this star to have, in increasing frequency, a dipole mode, a quadrupole mode, and a dipole mode. We have inferred the presence of the dipole modes the sidelobes.}
\label{fig:456673854}
\end{figure}

\begin{figure}
\centering
\includegraphics[width=\columnwidth]{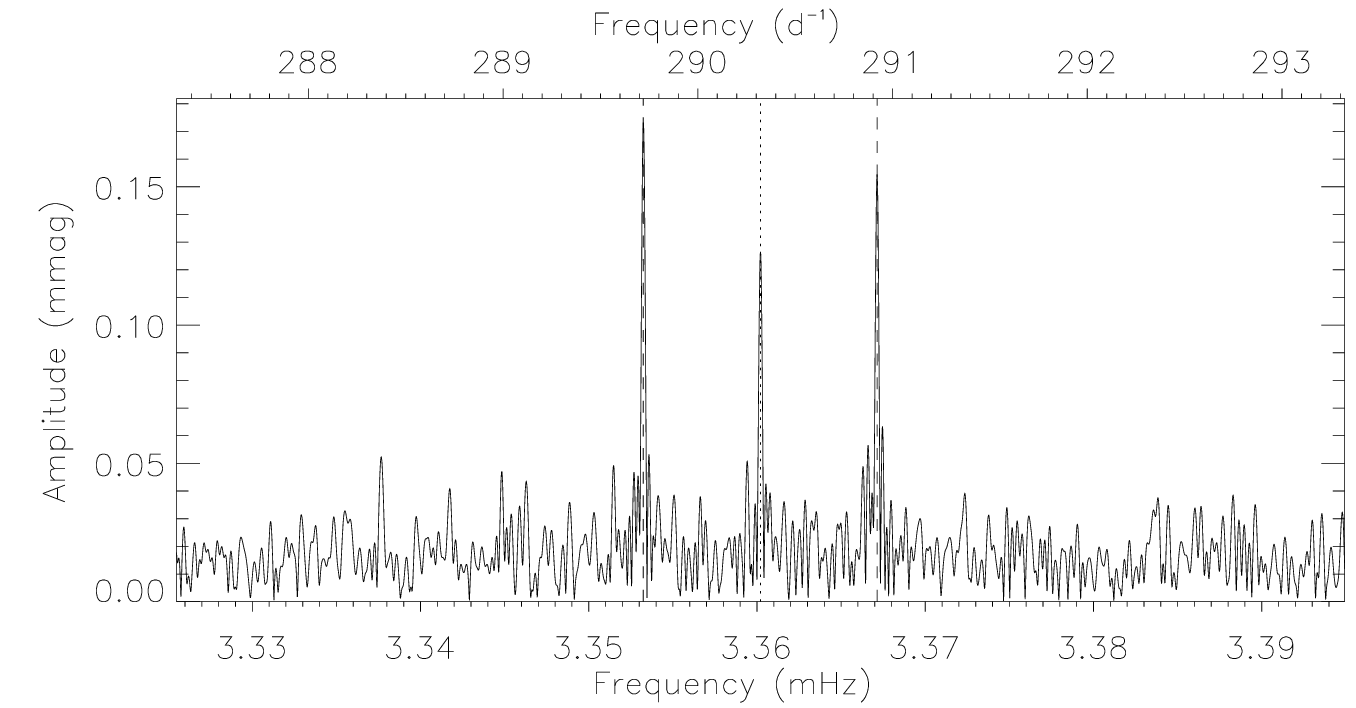}
\caption{Amplitude spectrum for TIC\,630844439. The vertical dotted line identifies the pulsation mode, with the dashed lines showing the rotational sidelobes.}
\label{fig:630844439}
\end{figure}

\begin{figure}
\centering
\includegraphics[width=\columnwidth]{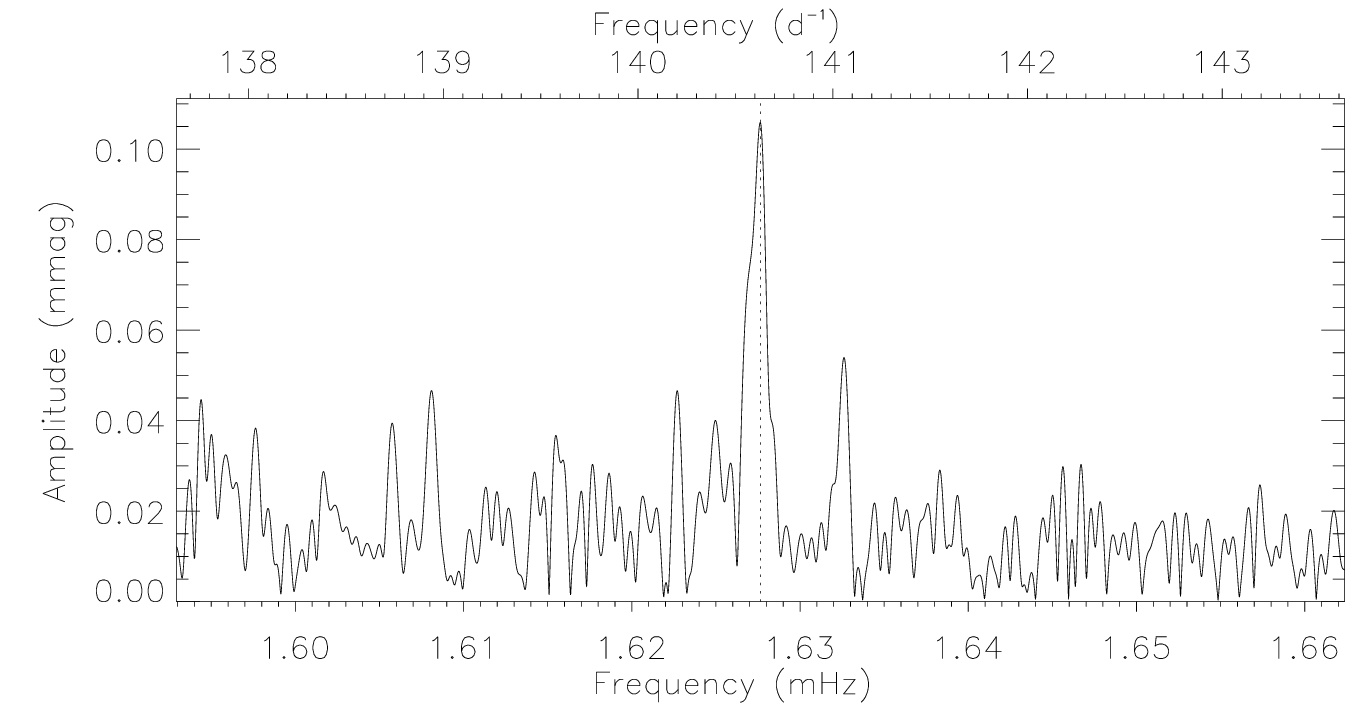}
\caption{Amplitude spectrum for TIC\,96855460. The vertical dotted line identifies the pulsation mode. The distorted shape is likely due to close, unresolved, sidelobes.}
\label{fig:96855460}
\end{figure}

\begin{figure}
\centering
\includegraphics[width=\columnwidth]{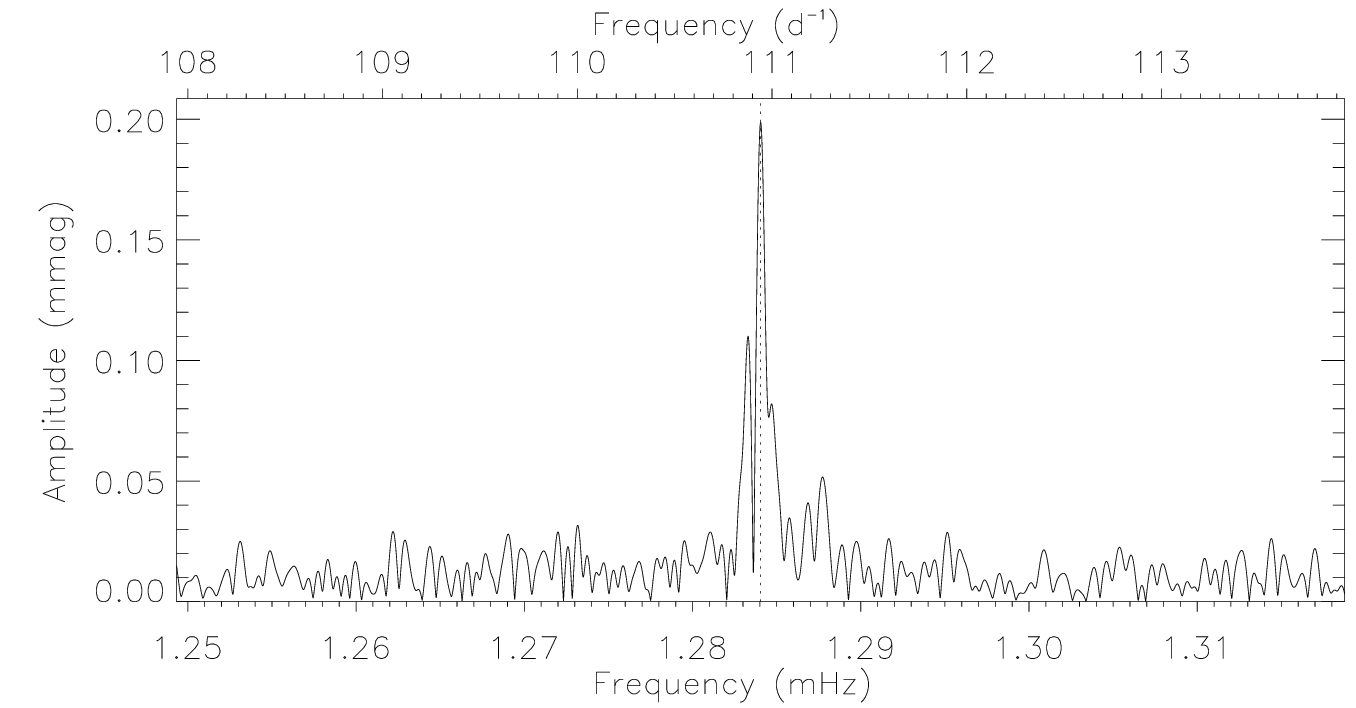}
\caption{Amplitude spectrum for TIC\,100196783. The vertical dotted line identifies the pulsation mode.}
\label{fig:100196783}
\end{figure}

\begin{figure}
\centering
\includegraphics[width=\columnwidth]{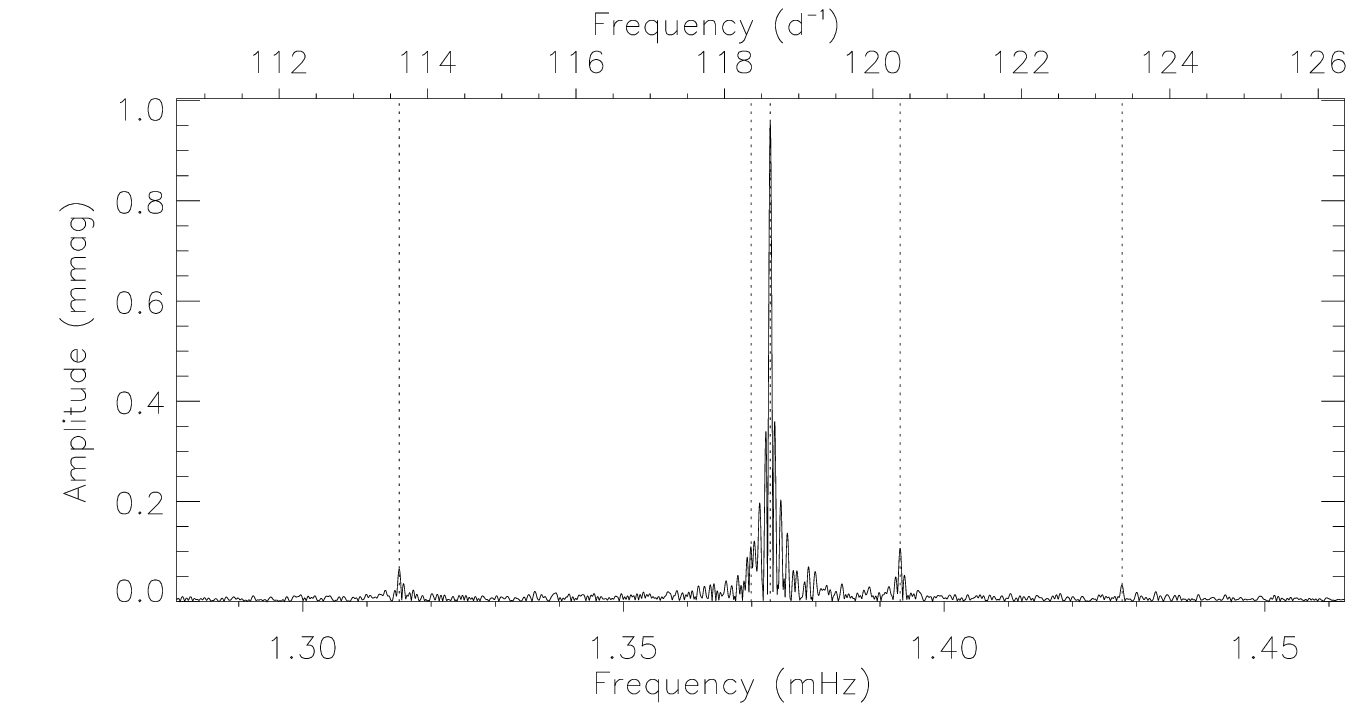}
\caption{Amplitude spectrum for TIC\,163587609. The vertical dotted lines identify the pulsation modes.}
\label{fig:163587609}
\end{figure}

\begin{figure}
\centering
\includegraphics[width=\columnwidth]{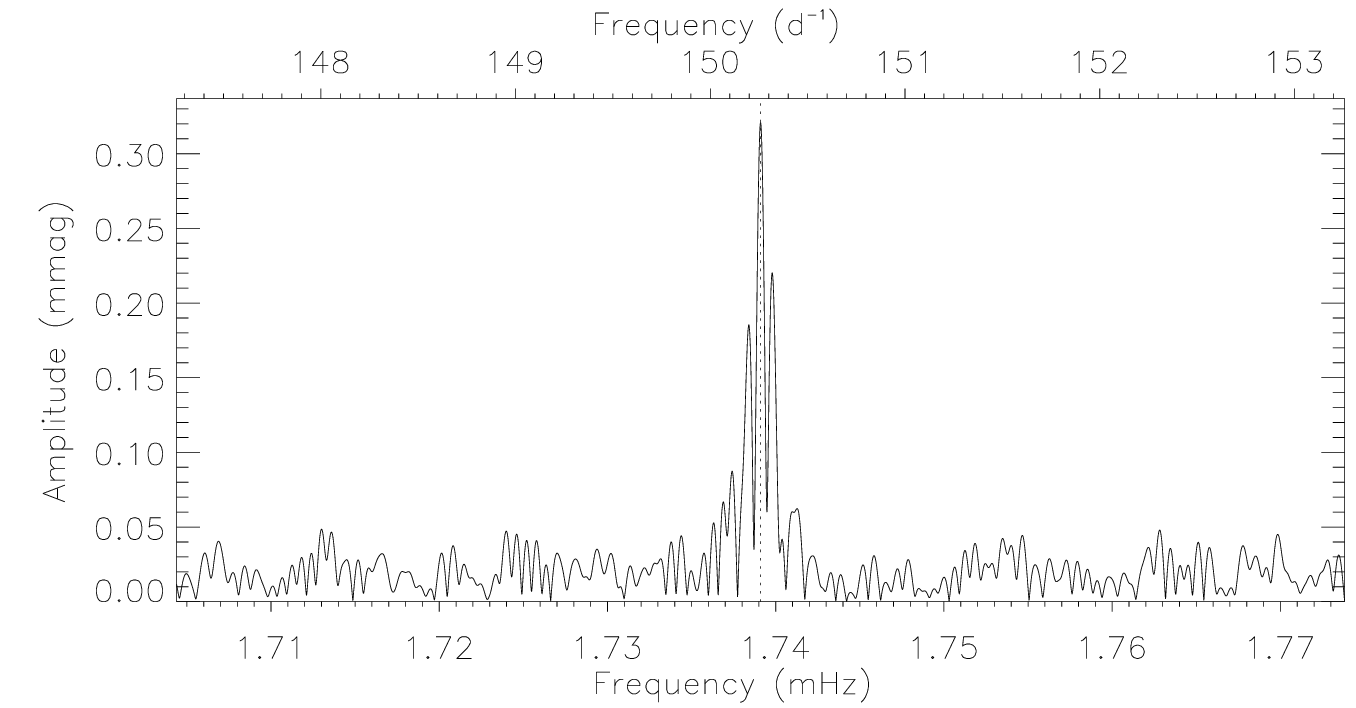}
\caption{Amplitude spectrum for TIC\,293265536. The vertical dotted line identifies the pulsation mode.}
\label{fig:293265536}
\end{figure}

\begin{figure}
\centering
\includegraphics[width=\columnwidth]{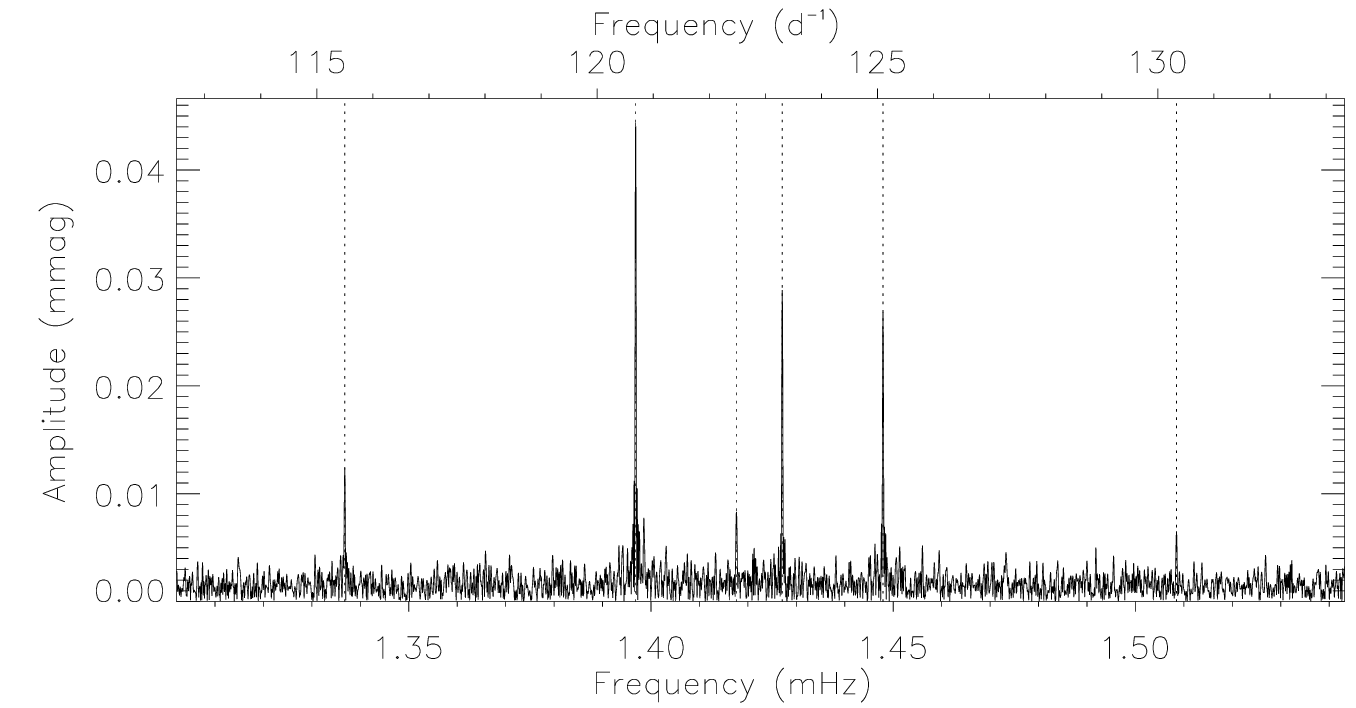}
\caption{Amplitude spectrum for TIC\,299000970. The vertical dotted lines identify the pulsation modes.}
\label{fig:299000970}
\end{figure}

\begin{figure}
\centering
\includegraphics[width=\columnwidth]{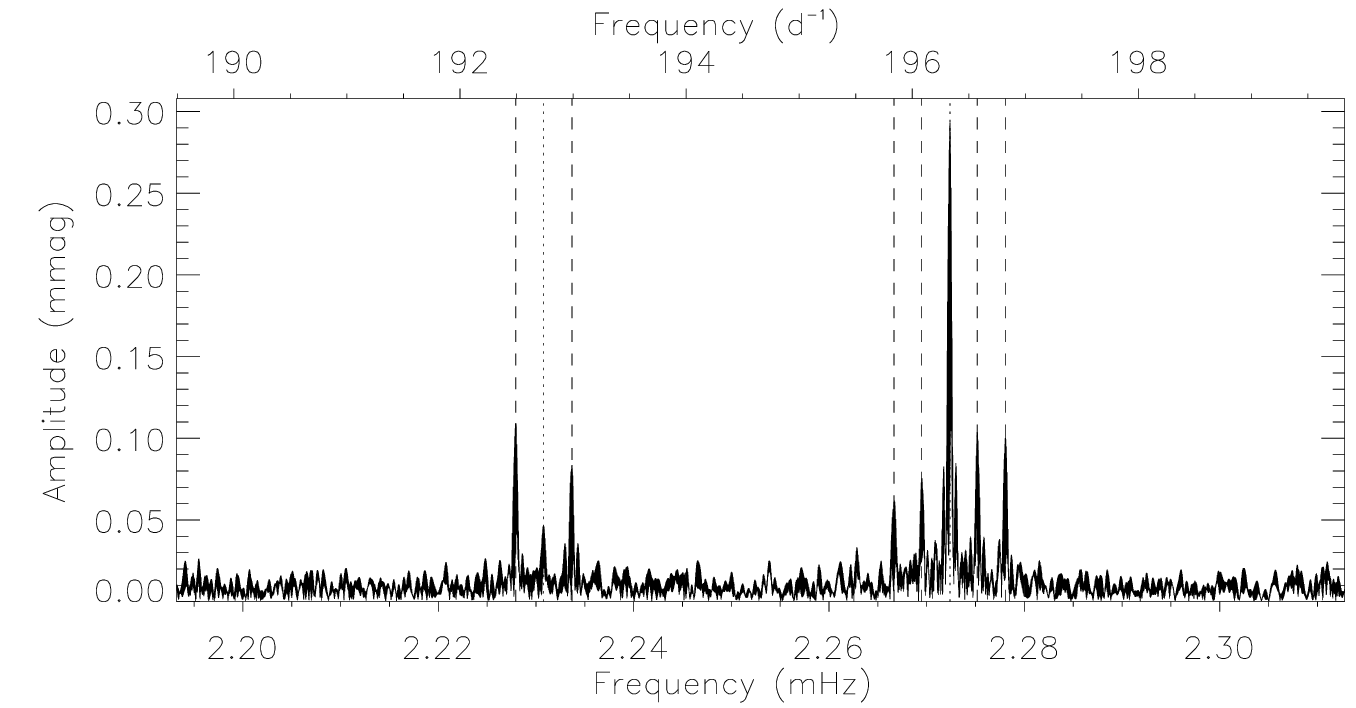}
\caption{Amplitude spectrum for TIC\,318007796. The vertical dotted lines identify the pulsation modes, with the dashed lines showing the rotational sidelobes.}
\label{fig:318007796}
\end{figure}

\begin{figure}
\centering
\includegraphics[width=\columnwidth]{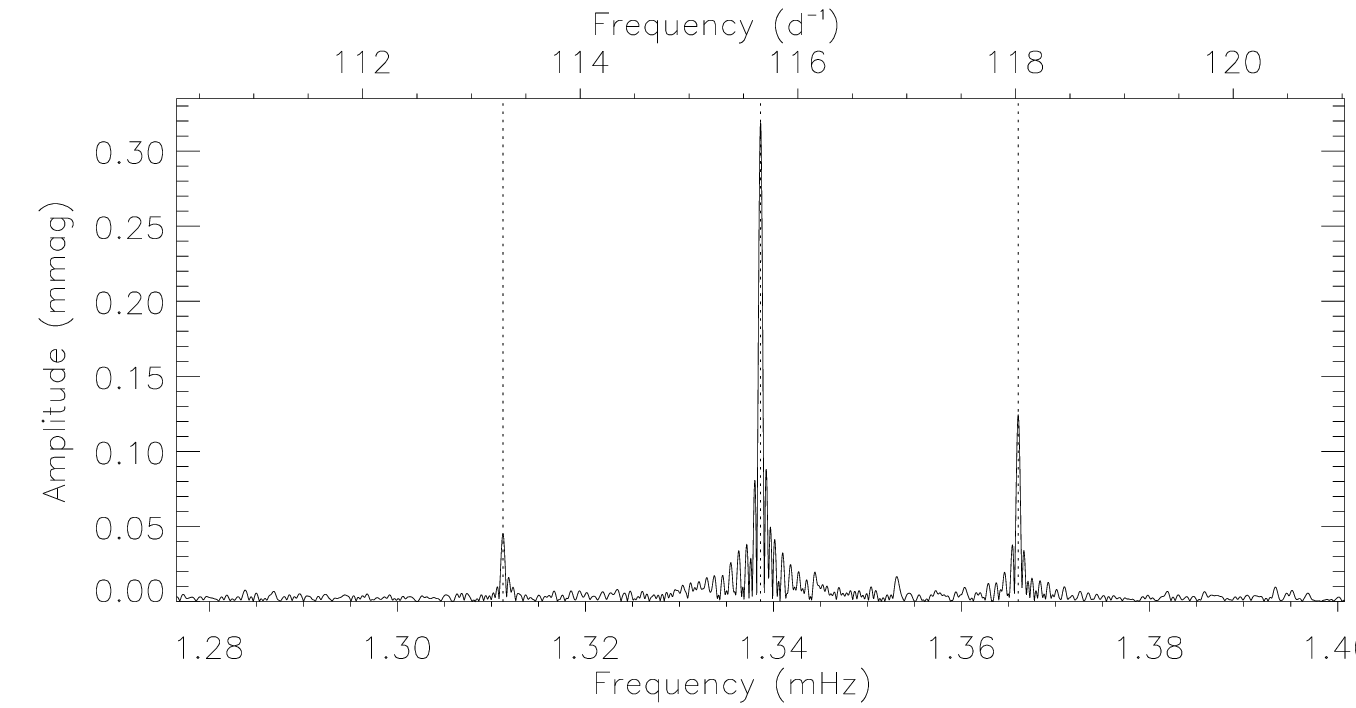}
\caption{Amplitude spectrum for TIC\,354619745. The vertical dotted lines identify the pulsation modes. }
\label{fig:354619745}
\end{figure}

\begin{figure}
\centering
\includegraphics[width=\columnwidth]{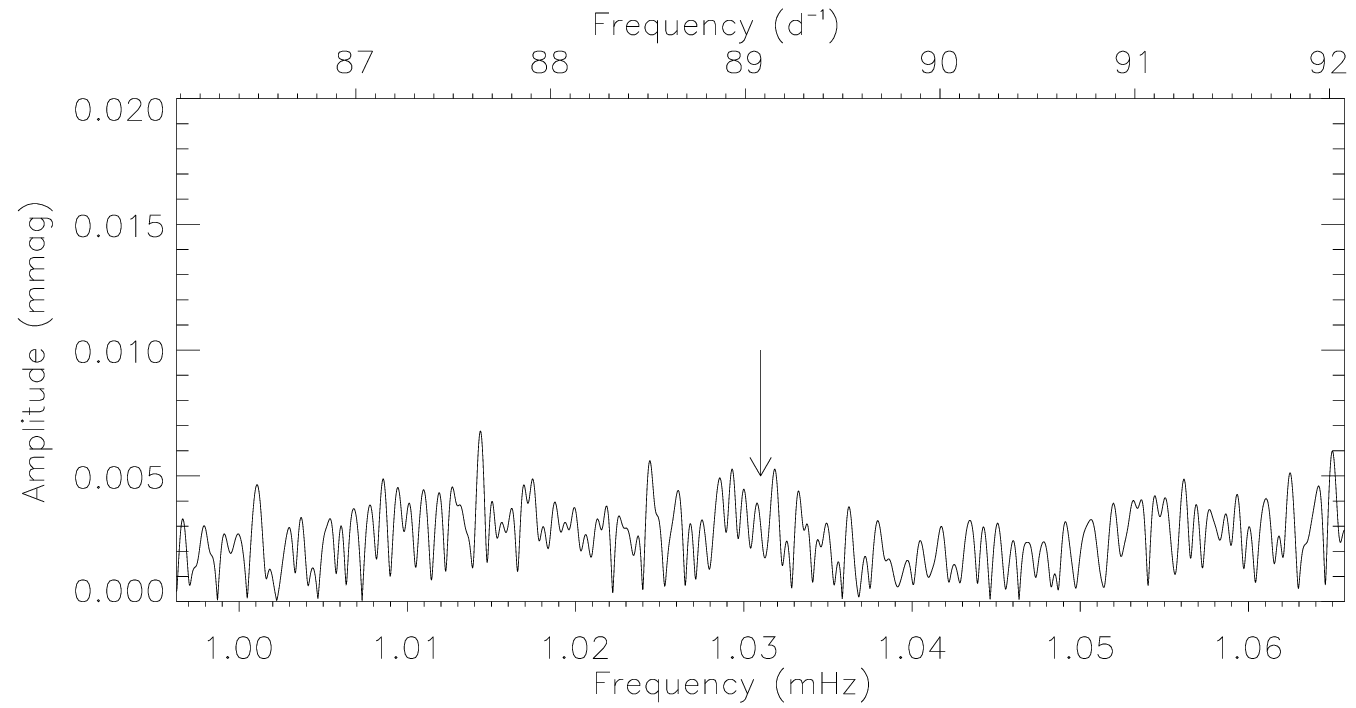}
\caption{Amplitude spectrum for TIC\,383521659 around the frequency range where pulsations were previously reported in spectroscopic data \citep{2007MNRAS.380..741K}. There is no indication of variability in the {\it TESS} data.}
\label{fig:383521659}
\end{figure}

\begin{figure}
\centering
\includegraphics[width=\columnwidth]{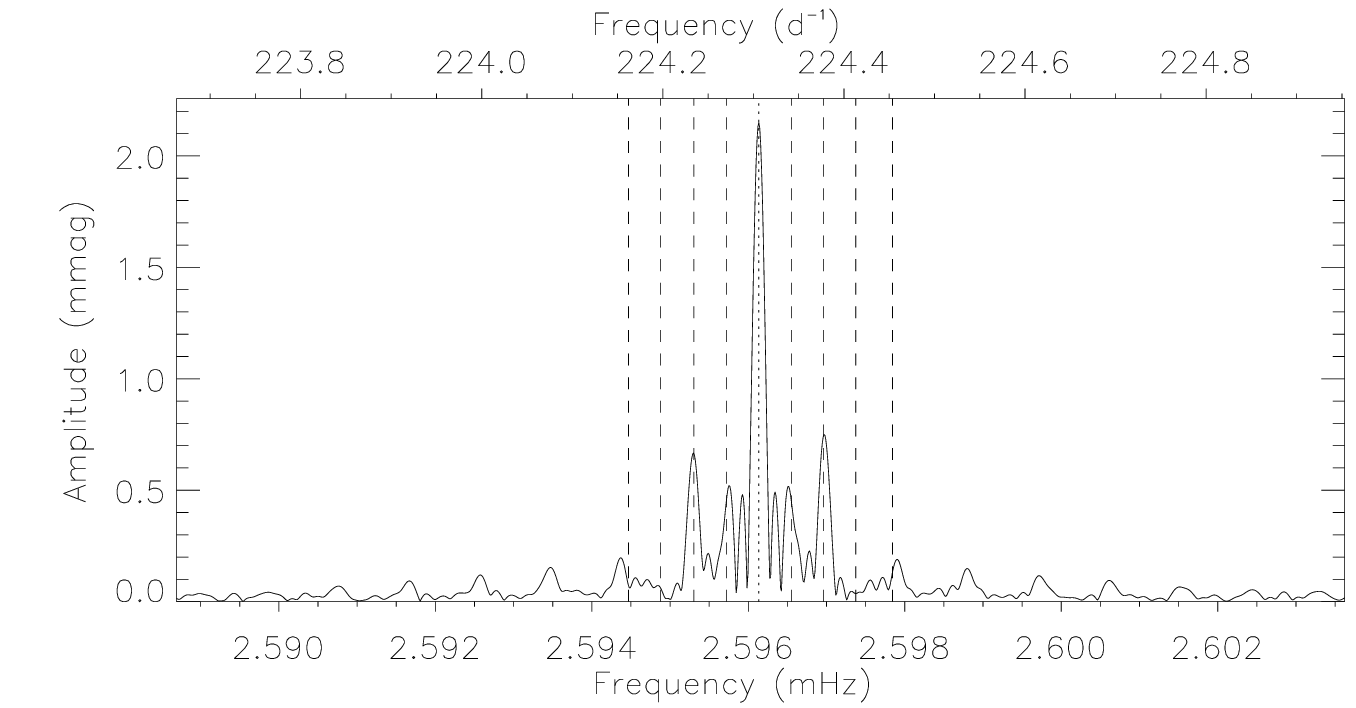}
\caption{Amplitude spectrum for TIC\,407929868. The vertical dotted line identifies the pulsation mode, while the dashed lines show the rotational sidelobes. In this view, the window function of the mode hides some of the lower-amplitude sidelobes.}
\label{fig:407929868}
\end{figure}

\begin{figure}
\centering
\includegraphics[width=\columnwidth]{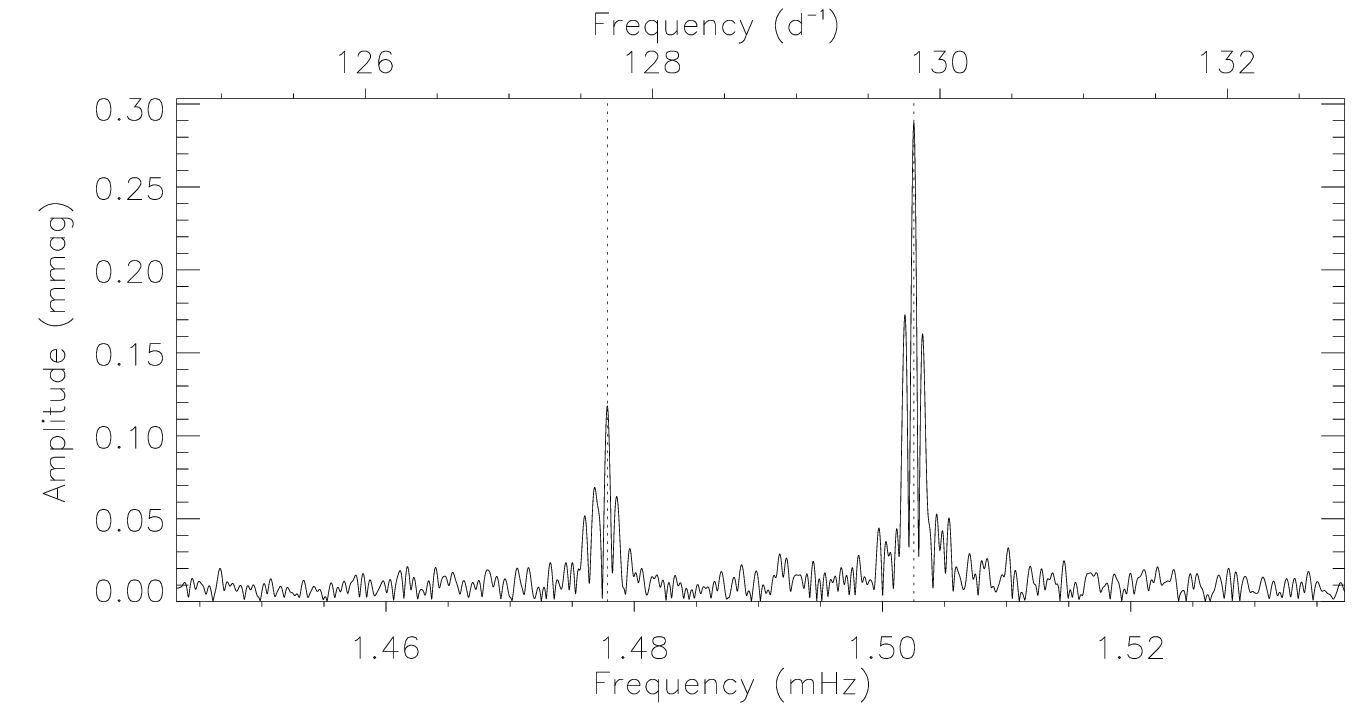}
\caption{Amplitude spectrum for TIC\,420687462. The vertical dotted lines identify the pulsation modes.}
\label{fig:420687462}
\end{figure}

\begin{figure}
\centering
\includegraphics[width=\columnwidth]{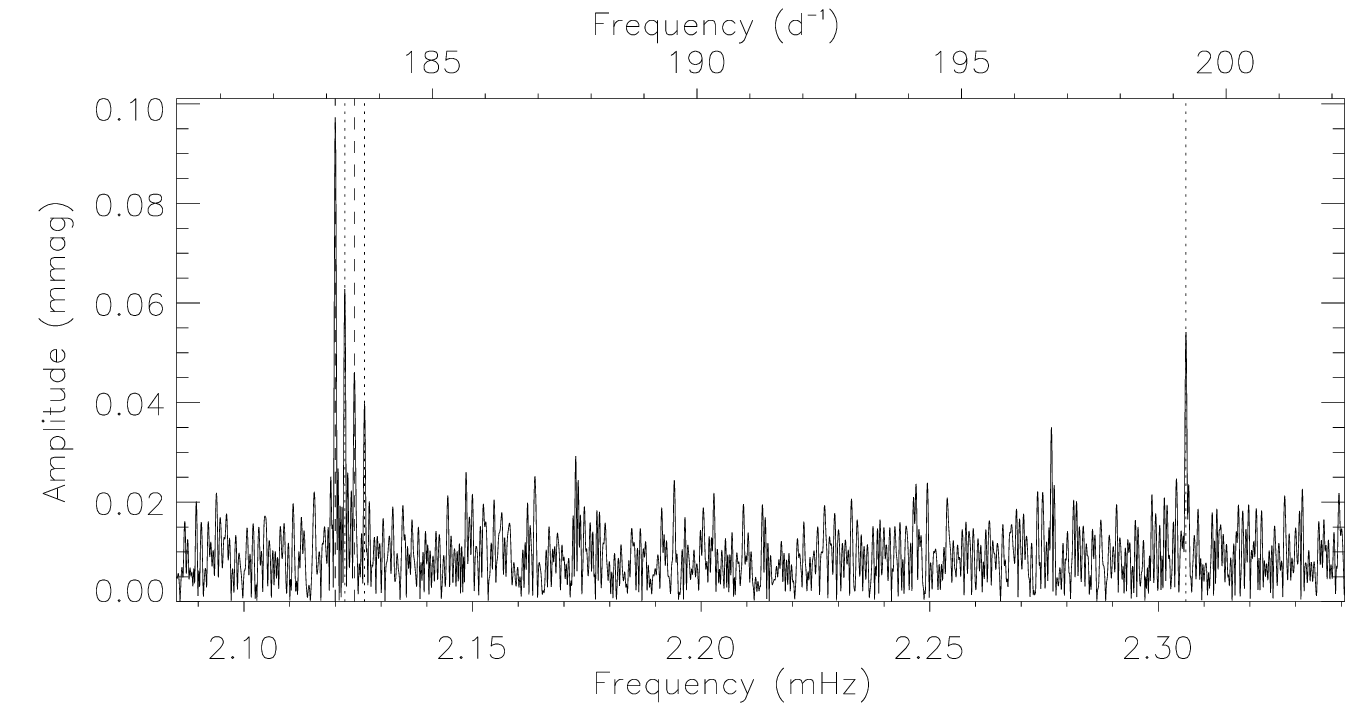}
\caption{Amplitude spectrum for TIC\,445543326. The vertical dotted lines identify the pulsation modes, while the dashed lines show the rotational sidelobes.}
\label{fig:445543326}
\end{figure}

\bsp
\label{lastpage}
\end{document}